\documentclass[reprint,amsmath,amssymb,aps,prb]{revtex4-2}

\usepackage[colorlinks,citecolor=blue,linkcolor=blue]{hyperref}
\usepackage{graphicx}
\usepackage{epstopdf}
\usepackage{amsmath,bm}
\usepackage{threeparttable}
\usepackage{multirow}
\usepackage{booktabs}
\usepackage{color}

\begin{document}

\title{Novel magnetic topological insulator FeBi$_{2}$Te$_{4}$ with controllable topological quantum phase}%
\author{Wen-Ti Guo$^{1,2}$}
\author{Ningjing Yang$^{3}$}%
\author{Zhigao Huang$^{1,2}$}%
\author{Jian-Min Zhang$^{1,2}$}%
\email[Corresponding author]{jmzhang@fjnu.edu.cn}
\affiliation{1 Fujian Provincial Key Laboratory of Quantum Manipulation and New Energy Materials, College of Physics and Energy, Fujian Normal University, Fuzhou 350117, China
}
\affiliation{2 Fujian Provincial Collaborative Innovation Center for Advanced High-Field Superconducting Materials and Engineering, Fuzhou, 350117, China
}
\affiliation{3 School of Physics Science and Technology, Kunming University, Kunming 650214, China}

\begin{abstract}
Here, we report a new intrinsic magnetic topological insulator FeBi$_{2}$Te$_{4}$ based on first-principles calculations and it can achieve a rich topological phase under pressure modulation. Without pressure, we predict that both FeBi$_{2}$Te$_{4}$ ferromagnetic and antiferromagnetic orders are non-trivial topological insulators. Furthermore, FeBi$_{2}$Te$_{4}$ of FM-$z$ order will undergo a series of phase transitions from topological insulator to semimetals and then to trivial insulator under pressure. Finally, we further clarify and verify topological phase transitions with low-energy effective model calculations. This topological phase transition process is attributed to the synergy of the magnetic moment and the spin-orbit coupling. The unique topological properties of FeBi$_{2}$Te$_{4}$ will be of great interest in driving the development of quantum effects.
\end{abstract}

\maketitle

\section{Introduction}
The combination of magnetism and topology will obtain many topological quantum materials, such as Weyl semimetal, Chern insulator, and axion insulator\cite{PhysRevB.81.245209,Tokura2019,Otrokov2019,Zhang2019,Li2019,Otrokov2019b,Liu2020,Deng2020}. Moreover, the effective ways to achieve topological and magnetic coupling are magnetic doping\cite{doi:10.1126/science.1187485,doi:10.1126/science.1189924, doi:10.1126/science.1234414,YunboOu2017}  and construction of magnetic/topological heterostructures\cite{PhysRevB.87.085431,Katmis2016,LI201977,Hirahara2017,Eremeev2018}. However, the above two approaches will correspondingly to problems leading to inhomogeneous magnetic regions and strong dependence on interfacial properties\cite{PhysRevB.87.085431,YunboOu2017,Wang2021}. 
Therefore, it is crucial to find a material that possesses magnetic and topological properties stoichiometrically.

Recently, an intrinsic magnetic topological insulator MnBi$_{2}$Te$_{4}$ has been reported, which possesses rich topological phases\cite{Otrokov2019,Zhang2019,Li2019,Otrokov2019b,Liu2020,PhysRevB.100.121104}. Its ferromagnetic (FM) and antiferromagnetic (AFM) orders are Weyl semimetal and axial insulator, respectively\cite{Zhang2019,Li2019}. 
Due to its typical van der Waals layered structure, MnBi$_{2}$Te$_{4}$ will exhibit Te-Bi-Te-Mn-Te-Mn-Te septuple layers (SLs) thin films. Further studies show that multiple heptad layered films of MnBi$_{2}$Te$_{4}$ have parity number dependence and may have quantum anomalous Hall effect and topological magnetoelectric effect, respectively\cite{Liu2020,PhysRevX.11.011003}. A series of multilayer MnBi$_{2m}$Te$_{1+3n}$ materials based on the MnBi$_{2}$Te$_{4}$ parent material derived by inserting different numbers of Bi$_{2}$Te$_{3}$ quintuple layers has also been widely investigated\cite{Sun2019,Yan2020,PhysRevB.102.035144,Klimovskikh2020,Hu2020,Hu2020b,Wu2020,Jo2020,Xu2020,Xie2020,Vidal2021,Jia2021,Chen2021,Lu2021}.
Although MnBi$_{2}$Te$_{4}$ has been a breakthrough in achieving novel physical phenomena, the experimental synthesis of the material is very demanding, slow-growing, and prone to sub-stability. In addition, the physical phenomena associated with MnBi$_{2}$Te$_{4}$ materials are hard to realize, especially since the observed temperature of the quantum anomalous Hall effect (QAHE) is extremely low\cite{Shi2020,Rani2019,Gong2019,Qi2020,He2020}. Although QAHE was observed in MnBi$_{2}$Te$_{4}$ thin film samples at 1.4 K\cite{Deng2020}, many experimental results failed to reach the quantum conductivity plateau\cite{Hu2020b,Wu2019,Vidal2019}, and no directly grown epitaxial films have been available to realize this effect\cite{Sun2021}. Therefore, the topological magnetic community has devoted their research to enhance the temperature for realizing QAHE. The observed temperature of QAHE can increase by doping\cite{PhysRevB.100.104409,PhysRevMaterials.4.064411,Liu2020,Deng2020,Hou2021,Zhang2021,Golias2021,Wimmer2021,doi:10.1021/acs.nanolett.0c05117,10.1093/nsr/nwaa089,Singh2021,Jiang2021,Watanabe2022}, building magnetic insulator/topological insulator heterostructures\cite{Qi2020,Zhu2020,Fu2020,Gao2021,Li2020,Yan2021}, and applying external magnetic fields\cite{10.1093/nsr/nwaa089,Yan2020,10.1093/nsr/nwaa089,Jo2020,PhysRevB.102.035144}. Sun \textit{et al}. established a topological phase diagram for triggering QAHE by tuning slab thickness and magnetization\cite{Sun2019}. However, the lack of stability of the MnBi$_{2}$Te$_{4}$ material has resulted in epitaxial films that are not yet directly grown\cite{Sun2021}. In disagreement with earlier theoretical and experimental results\cite{Otrokov2019}, later experimental results\cite{PhysRevX.9.041038,PhysRevX.9.041040,PhysRevB.101.161109} observed energy-gapless Dirac cones in the topological surface states of AFM MnBi$_{2}$Te$_{4}$. In addition, the dispersion relation of electrons in MnBi$_{2}$Te$_{4}$ is not consistent with the theoretical calculation\cite{doi:10.1021/acs.nanolett.0c00031}.

\begin{figure*}
	\begin{centering}
		\includegraphics[width=1.0\textwidth]{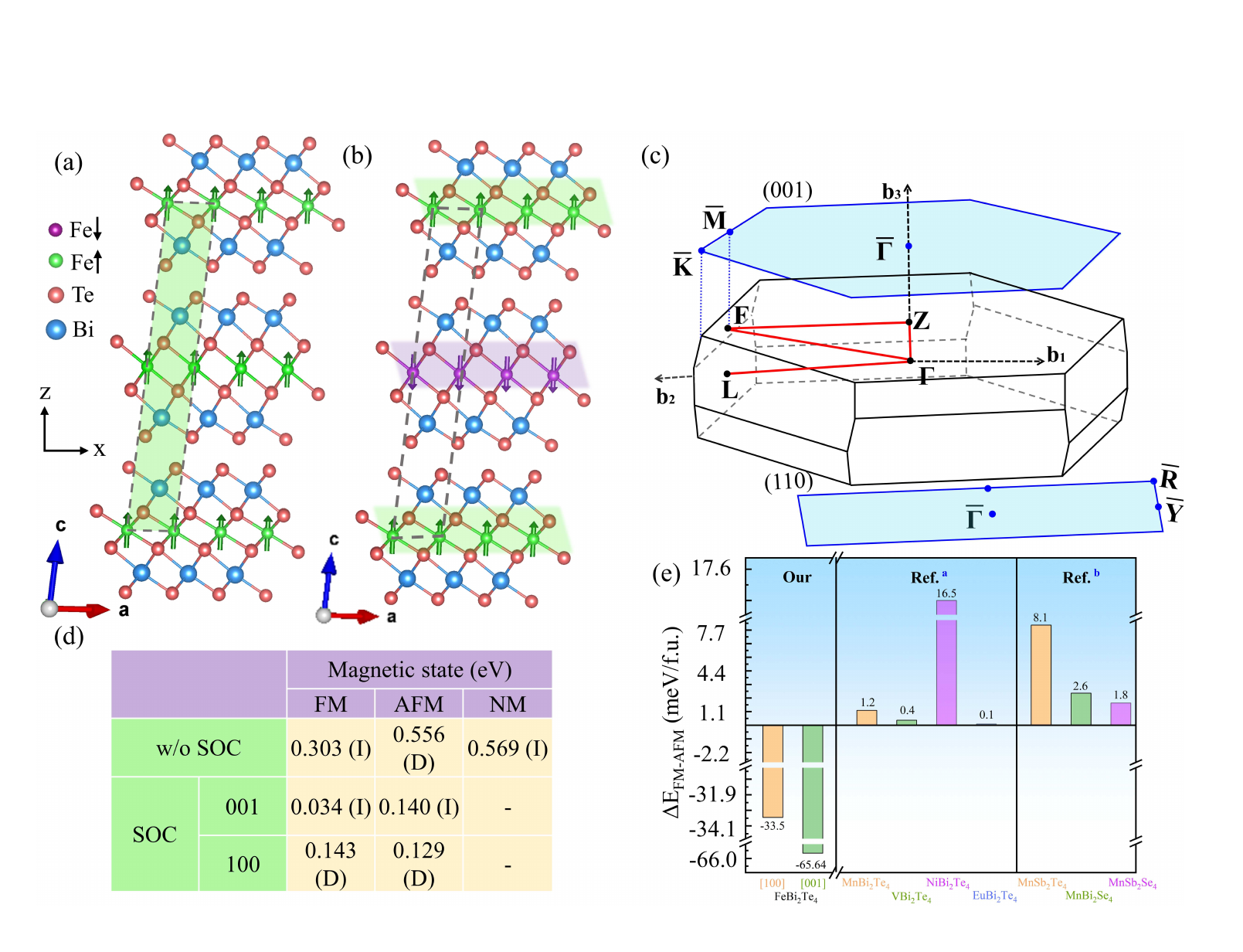}
		\par\end{centering}
	\centering{}\caption{\textbf{Crystal structures and energies of FeBi$_{2}$Te$_{4}$ in different orders.} Crystal structures of bulk FeBi$_{2}$Te$_{4}$ (a) FM ordering and (b) A-type AFM ordering. (c) The First Brillouin zone with (001) and (110) planes. (d) Comparison of the global band gap of the bulk FeBi$_{2}$Te$_{4}$ for different magnetic orders (FM, AFM and NM) with and without SOC, where the magnetic moment orientations [001] and [100] are calculated when SOC effects are considered. The direct band gap and indirect band gap are marked with D and I respectively in parentheses. (e) The relative energies of FM and AFM magnetic orders $\Delta$E$_{FM-AFM}$, where positive values indicate lower energy in the AFM order. To visualize the energy difference values close to 0 (like EuBi$_{2}$Te$_{4}$), we break the axes at certain positions. We considered AFM orders in both [100] and [001] directions, while the AFM states of the cited references a\cite{Li2019} and b\cite{Zhang2021} are along the [001] direction.}
	\label{1}
\end{figure*}
\begin{figure*}
	\begin{centering}
		\includegraphics[width=1.0\textwidth]{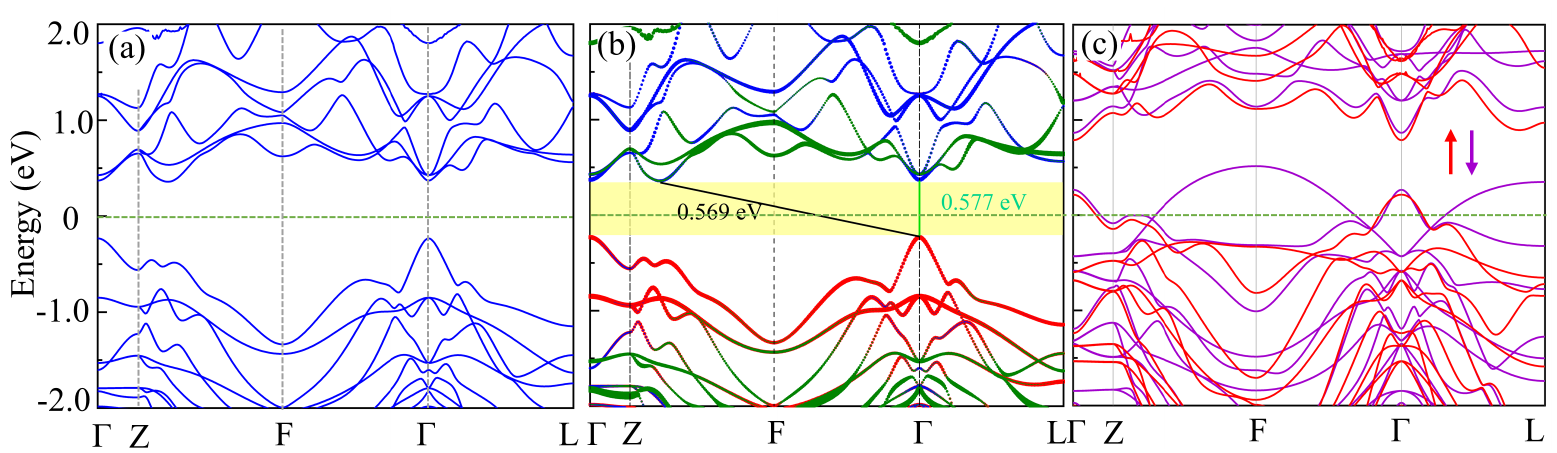}
		\par\end{centering}
	\centering{}\caption{\textbf{Band structures of FeBi$_{2}$Te$_{4}$. } (a) Band structure and (b) projected band structure of FeBi$_{2}$Te$_{4}$ in NM state. (c) Band structure without SOC of FeBi$_{2}$Te$_{4}$ in FM state.}
	\label{2}
\end{figure*}

\begin{figure*}
	\begin{centering}
		\includegraphics[width=0.65\textwidth]{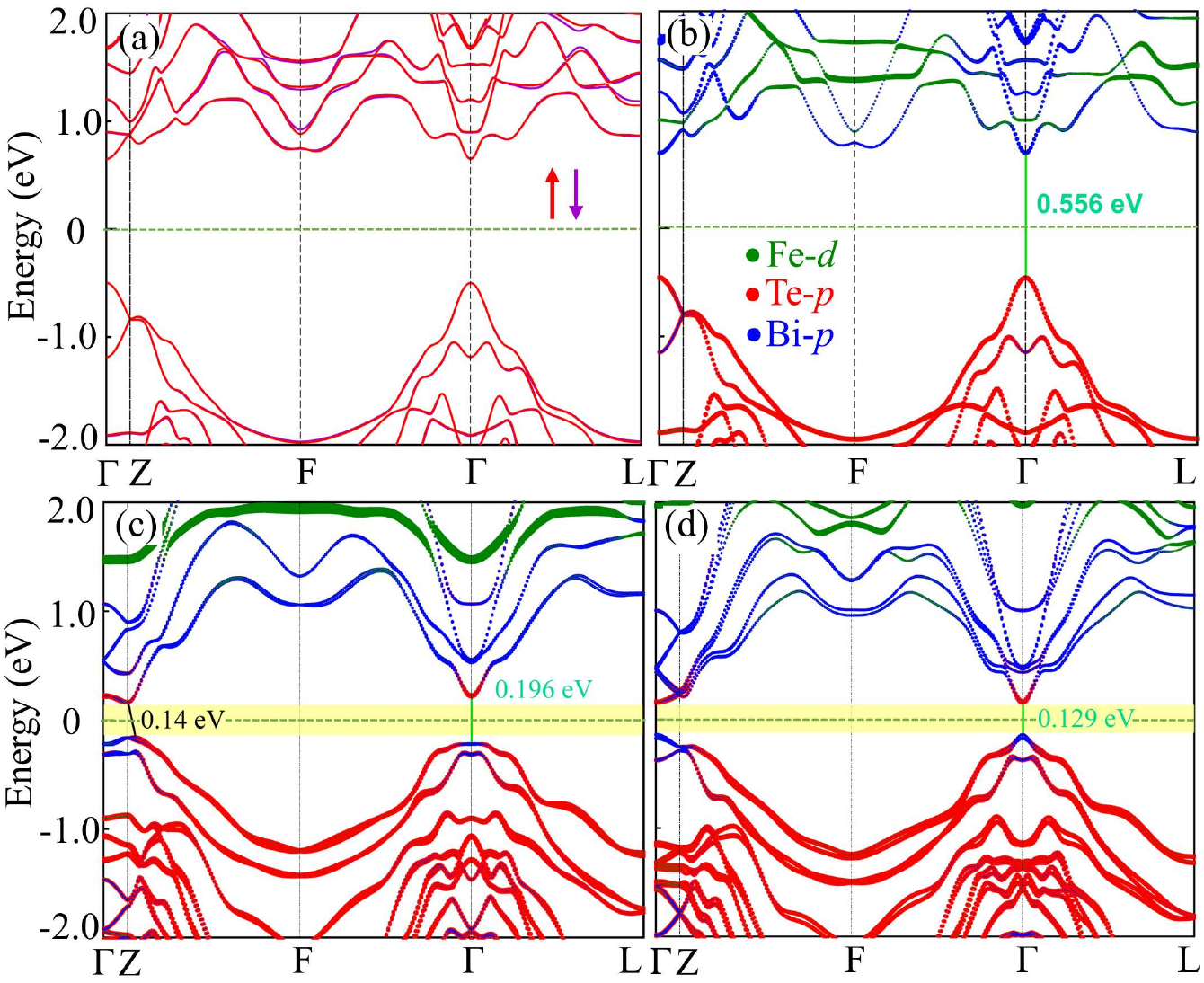}
		\par\end{centering}
	\centering{}\caption{\textbf{Band structures of FeBi$_{2}$Te$_{4}$ in AFM order.} (a) Bulk band structure and (b) projected band structure of AFM order FeBi$_{2}$Te$_{4}$ without SOC. Projected band structures of (c) AFM-$z$ and (d) AFM-$x$ orders. The black and green lines in the projected band structure indicate the global band gap and the local band gap at $\Gamma$, respectively, and in addition, the global band gap size is visualized by the yellow shading.}
	\label{3}
\end{figure*}

Meanwhile, the magnetic order of the MnBi$_{2}$Te$_{4}$ series materials is also controversial. For example, MnSb$_{2}$Te$_{4}$ may exhibit FM phase\cite{Wimmer2021}, AFM phase\cite{Huan2021,EREMEEV2017172,PhysRevB.100.104409,PhysRevMaterials.4.064411}, and a Weyl semimetal with ferrimagnetic properties\cite{PhysRevB.100.195103,Shi2020}. MnBi$_{6}$Te$_{10}$ was reported to be in the FM\cite{Xie2020} and AFM\cite{Jo2020} orders. Complex competing magnetic orders greatly hinder the application of intrinsically magnetic topological insulators. The same as MnBi$_{2}$Te$_{4}$, Li \textit{et al}. predicted that monolayer MBi$_{2}$Te$_{4}$ is an FM insulator for M=Ti, V, Mn, Ni, Eu, while it may present an unstable metallic state for M=Cr, Fe, Co\cite{Li2019}. Individual studies have also reported this series of materials one after another\cite{Li2020,Saxena2020,Zhu2020,doi:10.1021/acs.jpclett.1c02396,PhysRevB.105.214304}. Uplifting, it has been reported that FeBi$_{2}$Te$_{4}$ single crystal has been successfully synthesised\cite{Saxena2020}, which was shown to be stable. A study calculated the phonon dispersion spectrum, also confirming that FeBi$_{2}$Te$_{4}$ can be stably present\cite{PhysRevB.105.214304}. Meanwhile, Wang $et al.$ reported that FeBi$_{2}$Te$_{4}$ tends to form topologically trivial 120$^{\circ}$ non-collinear antiferromagnetic states. And they also propose the possibility of realising QAHE in the bilayer ferromagnetic FeBi$_{2}$Te$_{4}$\cite{10.1088/1674-1056/acd522}. However, as a new potential intrinsic magnetic topological insulator, the changes in the electronic structure and topological properties of the linear magnetic order FeBi$_{2}$Te$_{4}$ system under pressure are still unclear.

This work theoretically investigates the topological properties of different magnetic orders of FeBi$_{2}$Te$_{4}$ by first-principles calculation.
Firstly, we confirm that both AFM-$x$ and AFM-$z$ orders of FeBi$_{2}$Te$_{4}$ are intrinsic magnetic topological insulators with gapless surface states in (110) surface. In contrast, we find that modulating the in-plane to out-of-plane antiferromagnetism will break the Mirror symmetry and thus induce the (001) plane of FeBi$_{2}$Te$_{4}$ to acquire gapped surface states. Furthermore, the surface states of the FM-$x$ and FM-$z$ orders ((110) surface) are connected and separated from the conduction band, respectively. Remarkably, we find that magnetic ground state FM-$z$ ordering has a small bulk band gap and can be tuned to be a topological Weyl semimetal and normal insulator under slight pressure, which achieves the phase transition from topological insulator to topological semimetal then to a trivial insulator. Moreover, we further construct the low-energy effective model to study the essential reasons for the series of topological phase transitions under pressure regulation. These intriguing topological states in FeBi$_{2}$Te$_{4}$ not only deepen our understanding of magnetic topology physics but also can broaden its applications in different spin quantum devices.

\section{Methods}\label{sec:2}
First-principles calculations are performed in Vienna \textit{ab initio} simulation package (VASP) \cite{PhysRevB.48.13115,PhysRevB.54.11169} via using all-electron projected augmented wave (PAW) \cite{PhysRevB.50.17953} method and Perdew-Burke-Ernzerhof (PBE) type generalized gradient approximation (GGA) \cite{PhysRevLett.77.3865} exchange-correlation function. The valance wave functions are expanded on the plane-wave basis with a cutoff energy of 450 eV for energy calculations in the magnetic ground state. Our calculations, except for the nonmagnetic state, all take into account spin-orbit coupling (SOC). FeBi$_{2}$Te$_{4}$ has R$\bar3$m space group, and the optimized lattice constant is slightly smaller (\textit{a}= 4.325 \AA). 
For the 3$\textit{d}$ orbit of Fe, the on-site coulombic is chosen to be 4 eV and the magnetic moment is set to 4 $\mu_{B}$. 
We used the same kind of cell for all our co-linear magnetic state calculations (e.g., Fig. \ref{1}(a)-\ref{1}(b)). Furthermore, similar to MnBi$_{2}$Te$_{4}$, we chose the cell as a tilted conventional cell where the angle between the in-plane basis vectors ($a,b$) and the out-of-plane basis vectors ($c$) is not 90 $^{\circ}$.
The structures are relaxed with a conjugate-gradient algorithm till the energy on atoms is less than 1$\times$10$^{-5}$ eV in all our calculations. The $\Gamma$-centered Monkhorst-Pack $\textit{k}$-point mesh is chosen to 11$\times$11$\times$5 and the force on each atom is set to 1$\times$10$^{-4}$ for the energy difference calculation. 
The vdW correction of the DFT-D3\cite{10.1063/1.3382344} method is considered in all our calculations.
We reproduce band structure obtained by constructing a tight-binding model based on Maximum Local Wannier Functions (MLWFs)  \cite{PhysRevB.56.12847,souza2001maximally}. Based on the surface Green's function approach  \cite{lee1981simple1,lee1981simple2,sancho1984quick,sancho1985highly}, our calculations of Wannier Charge Center (WCC), surface states, spin-texture, and Fermi surfaces (FS) of the semi-infinite surfaces are performed via WannierTools package \cite{wu2018wanniertools}. Our topological classifications are also based on irvsp\cite{GAO2021107760} and a generated workflow website\cite{PhysRevB.106.035150}, which obtain Irreducible Representations of electronic states and compute symmetry indicators (SI) to determine the topological classification of magnetic materials.

\section{Results and discussion}\label{sec:3}

\begin{figure*}
	\centerline{\includegraphics[width=0.65\textwidth]{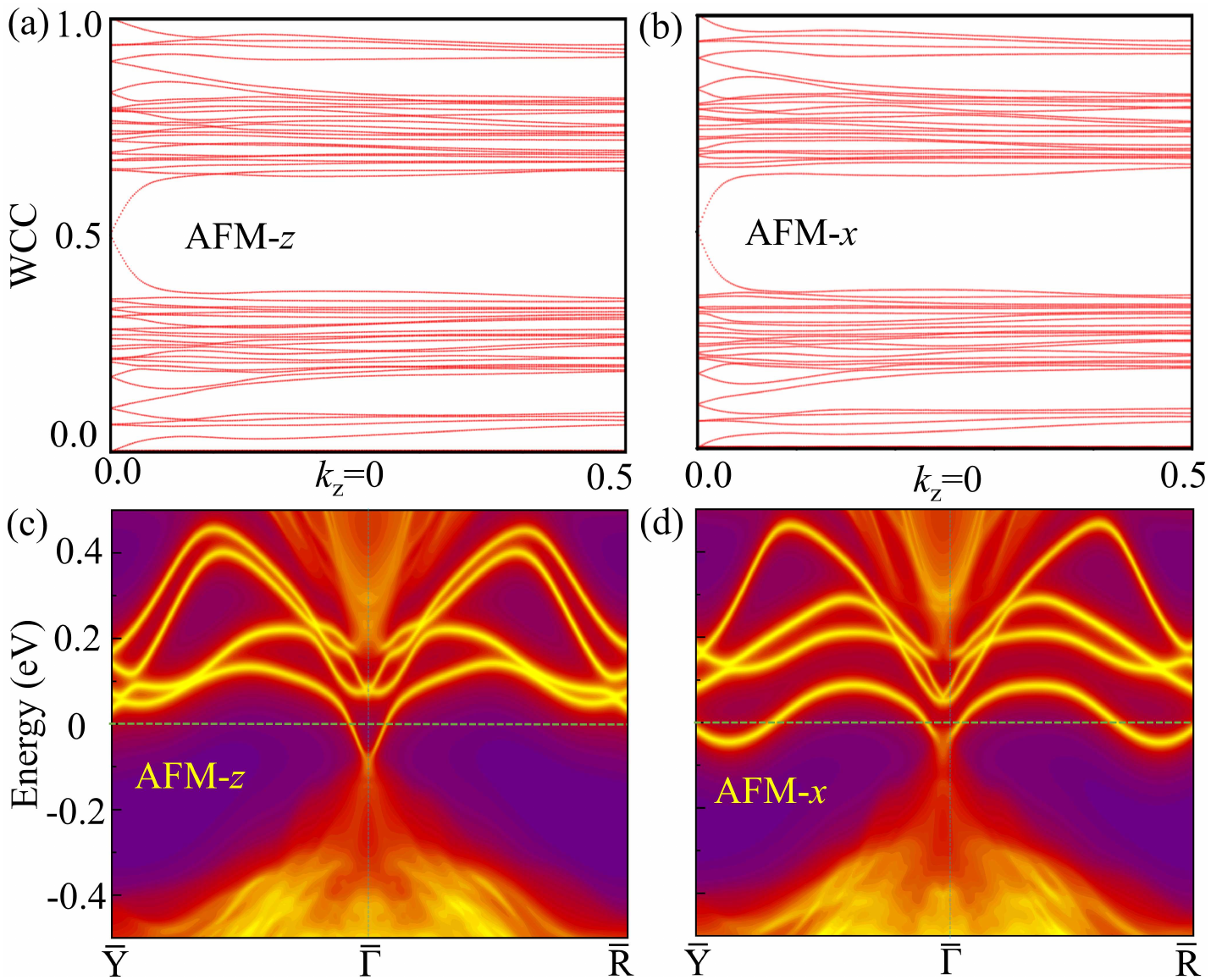}}
	\centering{}\caption{\textbf{Topological properties of AFM-$z$ and AFM-$x$ orders FeBi$_{2}$Te$_{4}$.} WCC calculation results for two configurations considering SOC for (a) AFM-$z$ and (b) AFM-$x$. Topological surface states in (110) termination of (c) AFM-$z$ and (d) AFM-$x$ orders.}
	\label{4}
\end{figure*}

\begin{figure*}
	\begin{centering}
		\centering{\includegraphics[width=0.65\textwidth]{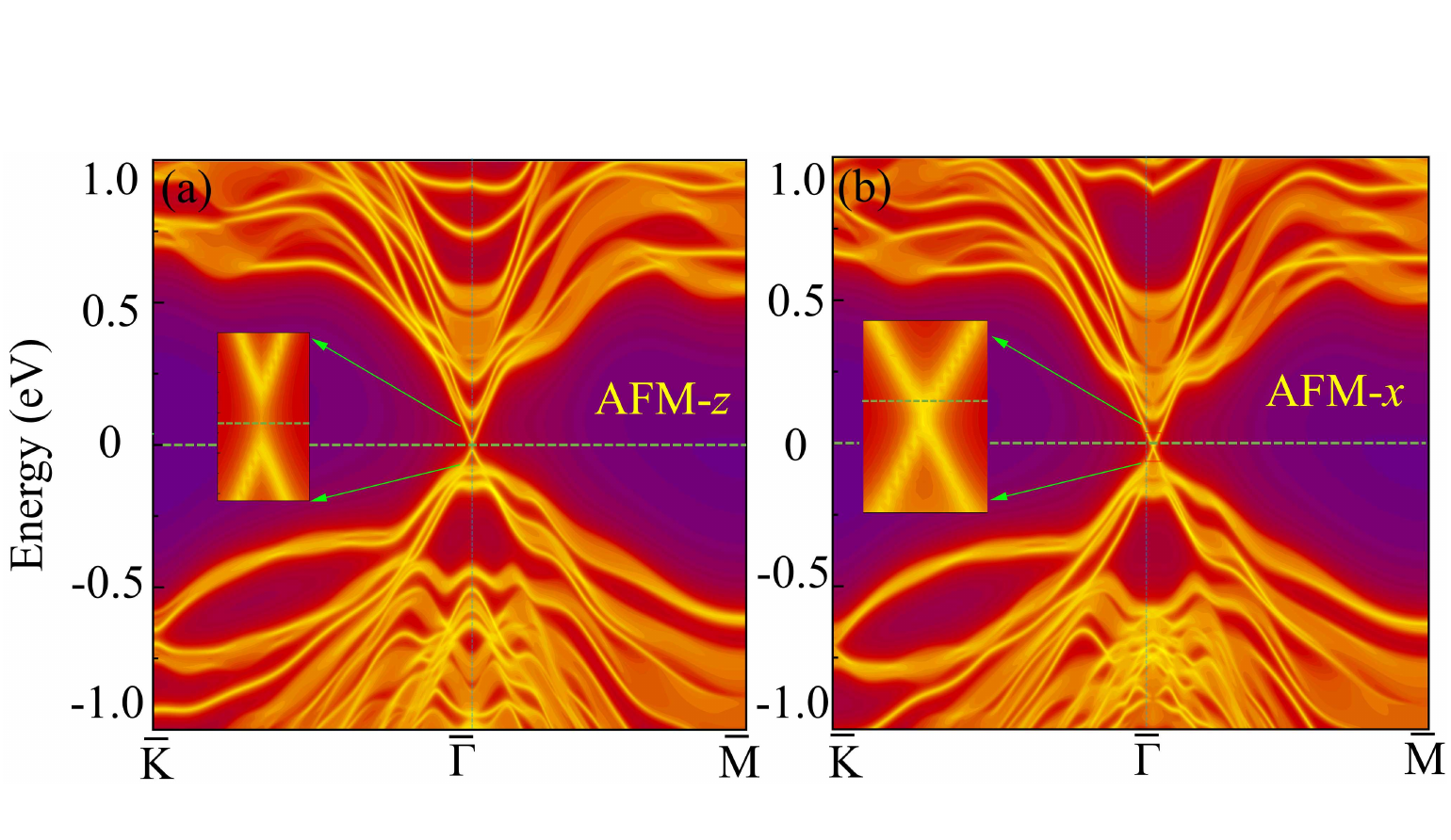}}
		\par\end{centering}
	\centering{}\caption{\textbf{Topological properties of AFM-$z$ (AFM-$x$) FeBi$_{2}$Te$_{4}$.}  Topological surface states in (001) termination of (c) AFM-$z$ and (d) AFM-$x$ orders.}
	\label{5}
\end{figure*}

\subsection{Stability and topological properties of the intrinsic magnetic topological insulator FeBi$_{2}$Te$_{4}$}
We first evaluate their stability for the different magnetic configurations of FeBi$_{2}$Te$_{4}$ (e.g. AFM-$x$, AFM-$z$, FM-$x$, and FM-$z$). Figures \ref{1}(a), \ref{1}(b), and \ref{1}(c) show the structures and their Brillouin zone of FeBi$_{2}$Te$_{4}$ with a magnetic order in FM and AFM, respectively. The arrows in the crystal structure indicate the magnetic moments along the $z$-direction for FM and AFM, respectively. The FM-$x$ and AFM-$x$ orders only require the magnetic moments along the $x$ direction. The schematic diagram of the magnetic moment settings in $z$ and $x$ directions is shown by the arrows on the left side of Fig. \ref{1}(a). All our antiferromagnetic sequences are calculated using the A-type antiferromagnetic configuration, which is the intra-layer ferromagnetic while inter-layer antiferromagnetic form. In both the magnetic configuration in the $z$ and $x$ directions, the interlayer exchange interaction ($\Delta$E$_{FM-AFM}$) is defined as the energy difference between the bulk structures of FM and AFM bulk per formula unit, respectively. As shown in Figs. \ref{1}(d) and \ref{1}(e), the band gaps and unit cell energy differences between the different magnetic orders of FeBi$_{2}$Te$_{4}$ were first calculated theoretically. Figure \ref{1}(d) reveals that the band gaps of both FM and AFM orders become narrower due to SOC, while the non-magnetic (NM) state without SOC has the largest indirect band gap as the black arrow shown in Fig. \ref{2}(b). When considering SOC, both FM and AFM orders will behave as direct and indirect band gaps if the [100] ($x$-direction) and [001] ($z$-direction) directions are set. Combined with Figs. \ref{1}(d) and \ref{1}(e), the FM-$z$ and FM-$x$ orders with lower energy in the FM and AFM orders, which are more stable, have indirect band gaps (0.034 eV) and direct band gaps (0.143 eV), respectively. As shown in Fig. \ref{1}(e), compared to MBi$_{2}$Te$_{4}$ (M=Mn, V, Ni, Eu)\cite{Li2019} and MnX$_{2}$Y$_{4}$ (X=Sb, Bi, Y=Se, Te)\cite{Zhang2021}, FeBi$_{2}$Te$_{4}$ tends to present an FM order in the [100] direction, which is about 33.5 meV/f.u. lower than the AFM order. Similarly, FeBi$_{2}$Te$_{4}$ prefers to form FM order in the [001] direction, which is quite different from the MnBi$_{2}$Te$_{4}$ with AFM-$z$ ground state. Furthermore, we find that the energy of the FM-$z$ magnetic ordering is 2.7 meV lower than that of the FM-$x$ magnetic ordering base on the same unit-cell calculation. This intrinsic ferromagnetism does not appear in the remaining materials enumerated in Fig. \ref{1}(e). The intrinsic ferromagnetism of FeBi$_{2}$Te$_{4}$ might facilitate the realization of physical phenomena such as QAHE. Notably, a critical recent study reported a new magnetic configuration of non-collinear frustrated 120 $^{\circ}$ AFM in the ground state.\cite{10.1088/1674-1056/acd522}. 

By investigating the electronic structure and topological properties of the bulk FeBi$_{2}$Te$_{4}$ AFM orders, we found that they both exhibit non-trivial topological features. From Figs. \ref{3}(a) and \ref{3}(b), the band gap (0.556 eV) of the FeBi$_{2}$Te$_{4}$ AFM order without SOC is slightly smaller than that of the NM state shown in \ref{2}(a) and \ref{2}(b). Their conduction and valence bands can be recognized as mainly contributed by Bi-\textit{p}/Fe-\textit{d} and Te-\textit{p}, respectively. When SOC effects are considered, as in Figs. \ref{3}(c) and \ref{3}(d), both AFM-$z$ and AFM-$x$ orders undergo band inversion (Bi-\textit{p} and Te-\textit{p}) near $\Gamma$, predicting a non-trivial topological property. AFM-$z$ has a larger direct band gap of 0.196 eV at $\Gamma$ than 0.129 eV of AFM-$x$ (green line in the projected band structure diagrams). Nevertheless, due to the mutual proximity of the conduction and valence bands near the Z high-symmetry point, AFM-$z$ presents an indirect band gap of 0.14 eV (yellow shading indicates the global band gap), different from the global direct band gap 0.16 eV at Z for MnBi$_{2}$Te$_{4}$\cite{Li2019}.

\begin{figure*}
	\begin{centering}
		\centering{\includegraphics[width=0.65\textwidth]{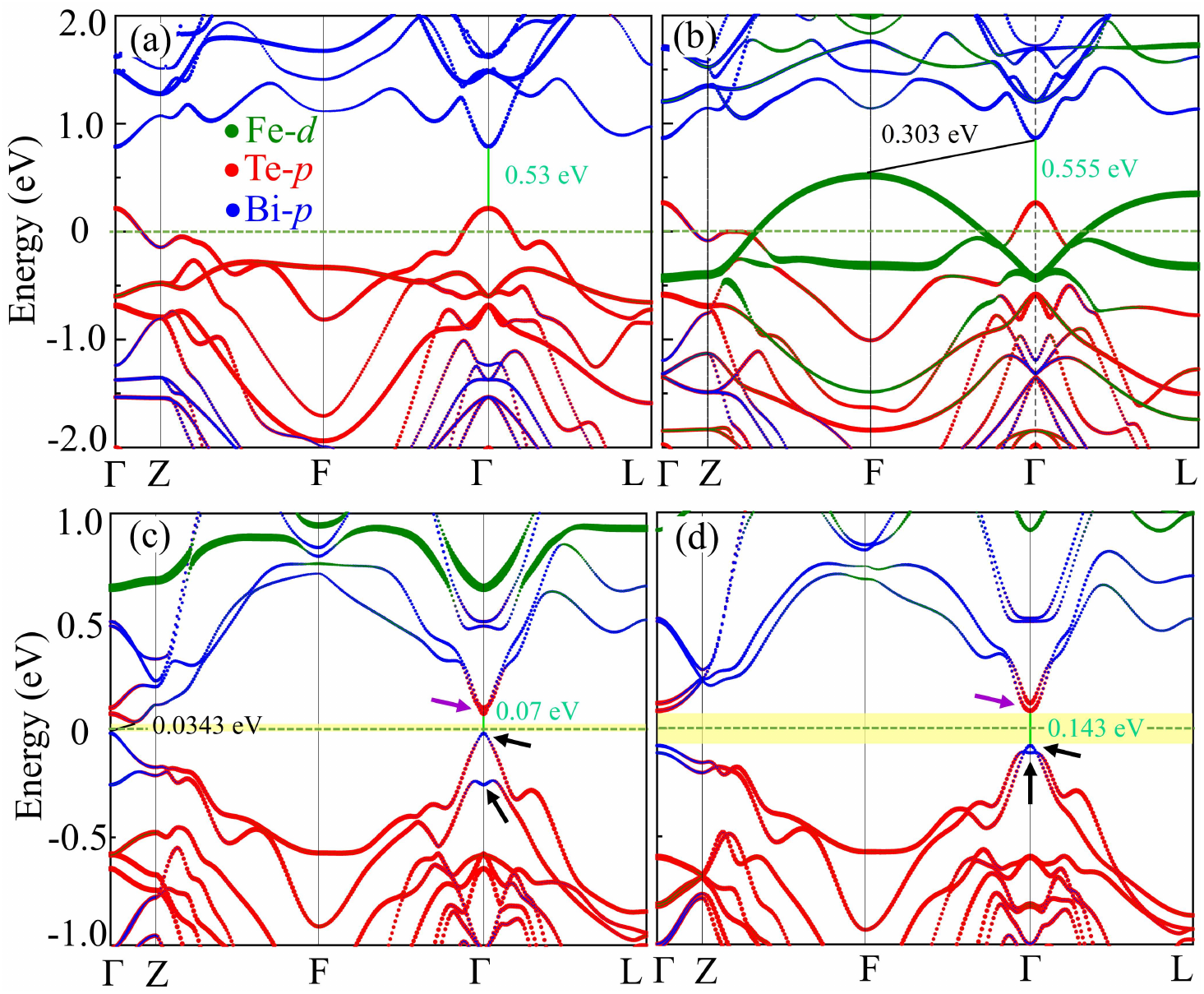}}
		\par\end{centering}
	\centering{}\caption{\textbf{Projected band structures. } (a) Spin-up and (b) spin-down of FM order FeBi$_{2}$Te$_{4}$ without SOC (spin polarization). Projected band structures with SOC of (c) FM-$z$ and (d) FM-$x$ orders. The black and green lines in the projected band structure indicate the global band gap and the local band gap at $\Gamma$, respectively, and in addition, the global band gap size is visualized by the yellow shading.}
	\label{6}
\end{figure*}

\begin{figure*}
	\begin{centering}
		\centering{\includegraphics[width=0.65\textwidth]{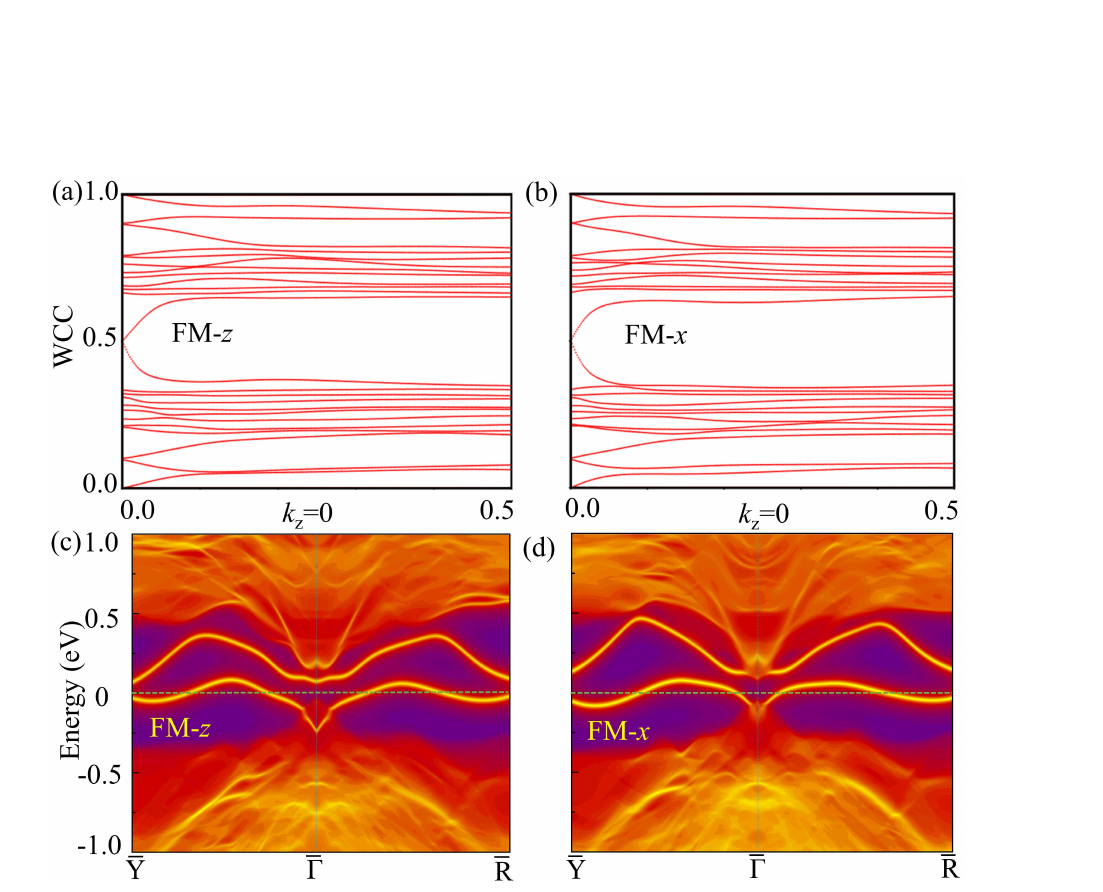}}
		\par\end{centering}
	\centering{}\caption{\textbf{Topological properties of FM-$z$(FM-$x$) FeBi$_{2}$Te$_{4}$. } WCC calculation results for two configurations considering SOC for (a) FM-$z$ and (b) FM-$x$. Topological surface states in (110) termination of (c) FM-$z$ and (d) FM-$x$ orders. }
	\label{7}
\end{figure*}
To confirm the non-trivial surface states of AFM FeBi$_{2}$Te$_{4}$, which are protected by the symmetry $\textit{S}=\Theta\tau_{1/2}$ ($\Theta$ represents time-reversal and $\tau_{1/2}$ represents a half magnetic-unit-cell translation), the $Z_{2}$ topological invariant are computed via Wannier Charge Centers (WCC) of six planes. The WCCs of AFM-$z$ and AFM-$x$ in the \textit{k}$_{z}$=0 plane (the plane perpendicular to the direction of $S$-symmetry) both have the topological number characteristic of Z$_{2}$=1, which means the AFM-$z$ and AFM-$x$ orders of FeBi$_{2}$Te$_{4}$ are non-trivial magnetic topological insulators. Furthermore, since the in-plane antiferromagnetic order of AFM-$x$ does not break the mirror symmetry of the \textit{k}$_{z}$=0 plane, it may be a potential mirror topological double insulator. We further calculated the surface states of FeBi$_{2}$Te$_{4}$ with AFM-$z$ and AFM-$x$ orders using a semi-infinite (110) surface, as shown in Figs. \ref{4}(c) and \ref{4}(d).
They exhibit gapless topological surface states protected by the \textit{S} symmetry. As for the semi-infinite (001) surface, AFM-$z$ appears as a gapped surface state, while the surface state of AFM-$x$ presents as a gapless feature, as shown in Fig. \ref{5}(a) and \ref{5}(b). As proposed by Liu's group\cite{PhysRevX.9.041038}, AFM-$x$ of FeBi$_{2}$Te$_{4}$ may be a dual topological insulator with both $S$-symmetry and mirror symmetry protection. For the FeBi$_{2}$Te$_{4}$ AFM order from in-plane and out-of-plane, the topological band gap will be opening as the breaking of Mirror symmetry M$_{x}$. In short, we can identify FeBi$_{2}$Te$_{4}$ with AFM orders as a typical intrinsic magnetic topological insulator.

In contrast to the AFM order, the spin-polarized band structure of the bulk FM order exhibits a metallic state, as shown in Figs. \ref{2}(c). Comparing the spin-up and spin-down projected band structures in Fig. \ref{6}(a) and \ref{6}(b), they possess a very close direct band gap at $\Gamma$. However, a bulge at the top of the valence band of the spin-down band structure between the Z-F-$\Gamma$ paths drives the global indirect band gap down to 0.303 eV. This bulge is mainly contributed by the Fe-3\textit{d} orbitals, thus inducing a spin magnetic moment of nearly 4 $\mu_{B}$. After the introduction of the SOC effect, both FM-$z$ (Fig. \ref{6}(c)) and FM-$x$ (Fig. \ref{6}(d)) exhibit topological insulator characteristics with band inversion near $\Gamma$. FM-$z$ order has an indirect band gap of about 0.034 eV, and its Fermi level is close to the valence band at $\Gamma$, unlike AFM-$z$, AFM-$x$, and FM-$x$ where the Fermi level is in the middle of the direct band gap at $\Gamma$. From FM-$z$ to FM-$x$, the valence bands at $\Gamma$ change from splitting to close to each other (black arrows at Figs. \ref{6}(c) and \ref{6}(d)), while the conduction bands change from degeneracy to splitting (purple arrows at Figs. \ref{6}(c) and \ref{6}(d)). We calculated the surface states in their (110) terminals to further observe their topological properties, as shown in Figs. \ref{7}(c) and \ref{7}(d). The WCC calculations, as shown in Figs. \ref{7}(a) and \ref{7}(b) confirm that FM-$z$ and FM-$x$ are topological insulators as well as the AFM orders.

\subsection{Pressure-induced topological phase transitions}

\begin{figure*}
	\begin{centering}
		\centering{\includegraphics[width=0.65\textwidth]{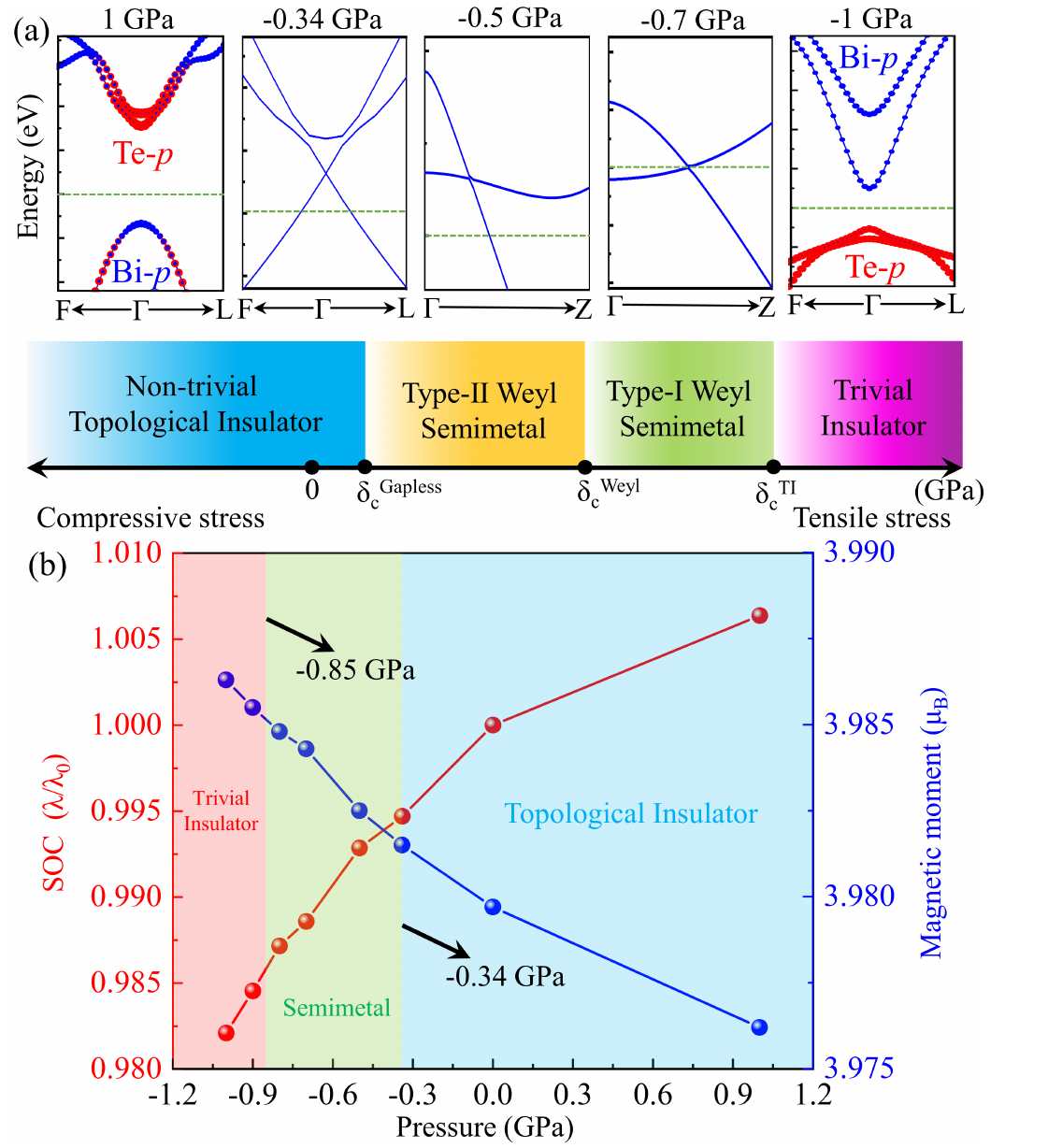}}
		\par\end{centering}
	\centering{}\caption{\textbf{Pressure induced a series of topological phase transitions in FM-$z$ order FeBi$_{2}$Te$_{4}$.} (a) FeBi$_{2}$Te$_{4}$ of the FM order undergoes multiple phase transitions under pressure: topological insulators, band crossing, type-$\uppercase\expandafter{\romannumeral1}$ Weyl semimetals, type-$\uppercase\expandafter{\romannumeral2}$ Weyl semimetals, and ordinary insulators. The upper panel shows the band structure of each phase at different pressures, and the lower panel shows the corresponding phase transition results. Where $\delta{_{c}^{Gapless}}=-0.34$ GPa, $\delta{_{c}^{Weyl}}=-0.6$ GPa, and $\delta{_{c}^{TI}}=-0.85$ GPa all represent the transition pressure of phase change. (b) The curves of SOC of Bi ions (take the quotient of the system containing pressure and the absence of pressure) and the magnetic moment of Fe ion along the $z$-direction as a function of pressure. Different color regions visually show different topological phases.}
	\label{8}
\end{figure*}

Pressure is an effective way of regulating the band structure of the material. A series of topological phase transitions are realized in the FM-$z$ magnetic order of FeBi$_{2}$Te$_{4}$ by pressure. As shown in Fig. \ref{8}(a), negative pressure induces the transformation of FeBi$_{2}$Te$_{4}$ with FM-$z$ magnetic order from a nontrivial topological insulator with band inversion to an ordinary insulator. The lower panel in Fig.\ref{8}(a) specifies the topological phase transition process under pressure regulation. The system undergoes five phases with increasing tension: the topological insulator, type-$\uppercase\expandafter{\romannumeral2}$ Weyl semimetal, type-$\uppercase\expandafter{\romannumeral1}$ Weyl semimetal, and trivial insulator. Bi ions in FeBi$_{2}$Te$_{4}$ have a significant SOC effect, and the magnetic moment of Fe ion is significant for the electronic structure properties near the Fermi level. Figure \ref{8}(b) shows the dependence of SOC and magnetic moment on pressure. As the pressure decreases to a negative value, the magnetic moment in the $z$-direction of the Fe ion gradually increases, inducing the system to transition from a topological insulator to a semimetal phase. At the same time, the SOC effect gradually weakened until the -0.85 GPa pressure was applied, and it transformed into a trivial insulator phase. Thus, we reveal that the topological quantum phase transition originates from the physical essence of the synergistic regulation of SOC and magnetic moments.

\begin{figure*}
	\begin{centering}
		\includegraphics[width=0.65\textwidth]{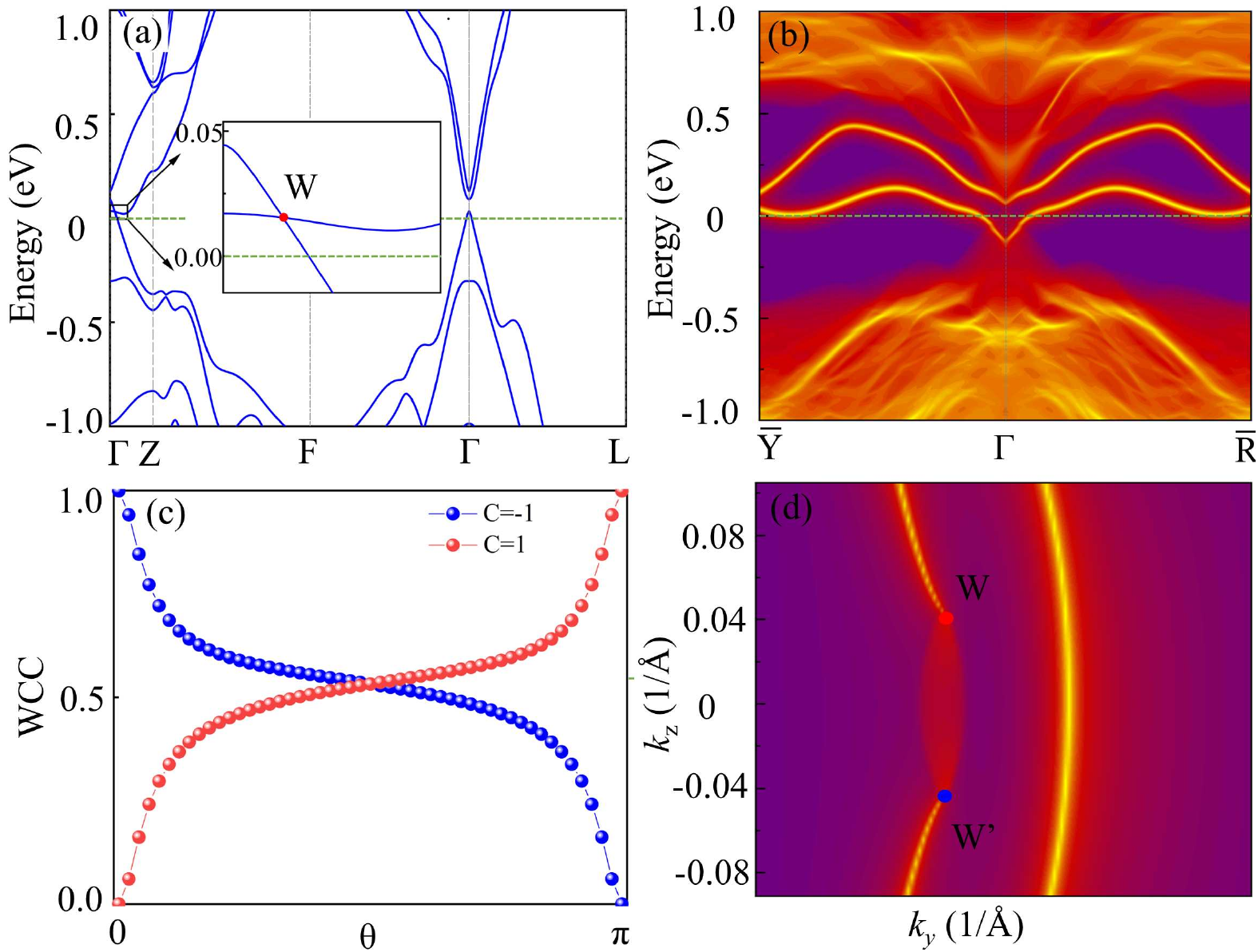}
		\par\end{centering}
	\centering{}\caption{\textbf{The existence of the type-$\uppercase\expandafter{\romannumeral2}$ Weyl semimetal phase.} (a) Band structure of FM-$z$ order under -0.5 GPa pressure, where the zoom-in diagram demonstrates the Weyl point W. (b) Topological surface states in (110) termination and (c) the motion of the sum of WCCs on a small sphere centered at W (C=1) and W' (C=-1) in momentum space. (d) Fermi arc (on the isoenergy plane of the Weyl points) of FM-$z$ order under -0.5 GPa pressure.}
	\label{9}
\end{figure*}

We transform the FM-$z$ of narrow gaped into the gapless type-$\uppercase\expandafter{\romannumeral2}$ Weyl semimetal state by applying tensile stress of -0.5 GPa, which is induced by interlayer orbital hybridizations, as shown in the band structure of $\Gamma$-Z line crossing each other in Fig. \ref{9}(a). The Weyl cone is slightly tilted along the $\Gamma$-Z line so that part of the electron pocket is below the cavity pocket, which is characteristic of type $\uppercase\expandafter{\romannumeral2}$ Weyl semimetals \cite{Soluyanov2015}.
In order to study its topological properties, we performed surface state calculations for the FM-$z$ system with -0.5 GPa action. The surface state at its (110) terminal (Fig. \ref{9}(b)) is slightly different from the one absent from the pressure in Fig. \ref{7}(c), which shows contact with the bulk band near the minimum of the conduction band. The Fermi arc exhibited by the energy contour at the Weyl point W in Fig. \ref{9}(d) is ample evidence of the fact that FM-$z$ can be tunable into Weyl semimetal. These two Weyl points (W/W'), very close to the Fermi level, are well separated in the momentum space. While convenient for experimental observation, its thin film may also facilitate the realization of QAHE. As the WCC in Fig. \ref{9}(c) shows, the W point is a monopole with a topological charge of +1 positioned in the momentum space,
corresponding to a Berry phase of 2$\pi$, and similarly, the W' point with a time-reversal relationship has the opposite topological charge, further proving that the system is a typical topological Weyl semimetal. Indeed, the latest study proposes that the bilayer ferromagnetic FeBi$_{2}$Te$_{4}$ will have a QAHE phenomenon, in line with our prediction\cite{10.1088/1674-1056/acd522}.

\begin{figure*}
	\begin{centering}
		\includegraphics[width=0.65\textwidth]{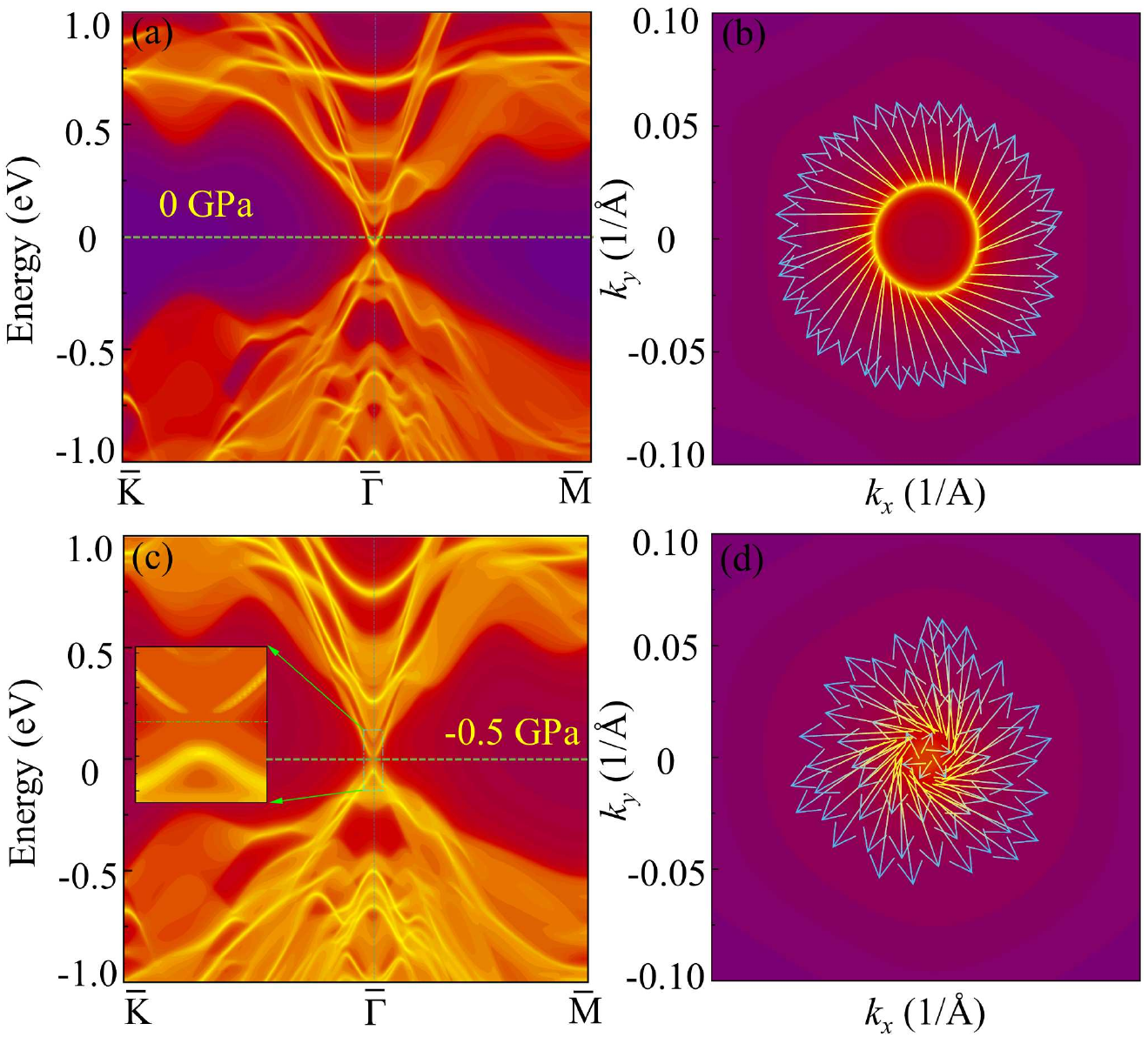}
		\par\end{centering}
	\centering{}\caption{\textbf{Comparison of topological properties of FM-$z$ FeBi$_{2}$Te$_{4}$ with or without pressure.} (a)/(c) Topological surface states in (001) termination and (b)/(d) spin textures (on the isoenergy plane of the Weyl points) of FM-$z$ order at the absence of pressure or under -0.5 GPa pressure. The zoom-in diagram in Fig. \ref{10}(c) is shows a gapped topological surface state.}
	\label{10}
\end{figure*}

In the absence of pressure, the FM-$z$ order of bulk FeBi$_{2}$Te$_{4}$ possesses a gaped (001) terminal surface state (Fig. \ref{10}(a)) and a highly helical spin texture (Fig. \ref{10}(b)). However, its Fermi level crosses the surface state (Fig. \ref{10}(a)) and presents a metallic character. Due to the tensile stress, the Fermi level evolves from crossing the surface state without applied pressure to residing in the surface band gap. At a pressure of -0.5 GPa, we achieved a phase transition of the topological surface state in the FM-$z$ order of the FeBi$_{2}$Te$_{4}$. Its spin texture in Fig. \ref{10}(d) also changes like a vortex. Moreover, FeBi$_{2}$Te$_{4}$ in the FM-$z$ order with -0.5 GPa pressure has a lower enthalpy of production than the ground state FM-$z$.

\begin{figure*}
	\begin{centering}
		\includegraphics[width=0.65\textwidth]{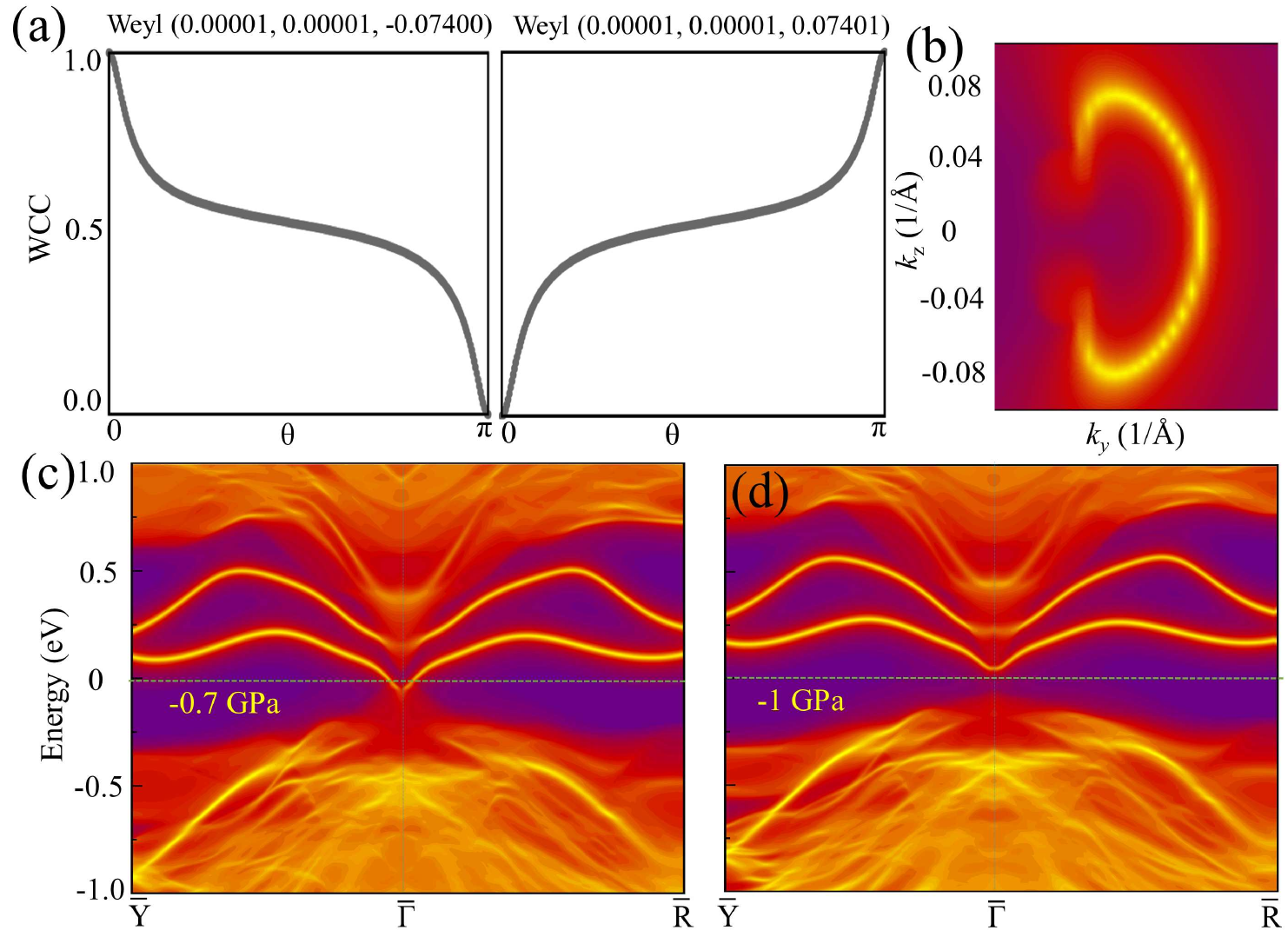}
		\par\end{centering}
	\centering{}\caption{\textbf{Type-$\uppercase\expandafter{\romannumeral1}$ Weyl semimetal state.}  (a) Motion of the sum of WCCs on a small sphere centered at $C=-1$ and $C=1$ in momentum space. (b) Fermi arc (on the isoenergy plane of the Weyl points) of FM-$z$ order under -0.7 GPa pressure. Topological surface states in (110) termination of (c) -0.7 GPa and (c) -1 GPa systems. 
	}
	\label{11}
\end{figure*}

Continuing to apply negative pressure to -0.7 GPa, the FeBi$_{2}$Te$_{4}$ FM-$z$ magnetic order system will be tuned to be a type-$\uppercase\expandafter{\romannumeral1}$ ideal Weyl semimetal. As shown in Figs. \ref{11}(a) and \ref{11}(b), the motion of the WCC sum over a small sphere centred at $C=-1$ and $C=1$ in the momentum space and the existence of the Fermi arc confirm that the system is a Weyl semimetal. The surface state of the (110) plane under -0.7 GPa pressure (Fig. \ref{11}(c)) is similar to that of -0.5 GPa (Fig. \ref{9}(b)), except that the surface state is shifted more towards the conduction band. When the pressure comes to -1 GPa, the nontrivial topological property of band inversion no longer exists, and the system transforms into a trivial insulator (see Fig. \ref{8}(a)). Its (110) plane surface state is no longer connected to the valence band, and the Fermi level is in the middle of the gapped surface state (see Fig. \ref{11}(d)). 

\begin{figure*}
	\begin{centering}
		\includegraphics[width=0.65\textwidth]{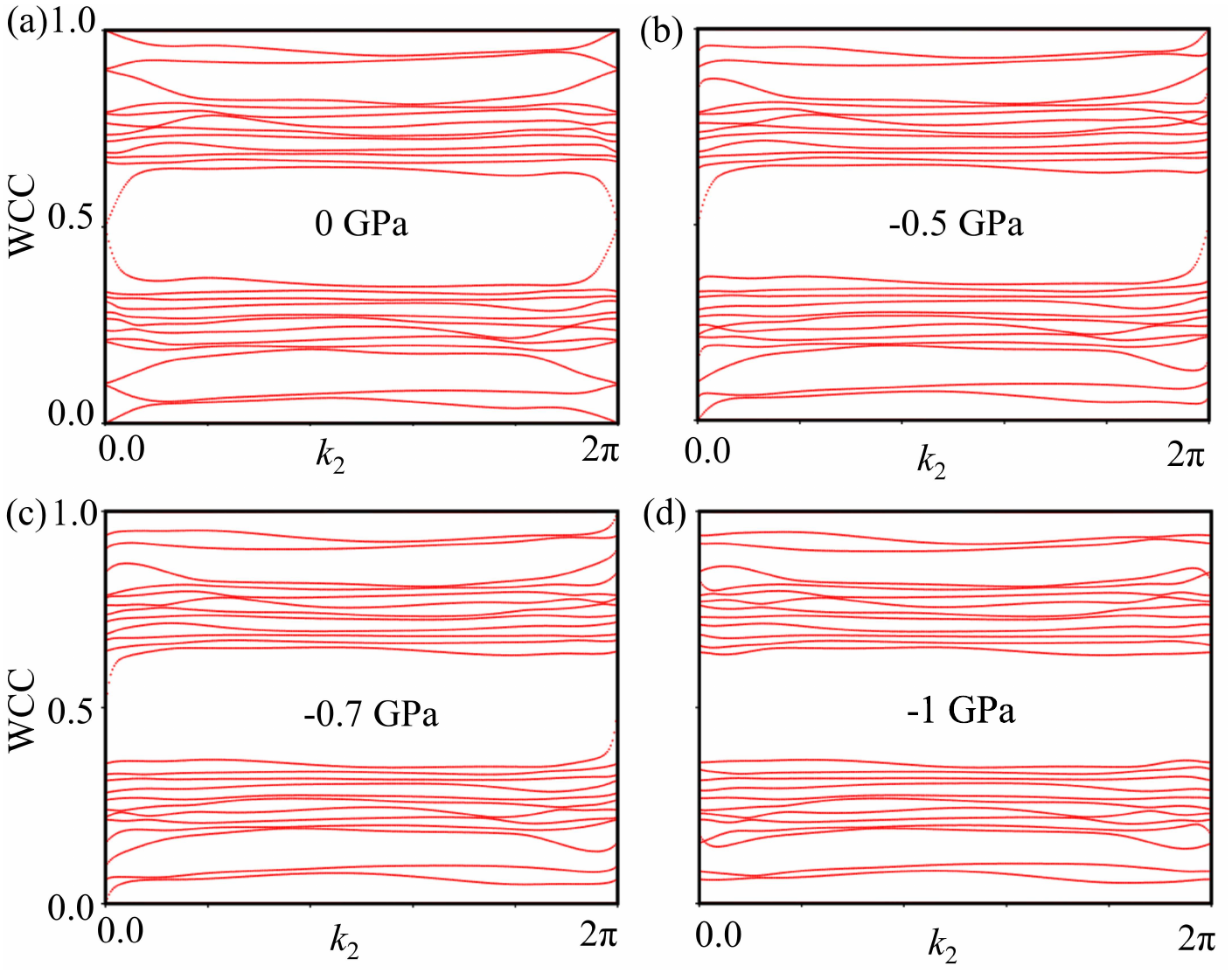}
		\par\end{centering}
	\centering{}\caption{WCCs of $k_2$-directed Wilson loops in the $k_z$=0 plane of FeBi$_{2}$Te$_{4}$ FM-$z$ order under different pressures.
	}
	\label{12}
\end{figure*}

Because the FM order breaks the time-reversal symmetry, it cannot be described by the $Z_2$ topological invariant. Here, we calculated the SI of the FM-$z$ configuration under pressure by the irvsp program\cite{PhysRevB.106.035150}. For the FM-$z$  order, its topological properties can be described by $Z_2$ $\times$ $Z_4$. At 0, -0.34 (-0.5,-0.7) and -1 GPa pressures, we obtained SIs for MSG1331 type-$\uppercase\expandafter{\romannumeral3}$ FeBi$_{2}$Te$_{4}$ with ($z_2$=0, $z_4$=3), ($z_2$=1,$z_4$=2) and ($z_2$=0,$z_4$=0), respectively. Based on the SI calculation results satisfying the Kane formula criterion and the WCC evolution curves given in Fig. \ref{12} under pressure modulation, we can determine the topological properties of the FeBi$_{2}$Te$_{4}$ FM-$z$ magnetic configuration. Without pressure, according to the WCC of FM-$z$  order FeBi$_{2}$Te$_{4}$ material and the even-number inversion-based $z_4$=2, we suggest FM as possessing a $Z_2$ $\times$ $Z_4$ high order topological insulator that will be tuned to a material with a set of Weyl nodal at special momenta. Indeed, we get two classes Weyl semimetal with SIs $z_2$=1, $z_4$=2. Although the topological classifications at -0.34,-0.5, and -0.7 GPa pressures are the same, they belong to the bandgap-closed critical regime, class $\uppercase\expandafter{\romannumeral2}$ Weyl semimetal, and class $\uppercase\expandafter{\romannumeral1}$ Weyl semimetal, respectively. Then a -1 GPa pressure will push the FM-$z$ system to be a normal insulator with  $z_2$=0, $z_4$=0. So when changing from positive to negative pressure, the FM-$z$  order FeBi$_{2}$Te$_{4}$ undergoes a topological phase transition from a strong topological insulator to a Weyl semimetal and then to a normal insulator. Thus, we achieve multiple topological phase transitions in FeBi$_{2}$Te$_{4}$ FM-$z$ ordering under pressure assistance.


\subsection{Low-energy effective model}

\begin{figure*}
	\begin{centering}
		\includegraphics[width=1.0\textwidth]{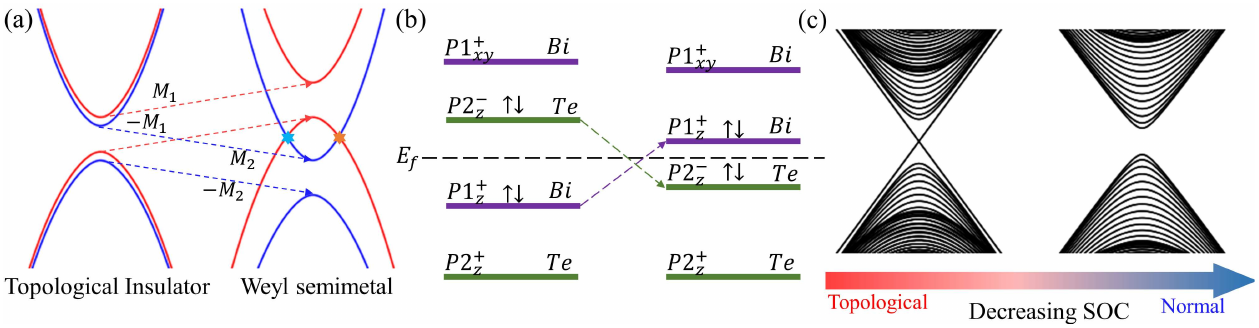}
		\par\end{centering}
	\centering{}\caption{\textbf{Low-energy effective model confirms topological quantum phase transition.} (a) Schematic of FM Weyl semimetal from NM insulator. As $M_{1,2}$ increases, a pair of Weyl points is generated. (b) Schematic diagram of the band inversion at $\Gamma$. (c) Schematic diagram of the phase transition for SOC reduction.
	}
	\label{13}
\end{figure*}
To understand the physical mechanisms embedded in the topological phase transition, we have further constructed a low-energy effective model for verification. Since the topological properties are mainly determined by the physical of the $\Gamma$ point, we can construct an effective Hamiltonian to characterize the system. We use the four-band Dirac model to describe the low-energy properties. This model was presented in \cite{TB1,Zhang2019} and has been well used to study the magnetoelectric properties of topological insulators. The Hamiltonian is written as Eq. (1).
\begin{equation}
\setlength{\arraycolsep}{1pt}
\begin{aligned}
\mathcal{H}(\bm{k}) = \varepsilon_{0}(k)+ &\\
\begin{pmatrix}
\mathcal{M}(k)+M_{1}\! & A_{1}k_{z} & 0 & -A_{2}k_{-}\\
A_{1}k_{z}  & -\mathcal{M}(k)+M_{2} & -A_{2}k_{-} & 0\\
0 & A_{2}k_{+} & \mathcal{M}(k)-M_{1} & -A_{1}k_{z}\\
A_{2}k_{+}  & 0 & -A_{1}k_{z} & -\mathcal{M}(k)-M_{2}\\
\end{pmatrix}
,
\end{aligned}
\end{equation}

where $k_{\pm} = k_{x} \pm ik_{y}$, $\varepsilon_{0}(k) = C +D_{1}k_{z}^{2} +D_{2}k_{\perp}^{2}$ and $M(k) = m - B_{1}k^{2}_{z} - B_{2}k^{2}_{\perp}$. In the basic of $(|P1^{+}_{z} ,\uparrow\rangle,|P2^{-}_{z} ,\uparrow\rangle,|P1^{+}_{z} ,\downarrow\rangle,|P2^{-}_{z} ,\downarrow\rangle)$, $P1^{+}_{z}$ and $P2^{-}_{z}$ represent the $p_{z}$ orbitals of bismuth and tellurium with positive ($+1$) and negative ($-1$) parity, respectively and $\uparrow/\downarrow$ represent the electron spin. $M_{1/2}$ represents the magnetization received in the ferromagnetic state. The parameters $A_{i}, B_{i}, C$, and $D_{i}$ characterize the hopping of the system, which are similar to those of MnBi$_{2}$Te$_{4}$. Under the action of tensile stress, the magnetization $M$ increases, and the system transitions from a topological insulator to a Weyl semimetal, as shown in Fig. \ref{13}(a). Meanwhile, note that the intrinsic spin-orbit coupling coefficient also changes, especially for Bi atoms, which are responsible for the effective mass term $m$ and induces the topological phase transition.

Next, we use finite thickness to show the qualitative characteristics of the phase transition. To make the model concise and general, we ignore $\varepsilon_{0}(k)$ because it does not affect the phase transition. The Hamiltonian Eq. (1) can be written in the standard form of the 3D massive Dirac model
\begin{equation}
	\begin{aligned}
		\mathcal{H}(\bm{k}) &= A_{2}(k_{x}\alpha_{x} + k_{y}\alpha_{y}) + A_{1}k_{z}\alpha_{z} + \mathcal{M}(k)\beta
	\end{aligned},
\end{equation}
where $\alpha_{i}(i=x, y, z)$ and $\beta$ are the Dirac matrix. As the mass parameter $m$ varies, the Hamiltonian quantity proves the phase transition from the topological phase $(m<0)$ to the one with a non-topological (trivial) phase $(m>0)$. Set to develop boundary conditions in the $z$-direction to simulate the finite-scale model and the $k_{z}$ must be replaced by gradient operator $i\partial_{z}$. So the Hamiltonian quantity in the plane can be written as $h_{\bot}(k) = A_{2}(k_{x}\alpha_{x} + k_{y}\alpha_{y}) + (m + 2B_{1} - B_{2}k_{\bot}^{2})\beta$, and the inter-plane hopping is $\triangle_{z} = -iA_{1}\alpha_{z}/2-B_{1}\beta$. When we let $m$ vary from complex to positive, the system also undergoes a topological phase transition, as shown in Fig. \ref{13}(c).

\section{Conclusion}\label{sec:4}
In conclusion, based on first-principles calculations, we have investigated the electronic structure and topological properties of FeBi$_{2}$Te$_{4}$ in different magnetic orders. Different with A-type AFM topological insulator MnBi$_{2}$Te$_{4}$, the ground state of FeBi$_{2}$Te$_{4}$ is out-plane ferromagnetic. The AFM-$x$ (AFM-$z$) and FM-$x$ (FM-$z$) orders are gapless and gapped non-trivial topological insulators, respectively. A topological phase transition can be induced in the surface states of the (001) plane of the FeBi$_{2}$Te$_{4}$ AFM order by changing the direction of the magnetic moment. That is to say that Mirror symmetry breaking induces the opening of the (001) surface band gap during the change of antiferromagnetism from in-plane to out-of-plane. When the pressure is absent, the FM-$z$ order has a Fermi level across the conduction band in the (001) terminal surface state, while the Fermi level under higher negative pressure will be between the gapped surface states, like -0.5 GPa. Notably, application of pressure induces a series topological phase transitions of the FM-$z$ ordering, from non-trivial topological insulators to band crossing metal state, type-$\uppercase\expandafter{\romannumeral2}$ Weyl semimetal and type-$\uppercase\expandafter{\romannumeral1}$ Weyl semimetal to ordinary insulators. Combined with low-energy effective model calculations, we find that the phase transition from topological insulators to Weyl semimetals is mainly due to the enhancement of the magnetic moment of the system. In addition, the phase transition from non-trivial topological insulators to trivial insulators critically depends on the weakening of the SOC effect.  Our results will not only provide opportunities for further research into quantum effects such as QAHE but also advance the design and application of magnetism topological quantum devices.

\section*{Author contributions}
J.-M. Zhang conceived and supervised this work. W.-T. Guo and N. Yang performed the calculations. W.-T. Guo, N. Yang and J.-M. Zhang wrote the manuscript. Z. Huang analyzed the datas and provided valuable and constructive suggestions.
\section*{Conflicts of interest}
There are no conflicts to declare.

\section*{Acknowledgements}
We acknowledge the financial support by the National Natural Science Foundation of China (No. 11874113) and the Natural Science Foundation of Fujian Province of China (No. 2020J02018). We thank Quansheng Wu for the guidance on calculating topological invariants and surface states in WannierTools.
The work was carried out at National Supercomputer Center in Tianjin,and the calculations were performed on TianHe-1(A).

\bibliography{Refs}

\begin{thebibliography}{85}%
\makeatletter
\providecommand \@ifxundefined [1]{%
 \@ifx{#1\undefined}
}%
\providecommand \@ifnum [1]{%
 \ifnum #1\expandafter \@firstoftwo
 \else \expandafter \@secondoftwo
 \fi
}%
\providecommand \@ifx [1]{%
 \ifx #1\expandafter \@firstoftwo
 \else \expandafter \@secondoftwo
 \fi
}%
\providecommand \natexlab [1]{#1}%
\providecommand \enquote  [1]{``#1''}%
\providecommand \bibnamefont  [1]{#1}%
\providecommand \bibfnamefont [1]{#1}%
\providecommand \citenamefont [1]{#1}%
\providecommand \href@noop [0]{\@secondoftwo}%
\providecommand \href [0]{\begingroup \@sanitize@url \@href}%
\providecommand \@href[1]{\@@startlink{#1}\@@href}%
\providecommand \@@href[1]{\endgroup#1\@@endlink}%
\providecommand \@sanitize@url [0]{\catcode `\\12\catcode `\$12\catcode
  `\&12\catcode `\#12\catcode `\^12\catcode `\_12\catcode `\%12\relax}%
\providecommand \@@startlink[1]{}%
\providecommand \@@endlink[0]{}%
\providecommand \url  [0]{\begingroup\@sanitize@url \@url }%
\providecommand \@url [1]{\endgroup\@href {#1}{\urlprefix }}%
\providecommand \urlprefix  [0]{URL }%
\providecommand \Eprint [0]{\href }%
\providecommand \doibase [0]{https://doi.org/}%
\providecommand \selectlanguage [0]{\@gobble}%
\providecommand \bibinfo  [0]{\@secondoftwo}%
\providecommand \bibfield  [0]{\@secondoftwo}%
\providecommand \translation [1]{[#1]}%
\providecommand \BibitemOpen [0]{}%
\providecommand \bibitemStop [0]{}%
\providecommand \bibitemNoStop [0]{.\EOS\space}%
\providecommand \EOS [0]{\spacefactor3000\relax}%
\providecommand \BibitemShut  [1]{\csname bibitem#1\endcsname}%
\let\auto@bib@innerbib\@empty
\bibitem [{\citenamefont {Mong}\ \emph {et~al.}(2010)\citenamefont {Mong},
  \citenamefont {Essin},\ and\ \citenamefont {Moore}}]{PhysRevB.81.245209}%
  \BibitemOpen
  \bibfield  {author} {\bibinfo {author} {\bibfnamefont {R.~S.~K.}\
  \bibnamefont {Mong}}, \bibinfo {author} {\bibfnamefont {A.~M.}\ \bibnamefont
  {Essin}},\ and\ \bibinfo {author} {\bibfnamefont {J.~E.}\ \bibnamefont
  {Moore}},\ }\bibfield  {title} {\bibinfo {title} {Antiferromagnetic
  topological insulators},\ }\href {https://doi.org/10.1103/PhysRevB.81.245209}
  {\bibfield  {journal} {\bibinfo  {journal} {Phys. Rev. B}\ }\textbf {\bibinfo
  {volume} {81}},\ \bibinfo {pages} {245209} (\bibinfo {year}
  {2010})}\BibitemShut {NoStop}%
\bibitem [{\citenamefont {Tokura}\ \emph {et~al.}(2019)\citenamefont {Tokura},
  \citenamefont {Yasuda},\ and\ \citenamefont {Tsukazaki}}]{Tokura2019}%
  \BibitemOpen
  \bibfield  {author} {\bibinfo {author} {\bibfnamefont {Y.}~\bibnamefont
  {Tokura}}, \bibinfo {author} {\bibfnamefont {K.}~\bibnamefont {Yasuda}},\
  and\ \bibinfo {author} {\bibfnamefont {A.}~\bibnamefont {Tsukazaki}},\
  }\bibfield  {title} {\bibinfo {title} {{Magnetic topological insulators}},\
  }\href {https://doi.org/10.1038/s42254-018-0011-5} {\bibfield  {journal}
  {\bibinfo  {journal} {Nat. Rev. Phys.}\ }\textbf {\bibinfo {volume} {1}},\
  \bibinfo {pages} {126} (\bibinfo {year} {2019})}\BibitemShut {NoStop}%
\bibitem [{\citenamefont {Otrokov}\ \emph
  {et~al.}(2019{\natexlab{a}})\citenamefont {Otrokov}, \citenamefont
  {Klimovskikh}, \citenamefont {Bentmann}, \citenamefont {Estyunin},
  \citenamefont {Zeugner}, \citenamefont {Aliev}, \citenamefont {Wolter},
  \citenamefont {Koroleva}, \citenamefont {Shikin}, \citenamefont {Hoffmann},
  \citenamefont {Rusinov}, \citenamefont {Eremeev}, \citenamefont {Kuznetsov},
  \citenamefont {Freyse}, \citenamefont {Amiraslanov}, \citenamefont {Babanly},
  \citenamefont {Mamedov}, \citenamefont {Abdullayev}, \citenamefont {Zverev},
  \citenamefont {Alfonsov}, \citenamefont {Kataev}, \citenamefont {Schwier},
  \citenamefont {Kumar}, \citenamefont {Kimura}, \citenamefont {Petaccia},
  \citenamefont {Santo}, \citenamefont {Vidal}, \citenamefont {Schatz},
  \citenamefont {Min}, \citenamefont {Moser}, \citenamefont {Peixoto},
  \citenamefont {Reinert}, \citenamefont {Ernst}, \citenamefont {Echenique},
  \citenamefont {Isaeva},\ and\ \citenamefont {Chulkov}}]{Otrokov2019}%
  \BibitemOpen
  \bibfield  {author} {\bibinfo {author} {\bibfnamefont {M.~M.}\ \bibnamefont
  {Otrokov}}, \bibinfo {author} {\bibfnamefont {I.~I.}\ \bibnamefont
  {Klimovskikh}}, \bibinfo {author} {\bibfnamefont {H.}~\bibnamefont
  {Bentmann}}, \bibinfo {author} {\bibfnamefont {D.}~\bibnamefont {Estyunin}},
  \bibinfo {author} {\bibfnamefont {A.}~\bibnamefont {Zeugner}}, \bibinfo
  {author} {\bibfnamefont {Z.~S.}\ \bibnamefont {Aliev}}, \bibinfo {author}
  {\bibfnamefont {A.~U.~B.}\ \bibnamefont {Wolter}}, \bibinfo {author}
  {\bibfnamefont {A.~V.}\ \bibnamefont {Koroleva}}, \bibinfo {author}
  {\bibfnamefont {A.~M.}\ \bibnamefont {Shikin}}, \bibinfo {author}
  {\bibfnamefont {M.}~\bibnamefont {Hoffmann}}, \bibinfo {author}
  {\bibfnamefont {I.~P.}\ \bibnamefont {Rusinov}}, \bibinfo {author}
  {\bibfnamefont {S.~V.}\ \bibnamefont {Eremeev}}, \bibinfo {author}
  {\bibfnamefont {V.~M.}\ \bibnamefont {Kuznetsov}}, \bibinfo {author}
  {\bibfnamefont {F.}~\bibnamefont {Freyse}}, \bibinfo {author} {\bibfnamefont
  {I.~R.}\ \bibnamefont {Amiraslanov}}, \bibinfo {author} {\bibfnamefont
  {M.~B.}\ \bibnamefont {Babanly}}, \bibinfo {author} {\bibfnamefont {N.~T.}\
  \bibnamefont {Mamedov}}, \bibinfo {author} {\bibfnamefont {N.~A.}\
  \bibnamefont {Abdullayev}}, \bibinfo {author} {\bibfnamefont {V.~N.}\
  \bibnamefont {Zverev}}, \bibinfo {author} {\bibfnamefont {A.}~\bibnamefont
  {Alfonsov}}, \bibinfo {author} {\bibfnamefont {V.}~\bibnamefont {Kataev}},
  \bibinfo {author} {\bibfnamefont {E.~F.}\ \bibnamefont {Schwier}}, \bibinfo
  {author} {\bibfnamefont {S.}~\bibnamefont {Kumar}}, \bibinfo {author}
  {\bibfnamefont {A.}~\bibnamefont {Kimura}}, \bibinfo {author} {\bibfnamefont
  {L.}~\bibnamefont {Petaccia}}, \bibinfo {author} {\bibfnamefont {G.~D.}\
  \bibnamefont {Santo}}, \bibinfo {author} {\bibfnamefont {R.~C.}\ \bibnamefont
  {Vidal}}, \bibinfo {author} {\bibfnamefont {S.}~\bibnamefont {Schatz}},
  \bibinfo {author} {\bibfnamefont {C.~H.}\ \bibnamefont {Min}}, \bibinfo
  {author} {\bibfnamefont {S.}~\bibnamefont {Moser}}, \bibinfo {author}
  {\bibfnamefont {T.~R.~F.}\ \bibnamefont {Peixoto}}, \bibinfo {author}
  {\bibfnamefont {F.}~\bibnamefont {Reinert}}, \bibinfo {author} {\bibfnamefont
  {A.}~\bibnamefont {Ernst}}, \bibinfo {author} {\bibfnamefont {P.~M.}\
  \bibnamefont {Echenique}}, \bibinfo {author} {\bibfnamefont {A.}~\bibnamefont
  {Isaeva}},\ and\ \bibinfo {author} {\bibfnamefont {E.~V.}\ \bibnamefont
  {Chulkov}},\ }\bibfield  {title} {\bibinfo {title} {{Prediction and
  observation of an antiferromagnetic topological insulator}},\ }\href
  {https://doi.org/10.1038/s41586-019-1840-9} {\bibfield  {journal} {\bibinfo
  {journal} {Nature}\ }\textbf {\bibinfo {volume} {576}},\ \bibinfo {pages}
  {416} (\bibinfo {year} {2019}{\natexlab{a}})}\BibitemShut {NoStop}%
\bibitem [{\citenamefont {Zhang}\ \emph {et~al.}(2019)\citenamefont {Zhang},
  \citenamefont {Shi}, \citenamefont {Zhu}, \citenamefont {Xing}, \citenamefont
  {Zhang},\ and\ \citenamefont {Wang}}]{Zhang2019}%
  \BibitemOpen
  \bibfield  {author} {\bibinfo {author} {\bibfnamefont {D.}~\bibnamefont
  {Zhang}}, \bibinfo {author} {\bibfnamefont {M.}~\bibnamefont {Shi}}, \bibinfo
  {author} {\bibfnamefont {T.}~\bibnamefont {Zhu}}, \bibinfo {author}
  {\bibfnamefont {D.}~\bibnamefont {Xing}}, \bibinfo {author} {\bibfnamefont
  {H.}~\bibnamefont {Zhang}},\ and\ \bibinfo {author} {\bibfnamefont
  {J.}~\bibnamefont {Wang}},\ }\bibfield  {title} {\bibinfo {title}
  {{Topological Axion States in the Magnetic Insulator
  ${\mathrm{MnBi}}_{2}{\mathrm{Te}}_{4}$ with the Quantized Magnetoelectric
  Effect}},\ }\href {https://doi.org/10.1103/PhysRevLett.122.206401} {\bibfield
   {journal} {\bibinfo  {journal} {Phys. Rev. Lett.}\ }\textbf {\bibinfo
  {volume} {122}},\ \bibinfo {pages} {206401} (\bibinfo {year} {2019})},\
  \Eprint {https://arxiv.org/abs/1808.08014} {arXiv:1808.08014} \BibitemShut
  {NoStop}%
\bibitem [{\citenamefont {Li}\ \emph {et~al.}(2019{\natexlab{a}})\citenamefont
  {Li}, \citenamefont {Li}, \citenamefont {Du}, \citenamefont {Wang},
  \citenamefont {Gu}, \citenamefont {Zhang}, \citenamefont {He}, \citenamefont
  {Duan},\ and\ \citenamefont {Xu}}]{Li2019}%
  \BibitemOpen
  \bibfield  {author} {\bibinfo {author} {\bibfnamefont {J.}~\bibnamefont
  {Li}}, \bibinfo {author} {\bibfnamefont {Y.}~\bibnamefont {Li}}, \bibinfo
  {author} {\bibfnamefont {S.}~\bibnamefont {Du}}, \bibinfo {author}
  {\bibfnamefont {Z.}~\bibnamefont {Wang}}, \bibinfo {author} {\bibfnamefont
  {B.~L.}\ \bibnamefont {Gu}}, \bibinfo {author} {\bibfnamefont {S.~C.}\
  \bibnamefont {Zhang}}, \bibinfo {author} {\bibfnamefont {K.}~\bibnamefont
  {He}}, \bibinfo {author} {\bibfnamefont {W.}~\bibnamefont {Duan}},\ and\
  \bibinfo {author} {\bibfnamefont {Y.}~\bibnamefont {Xu}},\ }\bibfield
  {title} {\bibinfo {title} {{Intrinsic magnetic topological insulators in van
  der Waals layered ${\mathrm{MnBi}}_{2}{\mathrm{Te}}_{4}$-family materials}},\
  }\href {https://doi.org/10.1126/sciadv.aaw5685} {\bibfield  {journal}
  {\bibinfo  {journal} {Sci. Adv.}\ }\textbf {\bibinfo {volume} {5}},\ \bibinfo
  {pages} {eaaw5685} (\bibinfo {year} {2019}{\natexlab{a}})},\ \Eprint
  {https://arxiv.org/abs/1808.08608} {arXiv:1808.08608} \BibitemShut {NoStop}%
\bibitem [{\citenamefont {Otrokov}\ \emph
  {et~al.}(2019{\natexlab{b}})\citenamefont {Otrokov}, \citenamefont {Rusinov},
  \citenamefont {Hoffmann}, \citenamefont {Vyazovskaya}, \citenamefont
  {Eremeev}, \citenamefont {Ernst}, \citenamefont {Echenique}, \citenamefont
  {Arnau},\ and\ \citenamefont {Chulkov}}]{Otrokov2019b}%
  \BibitemOpen
  \bibfield  {author} {\bibinfo {author} {\bibfnamefont {M.~M.}\ \bibnamefont
  {Otrokov}}, \bibinfo {author} {\bibfnamefont {I.~P.}\ \bibnamefont
  {Rusinov}}, \bibinfo {author} {\bibfnamefont {M.}~\bibnamefont {Hoffmann}},
  \bibinfo {author} {\bibfnamefont {A.~Y.}\ \bibnamefont {Vyazovskaya}},
  \bibinfo {author} {\bibfnamefont {S.~V.}\ \bibnamefont {Eremeev}}, \bibinfo
  {author} {\bibfnamefont {A.}~\bibnamefont {Ernst}}, \bibinfo {author}
  {\bibfnamefont {P.~M.}\ \bibnamefont {Echenique}}, \bibinfo {author}
  {\bibfnamefont {A.}~\bibnamefont {Arnau}},\ and\ \bibinfo {author}
  {\bibfnamefont {E.~V.}\ \bibnamefont {Chulkov}},\ }\bibfield  {title}
  {\bibinfo {title} {{Unique Thickness-Dependent Properties of the van der
  Waals Interlayer Antiferromagnet ${\mathrm{MnBi}}_{2}{\mathrm{Te}}_{4}$
  Films}},\ }\href {https://doi.org/10.1103/PhysRevLett.122.107202} {\bibfield
  {journal} {\bibinfo  {journal} {Phys. Rev. Lett.}\ }\textbf {\bibinfo
  {volume} {122}},\ \bibinfo {pages} {107202} (\bibinfo {year}
  {2019}{\natexlab{b}})}\BibitemShut {NoStop}%
\bibitem [{\citenamefont {Liu}\ \emph {et~al.}(2020)\citenamefont {Liu},
  \citenamefont {Wang}, \citenamefont {Li}, \citenamefont {Wu}, \citenamefont
  {Li}, \citenamefont {Li}, \citenamefont {He}, \citenamefont {Xu},
  \citenamefont {Zhang},\ and\ \citenamefont {Wang}}]{Liu2020}%
  \BibitemOpen
  \bibfield  {author} {\bibinfo {author} {\bibfnamefont {C.}~\bibnamefont
  {Liu}}, \bibinfo {author} {\bibfnamefont {Y.}~\bibnamefont {Wang}}, \bibinfo
  {author} {\bibfnamefont {H.}~\bibnamefont {Li}}, \bibinfo {author}
  {\bibfnamefont {Y.}~\bibnamefont {Wu}}, \bibinfo {author} {\bibfnamefont
  {Y.}~\bibnamefont {Li}}, \bibinfo {author} {\bibfnamefont {J.}~\bibnamefont
  {Li}}, \bibinfo {author} {\bibfnamefont {K.}~\bibnamefont {He}}, \bibinfo
  {author} {\bibfnamefont {Y.}~\bibnamefont {Xu}}, \bibinfo {author}
  {\bibfnamefont {J.}~\bibnamefont {Zhang}},\ and\ \bibinfo {author}
  {\bibfnamefont {Y.}~\bibnamefont {Wang}},\ }\bibfield  {title} {\bibinfo
  {title} {{Robust axion insulator and Chern insulator phases in a
  two-dimensional antiferromagnetic topological insulator}},\ }\href
  {https://doi.org/10.1038/s41563-019-0573-3} {\bibfield  {journal} {\bibinfo
  {journal} {Nat. Mater.}\ }\textbf {\bibinfo {volume} {19}},\ \bibinfo {pages}
  {522} (\bibinfo {year} {2020})}\BibitemShut {NoStop}%
\bibitem [{\citenamefont {Deng}\ \emph {et~al.}(2020)\citenamefont {Deng},
  \citenamefont {Yu}, \citenamefont {Shi}, \citenamefont {Guo}, \citenamefont
  {Xu}, \citenamefont {Wang}, \citenamefont {Chen},\ and\ \citenamefont
  {Zhang}}]{Deng2020}%
  \BibitemOpen
  \bibfield  {author} {\bibinfo {author} {\bibfnamefont {Y.}~\bibnamefont
  {Deng}}, \bibinfo {author} {\bibfnamefont {Y.}~\bibnamefont {Yu}}, \bibinfo
  {author} {\bibfnamefont {M.~Z.}\ \bibnamefont {Shi}}, \bibinfo {author}
  {\bibfnamefont {Z.}~\bibnamefont {Guo}}, \bibinfo {author} {\bibfnamefont
  {Z.}~\bibnamefont {Xu}}, \bibinfo {author} {\bibfnamefont {J.}~\bibnamefont
  {Wang}}, \bibinfo {author} {\bibfnamefont {X.~H.}\ \bibnamefont {Chen}},\
  and\ \bibinfo {author} {\bibfnamefont {Y.}~\bibnamefont {Zhang}},\ }\bibfield
   {title} {\bibinfo {title} {{Quantum anomalous Hall effect in intrinsic
  magnetic topological insulator ${\mathrm{MnBi}}_{2}{\mathrm{Te}}_{4}$}},\
  }\href {https://doi.org/10.1126/science.aax8156} {\bibfield  {journal}
  {\bibinfo  {journal} {Science}\ }\textbf {\bibinfo {volume} {367}},\ \bibinfo
  {pages} {895} (\bibinfo {year} {2020})}\BibitemShut {NoStop}%
\bibitem [{\citenamefont {Yu}\ \emph {et~al.}(2010)\citenamefont {Yu},
  \citenamefont {Zhang}, \citenamefont {Zhang}, \citenamefont {Zhang},
  \citenamefont {Dai},\ and\ \citenamefont
  {Fang}}]{doi:10.1126/science.1187485}%
  \BibitemOpen
  \bibfield  {author} {\bibinfo {author} {\bibfnamefont {R.}~\bibnamefont
  {Yu}}, \bibinfo {author} {\bibfnamefont {W.}~\bibnamefont {Zhang}}, \bibinfo
  {author} {\bibfnamefont {H.-J.}\ \bibnamefont {Zhang}}, \bibinfo {author}
  {\bibfnamefont {S.-C.}\ \bibnamefont {Zhang}}, \bibinfo {author}
  {\bibfnamefont {X.}~\bibnamefont {Dai}},\ and\ \bibinfo {author}
  {\bibfnamefont {Z.}~\bibnamefont {Fang}},\ }\bibfield  {title} {\bibinfo
  {title} {Quantized anomalous hall effect in magnetic topological
  insulators},\ }\href {https://doi.org/10.1126/science.1187485} {\bibfield
  {journal} {\bibinfo  {journal} {Science}\ }\textbf {\bibinfo {volume}
  {329}},\ \bibinfo {pages} {61} (\bibinfo {year} {2010})},\ \Eprint
  {https://arxiv.org/abs/https://www.science.org/doi/pdf/10.1126/science.1187485}
  {https://www.science.org/doi/pdf/10.1126/science.1187485} \BibitemShut
  {NoStop}%
\bibitem [{\citenamefont {Chen}\ \emph {et~al.}(2010)\citenamefont {Chen},
  \citenamefont {Chu}, \citenamefont {Analytis}, \citenamefont {Liu},
  \citenamefont {Igarashi}, \citenamefont {Kuo}, \citenamefont {Qi},
  \citenamefont {Mo}, \citenamefont {Moore}, \citenamefont {Lu}, \citenamefont
  {Hashimoto}, \citenamefont {Sasagawa}, \citenamefont {Zhang}, \citenamefont
  {Fisher}, \citenamefont {Hussain},\ and\ \citenamefont
  {Shen}}]{doi:10.1126/science.1189924}%
  \BibitemOpen
  \bibfield  {author} {\bibinfo {author} {\bibfnamefont {Y.~L.}\ \bibnamefont
  {Chen}}, \bibinfo {author} {\bibfnamefont {J.-H.}\ \bibnamefont {Chu}},
  \bibinfo {author} {\bibfnamefont {J.~G.}\ \bibnamefont {Analytis}}, \bibinfo
  {author} {\bibfnamefont {Z.~K.}\ \bibnamefont {Liu}}, \bibinfo {author}
  {\bibfnamefont {K.}~\bibnamefont {Igarashi}}, \bibinfo {author}
  {\bibfnamefont {H.-H.}\ \bibnamefont {Kuo}}, \bibinfo {author} {\bibfnamefont
  {X.~L.}\ \bibnamefont {Qi}}, \bibinfo {author} {\bibfnamefont {S.~K.}\
  \bibnamefont {Mo}}, \bibinfo {author} {\bibfnamefont {R.~G.}\ \bibnamefont
  {Moore}}, \bibinfo {author} {\bibfnamefont {D.~H.}\ \bibnamefont {Lu}},
  \bibinfo {author} {\bibfnamefont {M.}~\bibnamefont {Hashimoto}}, \bibinfo
  {author} {\bibfnamefont {T.}~\bibnamefont {Sasagawa}}, \bibinfo {author}
  {\bibfnamefont {S.~C.}\ \bibnamefont {Zhang}}, \bibinfo {author}
  {\bibfnamefont {I.~R.}\ \bibnamefont {Fisher}}, \bibinfo {author}
  {\bibfnamefont {Z.}~\bibnamefont {Hussain}},\ and\ \bibinfo {author}
  {\bibfnamefont {Z.~X.}\ \bibnamefont {Shen}},\ }\bibfield  {title} {\bibinfo
  {title} {Massive dirac fermion on the surface of a magnetically doped
  topological insulator},\ }\href {https://doi.org/10.1126/science.1189924}
  {\bibfield  {journal} {\bibinfo  {journal} {Science}\ }\textbf {\bibinfo
  {volume} {329}},\ \bibinfo {pages} {659} (\bibinfo {year} {2010})},\ \Eprint
  {https://arxiv.org/abs/https://www.science.org/doi/pdf/10.1126/science.1189924}
  {https://www.science.org/doi/pdf/10.1126/science.1189924} \BibitemShut
  {NoStop}%
\bibitem [{\citenamefont {Chang}\ \emph {et~al.}(2013)\citenamefont {Chang},
  \citenamefont {Zhang}, \citenamefont {Feng}, \citenamefont {Shen},
  \citenamefont {Zhang}, \citenamefont {Guo}, \citenamefont {Li}, \citenamefont
  {Ou}, \citenamefont {Wei}, \citenamefont {Wang}, \citenamefont {Ji},
  \citenamefont {Feng}, \citenamefont {Ji}, \citenamefont {Chen}, \citenamefont
  {Jia}, \citenamefont {Dai}, \citenamefont {Fang}, \citenamefont {Zhang},
  \citenamefont {He}, \citenamefont {Wang}, \citenamefont {Lu}, \citenamefont
  {Ma},\ and\ \citenamefont {Xue}}]{doi:10.1126/science.1234414}%
  \BibitemOpen
  \bibfield  {author} {\bibinfo {author} {\bibfnamefont {C.-Z.}\ \bibnamefont
  {Chang}}, \bibinfo {author} {\bibfnamefont {J.}~\bibnamefont {Zhang}},
  \bibinfo {author} {\bibfnamefont {X.}~\bibnamefont {Feng}}, \bibinfo {author}
  {\bibfnamefont {J.}~\bibnamefont {Shen}}, \bibinfo {author} {\bibfnamefont
  {Z.}~\bibnamefont {Zhang}}, \bibinfo {author} {\bibfnamefont
  {M.}~\bibnamefont {Guo}}, \bibinfo {author} {\bibfnamefont {K.}~\bibnamefont
  {Li}}, \bibinfo {author} {\bibfnamefont {Y.}~\bibnamefont {Ou}}, \bibinfo
  {author} {\bibfnamefont {P.}~\bibnamefont {Wei}}, \bibinfo {author}
  {\bibfnamefont {L.-L.}\ \bibnamefont {Wang}}, \bibinfo {author}
  {\bibfnamefont {Z.-Q.}\ \bibnamefont {Ji}}, \bibinfo {author} {\bibfnamefont
  {Y.}~\bibnamefont {Feng}}, \bibinfo {author} {\bibfnamefont {S.}~\bibnamefont
  {Ji}}, \bibinfo {author} {\bibfnamefont {X.}~\bibnamefont {Chen}}, \bibinfo
  {author} {\bibfnamefont {J.}~\bibnamefont {Jia}}, \bibinfo {author}
  {\bibfnamefont {X.}~\bibnamefont {Dai}}, \bibinfo {author} {\bibfnamefont
  {Z.}~\bibnamefont {Fang}}, \bibinfo {author} {\bibfnamefont {S.-C.}\
  \bibnamefont {Zhang}}, \bibinfo {author} {\bibfnamefont {K.}~\bibnamefont
  {He}}, \bibinfo {author} {\bibfnamefont {Y.}~\bibnamefont {Wang}}, \bibinfo
  {author} {\bibfnamefont {L.}~\bibnamefont {Lu}}, \bibinfo {author}
  {\bibfnamefont {X.-C.}\ \bibnamefont {Ma}},\ and\ \bibinfo {author}
  {\bibfnamefont {Q.-K.}\ \bibnamefont {Xue}},\ }\bibfield  {title} {\bibinfo
  {title} {Experimental observation of the quantum anomalous hall effect in a
  magnetic topological insulator},\ }\href
  {https://doi.org/10.1126/science.1234414} {\bibfield  {journal} {\bibinfo
  {journal} {Science}\ }\textbf {\bibinfo {volume} {340}},\ \bibinfo {pages}
  {167} (\bibinfo {year} {2013})},\ \Eprint
  {https://arxiv.org/abs/https://www.science.org/doi/pdf/10.1126/science.1234414}
  {https://www.science.org/doi/pdf/10.1126/science.1234414} \BibitemShut
  {NoStop}%
\bibitem [{\citenamefont {{Yunbo Ou, Chang Liu, Gaoyuan Jiang, Yang Feng,
  Dongyang Zhao, Weixiong Wu, Xiao-Xiao Wang, Wei Li, Canli Song, Li-Li Wang,
  Wenbo Wang, Weida Wu, Yayu Wang, Ke He, Xu-Cun Ma}}(2017)}]{YunboOu2017}%
  \BibitemOpen
  \bibfield  {author} {\bibinfo {author} {\bibfnamefont {Q.-K.~X.}\
  \bibnamefont {{Yunbo Ou, Chang Liu, Gaoyuan Jiang, Yang Feng, Dongyang Zhao,
  Weixiong Wu, Xiao-Xiao Wang, Wei Li, Canli Song, Li-Li Wang, Wenbo Wang,
  Weida Wu, Yayu Wang, Ke He, Xu-Cun Ma}}},\ }\bibfield  {title} {\bibinfo
  {title} {{Enhancing the Quantum Anomalous Hall Effect by Magnetic Codoping in
  a Topological Insulator}},\ }\href
  {https://doi.org/10.1002/adma.201703062.This} {\bibfield  {journal} {\bibinfo
   {journal} {Adv. Mater.}\ }\textbf {\bibinfo {volume} {30}},\ \bibinfo
  {pages} {1703062} (\bibinfo {year} {2017})}\BibitemShut {NoStop}%
\bibitem [{\citenamefont {Luo}\ and\ \citenamefont
  {Qi}(2013)}]{PhysRevB.87.085431}%
  \BibitemOpen
  \bibfield  {author} {\bibinfo {author} {\bibfnamefont {W.}~\bibnamefont
  {Luo}}\ and\ \bibinfo {author} {\bibfnamefont {X.-L.}\ \bibnamefont {Qi}},\
  }\bibfield  {title} {\bibinfo {title} {Massive dirac surface states in
  topological insulator/magnetic insulator heterostructures},\ }\href
  {https://doi.org/10.1103/PhysRevB.87.085431} {\bibfield  {journal} {\bibinfo
  {journal} {Phys. Rev. B}\ }\textbf {\bibinfo {volume} {87}},\ \bibinfo
  {pages} {085431} (\bibinfo {year} {2013})}\BibitemShut {NoStop}%
\bibitem [{\citenamefont {Katmis}\ \emph {et~al.}(2016)\citenamefont {Katmis},
  \citenamefont {Lauter}, \citenamefont {Nogueira}, \citenamefont {Assaf},
  \citenamefont {Jamer}, \citenamefont {Wei},\ and\ \citenamefont
  {Satpati}}]{Katmis2016}%
  \BibitemOpen
  \bibfield  {author} {\bibinfo {author} {\bibfnamefont {F.}~\bibnamefont
  {Katmis}}, \bibinfo {author} {\bibfnamefont {V.}~\bibnamefont {Lauter}},
  \bibinfo {author} {\bibfnamefont {F.~S.}\ \bibnamefont {Nogueira}}, \bibinfo
  {author} {\bibfnamefont {B.~A.}\ \bibnamefont {Assaf}}, \bibinfo {author}
  {\bibfnamefont {M.~E.}\ \bibnamefont {Jamer}}, \bibinfo {author}
  {\bibfnamefont {P.}~\bibnamefont {Wei}},\ and\ \bibinfo {author}
  {\bibfnamefont {B.}~\bibnamefont {Satpati}},\ }\bibfield  {title} {\bibinfo
  {title} {{A high-temperature ferromagnetic topological insulating phase by
  proximity coupling}},\ }\href {https://doi.org/10.1038/nature17635}
  {\bibfield  {journal} {\bibinfo  {journal} {Nature}\ }\textbf {\bibinfo
  {volume} {533}},\ \bibinfo {pages} {513} (\bibinfo {year}
  {2016})}\BibitemShut {NoStop}%
\bibitem [{\citenamefont {Li}\ \emph {et~al.}(2019{\natexlab{b}})\citenamefont
  {Li}, \citenamefont {Yu}, \citenamefont {Xu}, \citenamefont {Zhang},\ and\
  \citenamefont {Huang}}]{LI201977}%
  \BibitemOpen
  \bibfield  {author} {\bibinfo {author} {\bibfnamefont {P.}~\bibnamefont
  {Li}}, \bibinfo {author} {\bibfnamefont {J.}~\bibnamefont {Yu}}, \bibinfo
  {author} {\bibfnamefont {J.}~\bibnamefont {Xu}}, \bibinfo {author}
  {\bibfnamefont {L.}~\bibnamefont {Zhang}},\ and\ \bibinfo {author}
  {\bibfnamefont {K.}~\bibnamefont {Huang}},\ }\bibfield  {title} {\bibinfo
  {title} {{Topological-magnetic proximity effect in
  $Sb_{2}Te_{3}$/$CrI_{3}$(heterostructures}},\ }\href
  {https://doi.org/https://doi.org/10.1016/j.physb.2019.08.009} {\bibfield
  {journal} {\bibinfo  {journal} {Physica B}\ }\textbf {\bibinfo {volume}
  {573}},\ \bibinfo {pages} {77} (\bibinfo {year}
  {2019}{\natexlab{b}})}\BibitemShut {NoStop}%
\bibitem [{\citenamefont {Hirahara}\ \emph {et~al.}(2017)\citenamefont
  {Hirahara}, \citenamefont {Eremeev}, \citenamefont {Shirasawa}, \citenamefont
  {Okuyama}, \citenamefont {Kubo}, \citenamefont {Nakanishi}, \citenamefont
  {Akiyama}, \citenamefont {Takayama}, \citenamefont {Hajiri}, \citenamefont
  {Ideta}, \citenamefont {Matsunami}, \citenamefont {Sumida}, \citenamefont
  {Miyamoto}, \citenamefont {Takagi}, \citenamefont {Tanaka}, \citenamefont
  {Okuda}, \citenamefont {Yokoyama}, \citenamefont {Kimura}, \citenamefont
  {Hasegawa},\ and\ \citenamefont {Chulkov}}]{Hirahara2017}%
  \BibitemOpen
  \bibfield  {author} {\bibinfo {author} {\bibfnamefont {T.}~\bibnamefont
  {Hirahara}}, \bibinfo {author} {\bibfnamefont {S.~V.}\ \bibnamefont
  {Eremeev}}, \bibinfo {author} {\bibfnamefont {T.}~\bibnamefont {Shirasawa}},
  \bibinfo {author} {\bibfnamefont {Y.}~\bibnamefont {Okuyama}}, \bibinfo
  {author} {\bibfnamefont {T.}~\bibnamefont {Kubo}}, \bibinfo {author}
  {\bibfnamefont {R.}~\bibnamefont {Nakanishi}}, \bibinfo {author}
  {\bibfnamefont {R.}~\bibnamefont {Akiyama}}, \bibinfo {author} {\bibfnamefont
  {A.}~\bibnamefont {Takayama}}, \bibinfo {author} {\bibfnamefont
  {T.}~\bibnamefont {Hajiri}}, \bibinfo {author} {\bibfnamefont {S.~I.}\
  \bibnamefont {Ideta}}, \bibinfo {author} {\bibfnamefont {M.}~\bibnamefont
  {Matsunami}}, \bibinfo {author} {\bibfnamefont {K.}~\bibnamefont {Sumida}},
  \bibinfo {author} {\bibfnamefont {K.}~\bibnamefont {Miyamoto}}, \bibinfo
  {author} {\bibfnamefont {Y.}~\bibnamefont {Takagi}}, \bibinfo {author}
  {\bibfnamefont {K.}~\bibnamefont {Tanaka}}, \bibinfo {author} {\bibfnamefont
  {T.}~\bibnamefont {Okuda}}, \bibinfo {author} {\bibfnamefont
  {T.}~\bibnamefont {Yokoyama}}, \bibinfo {author} {\bibfnamefont {S.~I.}\
  \bibnamefont {Kimura}}, \bibinfo {author} {\bibfnamefont {S.}~\bibnamefont
  {Hasegawa}},\ and\ \bibinfo {author} {\bibfnamefont {E.~V.}\ \bibnamefont
  {Chulkov}},\ }\bibfield  {title} {\bibinfo {title} {{Large-Gap Magnetic
  Topological Heterostructure Formed by Subsurface Incorporation of a
  Ferromagnetic Layer}},\ }\href {https://doi.org/10.1021/acs.nanolett.7b00560}
  {\bibfield  {journal} {\bibinfo  {journal} {Nano Lett.}\ }\textbf {\bibinfo
  {volume} {17}},\ \bibinfo {pages} {3493} (\bibinfo {year} {2017})},\ \Eprint
  {https://arxiv.org/abs/1709.02004} {arXiv:1709.02004} \BibitemShut {NoStop}%
\bibitem [{\citenamefont {Eremeev}\ \emph {et~al.}(2018)\citenamefont
  {Eremeev}, \citenamefont {Otrokov},\ and\ \citenamefont
  {Chulkov}}]{Eremeev2018}%
  \BibitemOpen
  \bibfield  {author} {\bibinfo {author} {\bibfnamefont {S.~V.}\ \bibnamefont
  {Eremeev}}, \bibinfo {author} {\bibfnamefont {M.~M.}\ \bibnamefont
  {Otrokov}},\ and\ \bibinfo {author} {\bibfnamefont {E.~V.}\ \bibnamefont
  {Chulkov}},\ }\bibfield  {title} {\bibinfo {title} {{New Universal Type of
  Interface in the Magnetic Insulator/Topological Insulator
  Heterostructures}},\ }\href {https://doi.org/10.1021/acs.nanolett.8b03057}
  {\bibfield  {journal} {\bibinfo  {journal} {Nano Lett.}\ }\textbf {\bibinfo
  {volume} {18}},\ \bibinfo {pages} {6521} (\bibinfo {year}
  {2018})}\BibitemShut {NoStop}%
\bibitem [{\citenamefont {Wang}\ \emph {et~al.}(2021)\citenamefont {Wang},
  \citenamefont {Ge}, \citenamefont {Li}, \citenamefont {Liu}, \citenamefont
  {Xu},\ and\ \citenamefont {Wang}}]{Wang2021}%
  \BibitemOpen
  \bibfield  {author} {\bibinfo {author} {\bibfnamefont {P.}~\bibnamefont
  {Wang}}, \bibinfo {author} {\bibfnamefont {J.}~\bibnamefont {Ge}}, \bibinfo
  {author} {\bibfnamefont {J.}~\bibnamefont {Li}}, \bibinfo {author}
  {\bibfnamefont {Y.}~\bibnamefont {Liu}}, \bibinfo {author} {\bibfnamefont
  {Y.}~\bibnamefont {Xu}},\ and\ \bibinfo {author} {\bibfnamefont
  {J.}~\bibnamefont {Wang}},\ }\bibfield  {title} {\bibinfo {title} {{Intrinsic
  magnetic topological insulators}},\ }\href
  {https://doi.org/10.1016/j.xinn.2021.100098} {\bibfield  {journal} {\bibinfo
  {journal} {The Innovation}\ }\textbf {\bibinfo {volume} {2}},\ \bibinfo
  {pages} {100098} (\bibinfo {year} {2021})},\ \Eprint
  {https://arxiv.org/abs/2012.04258} {arXiv:2012.04258} \BibitemShut {NoStop}%
\bibitem [{\citenamefont {Vidal}\ \emph
  {et~al.}(2019{\natexlab{a}})\citenamefont {Vidal}, \citenamefont {Bentmann},
  \citenamefont {Peixoto}, \citenamefont {Zeugner}, \citenamefont {Moser},
  \citenamefont {Min}, \citenamefont {Schatz}, \citenamefont {Ki\ss{}ner},
  \citenamefont {\"Unzelmann}, \citenamefont {Fornari}, \citenamefont {Vasili},
  \citenamefont {Valvidares}, \citenamefont {Sakamoto}, \citenamefont {Mondal},
  \citenamefont {Fujii}, \citenamefont {Vobornik}, \citenamefont {Jung},
  \citenamefont {Cacho}, \citenamefont {Kim}, \citenamefont {Koch},
  \citenamefont {Jozwiak}, \citenamefont {Bostwick}, \citenamefont {Denlinger},
  \citenamefont {Rotenberg}, \citenamefont {Buck}, \citenamefont {Hoesch},
  \citenamefont {Diekmann}, \citenamefont {Rohlf}, \citenamefont {Kall\"ane},
  \citenamefont {Rossnagel}, \citenamefont {Otrokov}, \citenamefont {Chulkov},
  \citenamefont {Ruck}, \citenamefont {Isaeva},\ and\ \citenamefont
  {Reinert}}]{PhysRevB.100.121104}%
  \BibitemOpen
  \bibfield  {author} {\bibinfo {author} {\bibfnamefont {R.~C.}\ \bibnamefont
  {Vidal}}, \bibinfo {author} {\bibfnamefont {H.}~\bibnamefont {Bentmann}},
  \bibinfo {author} {\bibfnamefont {T.~R.~F.}\ \bibnamefont {Peixoto}},
  \bibinfo {author} {\bibfnamefont {A.}~\bibnamefont {Zeugner}}, \bibinfo
  {author} {\bibfnamefont {S.}~\bibnamefont {Moser}}, \bibinfo {author}
  {\bibfnamefont {C.-H.}\ \bibnamefont {Min}}, \bibinfo {author} {\bibfnamefont
  {S.}~\bibnamefont {Schatz}}, \bibinfo {author} {\bibfnamefont
  {K.}~\bibnamefont {Ki\ss{}ner}}, \bibinfo {author} {\bibfnamefont
  {M.}~\bibnamefont {\"Unzelmann}}, \bibinfo {author} {\bibfnamefont {C.~I.}\
  \bibnamefont {Fornari}}, \bibinfo {author} {\bibfnamefont {H.~B.}\
  \bibnamefont {Vasili}}, \bibinfo {author} {\bibfnamefont {M.}~\bibnamefont
  {Valvidares}}, \bibinfo {author} {\bibfnamefont {K.}~\bibnamefont
  {Sakamoto}}, \bibinfo {author} {\bibfnamefont {D.}~\bibnamefont {Mondal}},
  \bibinfo {author} {\bibfnamefont {J.}~\bibnamefont {Fujii}}, \bibinfo
  {author} {\bibfnamefont {I.}~\bibnamefont {Vobornik}}, \bibinfo {author}
  {\bibfnamefont {S.}~\bibnamefont {Jung}}, \bibinfo {author} {\bibfnamefont
  {C.}~\bibnamefont {Cacho}}, \bibinfo {author} {\bibfnamefont {T.~K.}\
  \bibnamefont {Kim}}, \bibinfo {author} {\bibfnamefont {R.~J.}\ \bibnamefont
  {Koch}}, \bibinfo {author} {\bibfnamefont {C.}~\bibnamefont {Jozwiak}},
  \bibinfo {author} {\bibfnamefont {A.}~\bibnamefont {Bostwick}}, \bibinfo
  {author} {\bibfnamefont {J.~D.}\ \bibnamefont {Denlinger}}, \bibinfo {author}
  {\bibfnamefont {E.}~\bibnamefont {Rotenberg}}, \bibinfo {author}
  {\bibfnamefont {J.}~\bibnamefont {Buck}}, \bibinfo {author} {\bibfnamefont
  {M.}~\bibnamefont {Hoesch}}, \bibinfo {author} {\bibfnamefont
  {F.}~\bibnamefont {Diekmann}}, \bibinfo {author} {\bibfnamefont
  {S.}~\bibnamefont {Rohlf}}, \bibinfo {author} {\bibfnamefont
  {M.}~\bibnamefont {Kall\"ane}}, \bibinfo {author} {\bibfnamefont
  {K.}~\bibnamefont {Rossnagel}}, \bibinfo {author} {\bibfnamefont {M.~M.}\
  \bibnamefont {Otrokov}}, \bibinfo {author} {\bibfnamefont {E.~V.}\
  \bibnamefont {Chulkov}}, \bibinfo {author} {\bibfnamefont {M.}~\bibnamefont
  {Ruck}}, \bibinfo {author} {\bibfnamefont {A.}~\bibnamefont {Isaeva}},\ and\
  \bibinfo {author} {\bibfnamefont {F.}~\bibnamefont {Reinert}},\ }\bibfield
  {title} {\bibinfo {title} {{Surface states and Rashba-type spin polarization
  in antiferromagnetic $MnBi_{2}Te_{4}$(0001)}},\ }\href
  {https://doi.org/10.1103/PhysRevB.100.121104} {\bibfield  {journal} {\bibinfo
   {journal} {Phys. Rev. B}\ }\textbf {\bibinfo {volume} {100}},\ \bibinfo
  {pages} {121104} (\bibinfo {year} {2019}{\natexlab{a}})}\BibitemShut
  {NoStop}%
\bibitem [{\citenamefont {Yang}\ \emph {et~al.}(2021)\citenamefont {Yang},
  \citenamefont {Xu}, \citenamefont {Zhu}, \citenamefont {Niu}, \citenamefont
  {Xu}, \citenamefont {Peng}, \citenamefont {Cheng}, \citenamefont {Jia},
  \citenamefont {Huang}, \citenamefont {Xu}, \citenamefont {Lu},\ and\
  \citenamefont {Ye}}]{PhysRevX.11.011003}%
  \BibitemOpen
  \bibfield  {author} {\bibinfo {author} {\bibfnamefont {S.}~\bibnamefont
  {Yang}}, \bibinfo {author} {\bibfnamefont {X.}~\bibnamefont {Xu}}, \bibinfo
  {author} {\bibfnamefont {Y.}~\bibnamefont {Zhu}}, \bibinfo {author}
  {\bibfnamefont {R.}~\bibnamefont {Niu}}, \bibinfo {author} {\bibfnamefont
  {C.}~\bibnamefont {Xu}}, \bibinfo {author} {\bibfnamefont {Y.}~\bibnamefont
  {Peng}}, \bibinfo {author} {\bibfnamefont {X.}~\bibnamefont {Cheng}},
  \bibinfo {author} {\bibfnamefont {X.}~\bibnamefont {Jia}}, \bibinfo {author}
  {\bibfnamefont {Y.}~\bibnamefont {Huang}}, \bibinfo {author} {\bibfnamefont
  {X.}~\bibnamefont {Xu}}, \bibinfo {author} {\bibfnamefont {J.}~\bibnamefont
  {Lu}},\ and\ \bibinfo {author} {\bibfnamefont {Y.}~\bibnamefont {Ye}},\
  }\bibfield  {title} {\bibinfo {title} {{Odd-Even Layer-Number Effect and
  Layer-Dependent Magnetic Phase Diagrams in
  ${\mathrm{MnBi}}_{2}{\mathrm{Te}}_{4}$}},\ }\href
  {https://doi.org/10.1103/PhysRevX.11.011003} {\bibfield  {journal} {\bibinfo
  {journal} {Phys. Rev. X}\ }\textbf {\bibinfo {volume} {11}},\ \bibinfo
  {pages} {011003} (\bibinfo {year} {2021})}\BibitemShut {NoStop}%
\bibitem [{\citenamefont {Sun}\ \emph {et~al.}(2019)\citenamefont {Sun},
  \citenamefont {Xia}, \citenamefont {Chen}, \citenamefont {Zhang},
  \citenamefont {Liu}, \citenamefont {Yao}, \citenamefont {Tang}, \citenamefont
  {Zhao}, \citenamefont {Xu},\ and\ \citenamefont {Liu}}]{Sun2019}%
  \BibitemOpen
  \bibfield  {author} {\bibinfo {author} {\bibfnamefont {H.}~\bibnamefont
  {Sun}}, \bibinfo {author} {\bibfnamefont {B.}~\bibnamefont {Xia}}, \bibinfo
  {author} {\bibfnamefont {Z.}~\bibnamefont {Chen}}, \bibinfo {author}
  {\bibfnamefont {Y.}~\bibnamefont {Zhang}}, \bibinfo {author} {\bibfnamefont
  {P.}~\bibnamefont {Liu}}, \bibinfo {author} {\bibfnamefont {Q.}~\bibnamefont
  {Yao}}, \bibinfo {author} {\bibfnamefont {H.}~\bibnamefont {Tang}}, \bibinfo
  {author} {\bibfnamefont {Y.}~\bibnamefont {Zhao}}, \bibinfo {author}
  {\bibfnamefont {H.}~\bibnamefont {Xu}},\ and\ \bibinfo {author}
  {\bibfnamefont {Q.}~\bibnamefont {Liu}},\ }\bibfield  {title} {\bibinfo
  {title} {{Rational Design Principles of the Quantum Anomalous Hall Effect in
  Superlatticelike Magnetic Topological Insulators}},\ }\href
  {https://doi.org/10.1103/PhysRevLett.123.096401} {\bibfield  {journal}
  {\bibinfo  {journal} {Phys. Rev. Lett.}\ }\textbf {\bibinfo {volume} {123}},\
  \bibinfo {pages} {096401} (\bibinfo {year} {2019})},\ \Eprint
  {https://arxiv.org/abs/1905.12208} {arXiv:1905.12208} \BibitemShut {NoStop}%
\bibitem [{\citenamefont {Yan}\ \emph {et~al.}(2020)\citenamefont {Yan},
  \citenamefont {Liu}, \citenamefont {Parker}, \citenamefont {Wu},
  \citenamefont {Aczel}, \citenamefont {Matsuda}, \citenamefont {McGuire},\
  and\ \citenamefont {Sales}}]{Yan2020}%
  \BibitemOpen
  \bibfield  {author} {\bibinfo {author} {\bibfnamefont {J.~Q.}\ \bibnamefont
  {Yan}}, \bibinfo {author} {\bibfnamefont {Y.~H.}\ \bibnamefont {Liu}},
  \bibinfo {author} {\bibfnamefont {D.~S.}\ \bibnamefont {Parker}}, \bibinfo
  {author} {\bibfnamefont {Y.}~\bibnamefont {Wu}}, \bibinfo {author}
  {\bibfnamefont {A.~A.}\ \bibnamefont {Aczel}}, \bibinfo {author}
  {\bibfnamefont {M.}~\bibnamefont {Matsuda}}, \bibinfo {author} {\bibfnamefont
  {M.~A.}\ \bibnamefont {McGuire}},\ and\ \bibinfo {author} {\bibfnamefont
  {B.~C.}\ \bibnamefont {Sales}},\ }\bibfield  {title} {\bibinfo {title}
  {{A-type antiferromagnetic order in ${\mathrm{MnBi}}_{4}{\mathrm{Te}}_{7}$
  and ${\mathrm{MnBi}}_{6}{\mathrm{Te}}_{10}$ single crystals}},\ }\href
  {https://doi.org/10.1103/PhysRevMaterials.4.054202} {\bibfield  {journal}
  {\bibinfo  {journal} {Phys. Rev. Mater.}\ }\textbf {\bibinfo {volume} {4}},\
  \bibinfo {pages} {054202} (\bibinfo {year} {2020})},\ \Eprint
  {https://arxiv.org/abs/1910.06273} {arXiv:1910.06273} \BibitemShut {NoStop}%
\bibitem [{\citenamefont {Tian}\ \emph {et~al.}(2020)\citenamefont {Tian},
  \citenamefont {Gao}, \citenamefont {Nie}, \citenamefont {Qian}, \citenamefont
  {Gong}, \citenamefont {Fu}, \citenamefont {Li}, \citenamefont {Fan},
  \citenamefont {Zhang}, \citenamefont {Kondo}, \citenamefont {Shin},
  \citenamefont {Adell}, \citenamefont {Fedderwitz}, \citenamefont {Ding},
  \citenamefont {Wang}, \citenamefont {Qian},\ and\ \citenamefont
  {Lei}}]{PhysRevB.102.035144}%
  \BibitemOpen
  \bibfield  {author} {\bibinfo {author} {\bibfnamefont {S.}~\bibnamefont
  {Tian}}, \bibinfo {author} {\bibfnamefont {S.}~\bibnamefont {Gao}}, \bibinfo
  {author} {\bibfnamefont {S.}~\bibnamefont {Nie}}, \bibinfo {author}
  {\bibfnamefont {Y.}~\bibnamefont {Qian}}, \bibinfo {author} {\bibfnamefont
  {C.}~\bibnamefont {Gong}}, \bibinfo {author} {\bibfnamefont {Y.}~\bibnamefont
  {Fu}}, \bibinfo {author} {\bibfnamefont {H.}~\bibnamefont {Li}}, \bibinfo
  {author} {\bibfnamefont {W.}~\bibnamefont {Fan}}, \bibinfo {author}
  {\bibfnamefont {P.}~\bibnamefont {Zhang}}, \bibinfo {author} {\bibfnamefont
  {T.}~\bibnamefont {Kondo}}, \bibinfo {author} {\bibfnamefont
  {S.}~\bibnamefont {Shin}}, \bibinfo {author} {\bibfnamefont {J.}~\bibnamefont
  {Adell}}, \bibinfo {author} {\bibfnamefont {H.}~\bibnamefont {Fedderwitz}},
  \bibinfo {author} {\bibfnamefont {H.}~\bibnamefont {Ding}}, \bibinfo {author}
  {\bibfnamefont {Z.}~\bibnamefont {Wang}}, \bibinfo {author} {\bibfnamefont
  {T.}~\bibnamefont {Qian}},\ and\ \bibinfo {author} {\bibfnamefont
  {H.}~\bibnamefont {Lei}},\ }\bibfield  {title} {\bibinfo {title} {{Magnetic
  topological insulator $MnBi_{6}Te_{10}$ with a zero-field ferromagnetic state
  and gapped Dirac surface states}},\ }\href
  {https://doi.org/10.1103/PhysRevB.102.035144} {\bibfield  {journal} {\bibinfo
   {journal} {Phys. Rev. B}\ }\textbf {\bibinfo {volume} {102}},\ \bibinfo
  {pages} {035144} (\bibinfo {year} {2020})}\BibitemShut {NoStop}%
\bibitem [{\citenamefont {Klimovskikh}\ \emph {et~al.}(2020)\citenamefont
  {Klimovskikh}, \citenamefont {Otrokov}, \citenamefont {Estyunin},
  \citenamefont {Eremeev}, \citenamefont {Filnov}, \citenamefont {Koroleva},
  \citenamefont {Shevchenko}, \citenamefont {Voroshnin}, \citenamefont
  {Rybkin}, \citenamefont {Rusinov}, \citenamefont {Blanco-Rey}, \citenamefont
  {Hoffmann}, \citenamefont {Aliev}, \citenamefont {Babanly}, \citenamefont
  {Amiraslanov}, \citenamefont {Abdullayev}, \citenamefont {Zverev},
  \citenamefont {Kimura}, \citenamefont {Tereshchenko}, \citenamefont {Kokh},
  \citenamefont {Petaccia}, \citenamefont {{Di Santo}}, \citenamefont {Ernst},
  \citenamefont {Echenique}, \citenamefont {Mamedov}, \citenamefont {Shikin},\
  and\ \citenamefont {Chulkov}}]{Klimovskikh2020}%
  \BibitemOpen
  \bibfield  {author} {\bibinfo {author} {\bibfnamefont {I.~I.}\ \bibnamefont
  {Klimovskikh}}, \bibinfo {author} {\bibfnamefont {M.~M.}\ \bibnamefont
  {Otrokov}}, \bibinfo {author} {\bibfnamefont {D.}~\bibnamefont {Estyunin}},
  \bibinfo {author} {\bibfnamefont {S.~V.}\ \bibnamefont {Eremeev}}, \bibinfo
  {author} {\bibfnamefont {S.~O.}\ \bibnamefont {Filnov}}, \bibinfo {author}
  {\bibfnamefont {A.}~\bibnamefont {Koroleva}}, \bibinfo {author}
  {\bibfnamefont {E.}~\bibnamefont {Shevchenko}}, \bibinfo {author}
  {\bibfnamefont {V.}~\bibnamefont {Voroshnin}}, \bibinfo {author}
  {\bibfnamefont {A.~G.}\ \bibnamefont {Rybkin}}, \bibinfo {author}
  {\bibfnamefont {I.~P.}\ \bibnamefont {Rusinov}}, \bibinfo {author}
  {\bibfnamefont {M.}~\bibnamefont {Blanco-Rey}}, \bibinfo {author}
  {\bibfnamefont {M.}~\bibnamefont {Hoffmann}}, \bibinfo {author}
  {\bibfnamefont {Z.~S.}\ \bibnamefont {Aliev}}, \bibinfo {author}
  {\bibfnamefont {M.~B.}\ \bibnamefont {Babanly}}, \bibinfo {author}
  {\bibfnamefont {I.~R.}\ \bibnamefont {Amiraslanov}}, \bibinfo {author}
  {\bibfnamefont {N.~A.}\ \bibnamefont {Abdullayev}}, \bibinfo {author}
  {\bibfnamefont {V.~N.}\ \bibnamefont {Zverev}}, \bibinfo {author}
  {\bibfnamefont {A.}~\bibnamefont {Kimura}}, \bibinfo {author} {\bibfnamefont
  {O.~E.}\ \bibnamefont {Tereshchenko}}, \bibinfo {author} {\bibfnamefont
  {K.~A.}\ \bibnamefont {Kokh}}, \bibinfo {author} {\bibfnamefont
  {L.}~\bibnamefont {Petaccia}}, \bibinfo {author} {\bibfnamefont
  {G.}~\bibnamefont {{Di Santo}}}, \bibinfo {author} {\bibfnamefont
  {A.}~\bibnamefont {Ernst}}, \bibinfo {author} {\bibfnamefont {P.~M.}\
  \bibnamefont {Echenique}}, \bibinfo {author} {\bibfnamefont {N.~T.}\
  \bibnamefont {Mamedov}}, \bibinfo {author} {\bibfnamefont {A.~M.}\
  \bibnamefont {Shikin}},\ and\ \bibinfo {author} {\bibfnamefont {E.~V.}\
  \bibnamefont {Chulkov}},\ }\bibfield  {title} {\bibinfo {title} {{Tunable
  3D/2D magnetism in the
  (${\mathrm{MnBi}}_{2}{\mathrm{Te}}_{4}$)(${\mathrm{Bi}}_{2}{\mathrm{Te}}_{3}$)m
  topological insulators family}},\ }\href
  {https://doi.org/10.1038/s41535-020-00255-9} {\bibfield  {journal} {\bibinfo
  {journal} {npj Quantum Mater.}\ }\textbf {\bibinfo {volume} {5}},\ \bibinfo
  {pages} {54} (\bibinfo {year} {2020})}\BibitemShut {NoStop}%
\bibitem [{\citenamefont {Hu}\ \emph {et~al.}(2020{\natexlab{a}})\citenamefont
  {Hu}, \citenamefont {Ding}, \citenamefont {Gordon}, \citenamefont {Ghosh},
  \citenamefont {Tien}, \citenamefont {Li}, \citenamefont {{Garrison Linn}},
  \citenamefont {Lien}, \citenamefont {Huang}, \citenamefont {Mackey},
  \citenamefont {Liu}, \citenamefont {{Sreenivasa Reddy}}, \citenamefont
  {Singh}, \citenamefont {Agarwal}, \citenamefont {Bansil}, \citenamefont
  {Song}, \citenamefont {Li}, \citenamefont {Xu}, \citenamefont {Lin},
  \citenamefont {Cao}, \citenamefont {Chang}, \citenamefont {Dessau},\ and\
  \citenamefont {Ni}}]{Hu2020}%
  \BibitemOpen
  \bibfield  {author} {\bibinfo {author} {\bibfnamefont {C.}~\bibnamefont
  {Hu}}, \bibinfo {author} {\bibfnamefont {L.}~\bibnamefont {Ding}}, \bibinfo
  {author} {\bibfnamefont {K.~N.}\ \bibnamefont {Gordon}}, \bibinfo {author}
  {\bibfnamefont {B.}~\bibnamefont {Ghosh}}, \bibinfo {author} {\bibfnamefont
  {H.~J.}\ \bibnamefont {Tien}}, \bibinfo {author} {\bibfnamefont
  {H.}~\bibnamefont {Li}}, \bibinfo {author} {\bibfnamefont {A.}~\bibnamefont
  {{Garrison Linn}}}, \bibinfo {author} {\bibfnamefont {S.~W.}\ \bibnamefont
  {Lien}}, \bibinfo {author} {\bibfnamefont {C.~Y.}\ \bibnamefont {Huang}},
  \bibinfo {author} {\bibfnamefont {S.}~\bibnamefont {Mackey}}, \bibinfo
  {author} {\bibfnamefont {J.}~\bibnamefont {Liu}}, \bibinfo {author}
  {\bibfnamefont {P.~V.}\ \bibnamefont {{Sreenivasa Reddy}}}, \bibinfo {author}
  {\bibfnamefont {B.}~\bibnamefont {Singh}}, \bibinfo {author} {\bibfnamefont
  {A.}~\bibnamefont {Agarwal}}, \bibinfo {author} {\bibfnamefont
  {A.}~\bibnamefont {Bansil}}, \bibinfo {author} {\bibfnamefont
  {M.}~\bibnamefont {Song}}, \bibinfo {author} {\bibfnamefont {D.}~\bibnamefont
  {Li}}, \bibinfo {author} {\bibfnamefont {S.~Y.}\ \bibnamefont {Xu}}, \bibinfo
  {author} {\bibfnamefont {H.}~\bibnamefont {Lin}}, \bibinfo {author}
  {\bibfnamefont {H.}~\bibnamefont {Cao}}, \bibinfo {author} {\bibfnamefont
  {T.~R.}\ \bibnamefont {Chang}}, \bibinfo {author} {\bibfnamefont
  {D.}~\bibnamefont {Dessau}},\ and\ \bibinfo {author} {\bibfnamefont
  {N.}~\bibnamefont {Ni}},\ }\bibfield  {title} {\bibinfo {title} {{Realization
  of an intrinsic ferromagnetic topological state in
  ${\mathrm{MnBi}}_{8}{\mathrm{Te}}_{13}$}},\ }\href
  {https://doi.org/10.1126/sciadv.aba4275} {\bibfield  {journal} {\bibinfo
  {journal} {Sci. Adv.}\ }\textbf {\bibinfo {volume} {6}},\ \bibinfo {pages}
  {eaba4275} (\bibinfo {year} {2020}{\natexlab{a}})}\BibitemShut {NoStop}%
\bibitem [{\citenamefont {Hu}\ \emph {et~al.}(2020{\natexlab{b}})\citenamefont
  {Hu}, \citenamefont {Gordon}, \citenamefont {Liu}, \citenamefont {Liu},
  \citenamefont {Zhou}, \citenamefont {Hao}, \citenamefont {Narayan},
  \citenamefont {Emmanouilidou}, \citenamefont {Sun}, \citenamefont {Liu},
  \citenamefont {Brawer}, \citenamefont {Ramirez}, \citenamefont {Ding},
  \citenamefont {Cao}, \citenamefont {Liu}, \citenamefont {Dessau},\ and\
  \citenamefont {Ni}}]{Hu2020b}%
  \BibitemOpen
  \bibfield  {author} {\bibinfo {author} {\bibfnamefont {C.}~\bibnamefont
  {Hu}}, \bibinfo {author} {\bibfnamefont {K.~N.}\ \bibnamefont {Gordon}},
  \bibinfo {author} {\bibfnamefont {P.}~\bibnamefont {Liu}}, \bibinfo {author}
  {\bibfnamefont {J.}~\bibnamefont {Liu}}, \bibinfo {author} {\bibfnamefont
  {X.}~\bibnamefont {Zhou}}, \bibinfo {author} {\bibfnamefont {P.}~\bibnamefont
  {Hao}}, \bibinfo {author} {\bibfnamefont {D.}~\bibnamefont {Narayan}},
  \bibinfo {author} {\bibfnamefont {E.}~\bibnamefont {Emmanouilidou}}, \bibinfo
  {author} {\bibfnamefont {H.}~\bibnamefont {Sun}}, \bibinfo {author}
  {\bibfnamefont {Y.}~\bibnamefont {Liu}}, \bibinfo {author} {\bibfnamefont
  {H.}~\bibnamefont {Brawer}}, \bibinfo {author} {\bibfnamefont {A.~P.}\
  \bibnamefont {Ramirez}}, \bibinfo {author} {\bibfnamefont {L.}~\bibnamefont
  {Ding}}, \bibinfo {author} {\bibfnamefont {H.}~\bibnamefont {Cao}}, \bibinfo
  {author} {\bibfnamefont {Q.}~\bibnamefont {Liu}}, \bibinfo {author}
  {\bibfnamefont {D.}~\bibnamefont {Dessau}},\ and\ \bibinfo {author}
  {\bibfnamefont {N.}~\bibnamefont {Ni}},\ }\bibfield  {title} {\bibinfo
  {title} {{A van der Waals antiferromagnetic topological insulator with weak
  interlayer magnetic coupling}},\ }\href
  {https://doi.org/10.1038/s41467-019-13814-x} {\bibfield  {journal} {\bibinfo
  {journal} {Nat. Commun.}\ }\textbf {\bibinfo {volume} {11}},\ \bibinfo
  {pages} {97} (\bibinfo {year} {2020}{\natexlab{b}})},\ \Eprint
  {https://arxiv.org/abs/1905.02154} {arXiv:1905.02154} \BibitemShut {NoStop}%
\bibitem [{\citenamefont {Wu}\ \emph {et~al.}(2020)\citenamefont {Wu},
  \citenamefont {Li}, \citenamefont {Ma}, \citenamefont {Zhang}, \citenamefont
  {Liu}, \citenamefont {Zhou}, \citenamefont {Shao}, \citenamefont {Wang},
  \citenamefont {Hao}, \citenamefont {Feng}, \citenamefont {Schwier},
  \citenamefont {Kumar}, \citenamefont {Sun}, \citenamefont {Liu},
  \citenamefont {Shimada}, \citenamefont {Miyamoto}, \citenamefont {Okuda},
  \citenamefont {Wang}, \citenamefont {Xie}, \citenamefont {Chen},
  \citenamefont {Liu}, \citenamefont {Liu},\ and\ \citenamefont
  {Zhao}}]{Wu2020}%
  \BibitemOpen
  \bibfield  {author} {\bibinfo {author} {\bibfnamefont {X.}~\bibnamefont
  {Wu}}, \bibinfo {author} {\bibfnamefont {J.}~\bibnamefont {Li}}, \bibinfo
  {author} {\bibfnamefont {X.~M.}\ \bibnamefont {Ma}}, \bibinfo {author}
  {\bibfnamefont {Y.}~\bibnamefont {Zhang}}, \bibinfo {author} {\bibfnamefont
  {Y.}~\bibnamefont {Liu}}, \bibinfo {author} {\bibfnamefont {C.~S.}\
  \bibnamefont {Zhou}}, \bibinfo {author} {\bibfnamefont {J.}~\bibnamefont
  {Shao}}, \bibinfo {author} {\bibfnamefont {Q.}~\bibnamefont {Wang}}, \bibinfo
  {author} {\bibfnamefont {Y.~J.}\ \bibnamefont {Hao}}, \bibinfo {author}
  {\bibfnamefont {Y.}~\bibnamefont {Feng}}, \bibinfo {author} {\bibfnamefont
  {E.~F.}\ \bibnamefont {Schwier}}, \bibinfo {author} {\bibfnamefont
  {S.}~\bibnamefont {Kumar}}, \bibinfo {author} {\bibfnamefont
  {H.}~\bibnamefont {Sun}}, \bibinfo {author} {\bibfnamefont {P.}~\bibnamefont
  {Liu}}, \bibinfo {author} {\bibfnamefont {K.}~\bibnamefont {Shimada}},
  \bibinfo {author} {\bibfnamefont {K.}~\bibnamefont {Miyamoto}}, \bibinfo
  {author} {\bibfnamefont {T.}~\bibnamefont {Okuda}}, \bibinfo {author}
  {\bibfnamefont {K.}~\bibnamefont {Wang}}, \bibinfo {author} {\bibfnamefont
  {M.}~\bibnamefont {Xie}}, \bibinfo {author} {\bibfnamefont {C.}~\bibnamefont
  {Chen}}, \bibinfo {author} {\bibfnamefont {Q.}~\bibnamefont {Liu}}, \bibinfo
  {author} {\bibfnamefont {C.}~\bibnamefont {Liu}},\ and\ \bibinfo {author}
  {\bibfnamefont {Y.}~\bibnamefont {Zhao}},\ }\bibfield  {title} {\bibinfo
  {title} {{Distinct Topological Surface States on the Two Terminations of
  ${\mathrm{MnBi}}_{4}{\mathrm{Te}}_{7}$}},\ }\href
  {https://doi.org/10.1103/PhysRevX.10.031013} {\bibfield  {journal} {\bibinfo
  {journal} {Phys. Rev. X}\ }\textbf {\bibinfo {volume} {10}},\ \bibinfo
  {pages} {031013} (\bibinfo {year} {2020})},\ \Eprint
  {https://arxiv.org/abs/2002.00320} {arXiv:2002.00320} \BibitemShut {NoStop}%
\bibitem [{\citenamefont {Jo}\ \emph {et~al.}(2020)\citenamefont {Jo},
  \citenamefont {Wang}, \citenamefont {Slager}, \citenamefont {Yan},
  \citenamefont {Wu}, \citenamefont {Lee}, \citenamefont {Schrunk},
  \citenamefont {Vishwanath},\ and\ \citenamefont {Kaminski}}]{Jo2020}%
  \BibitemOpen
  \bibfield  {author} {\bibinfo {author} {\bibfnamefont {N.~H.}\ \bibnamefont
  {Jo}}, \bibinfo {author} {\bibfnamefont {L.-L.}\ \bibnamefont {Wang}},
  \bibinfo {author} {\bibfnamefont {R.-J.}\ \bibnamefont {Slager}}, \bibinfo
  {author} {\bibfnamefont {J.}~\bibnamefont {Yan}}, \bibinfo {author}
  {\bibfnamefont {Y.}~\bibnamefont {Wu}}, \bibinfo {author} {\bibfnamefont
  {K.}~\bibnamefont {Lee}}, \bibinfo {author} {\bibfnamefont {B.}~\bibnamefont
  {Schrunk}}, \bibinfo {author} {\bibfnamefont {A.}~\bibnamefont
  {Vishwanath}},\ and\ \bibinfo {author} {\bibfnamefont {A.}~\bibnamefont
  {Kaminski}},\ }\bibfield  {title} {\bibinfo {title} {{Intrinsic axion
  insulating behavior in antiferromagnetic $MnBi_{6}Te_{10}$}},\ }\href
  {https://doi.org/10.1103/PhysRevB.102.045130} {\bibfield  {journal} {\bibinfo
   {journal} {Phys. Rev. B}\ }\textbf {\bibinfo {volume} {102}},\ \bibinfo
  {pages} {045130} (\bibinfo {year} {2020})}\BibitemShut {NoStop}%
\bibitem [{\citenamefont {Xu}\ \emph {et~al.}(2020)\citenamefont {Xu},
  \citenamefont {Mao}, \citenamefont {Wang}, \citenamefont {Li}, \citenamefont
  {Chen}, \citenamefont {Xia}, \citenamefont {Li}, \citenamefont {Pei},
  \citenamefont {Zhang}, \citenamefont {Zheng}, \citenamefont {Huang},
  \citenamefont {Zhang}, \citenamefont {Cui}, \citenamefont {Liang},
  \citenamefont {Xia}, \citenamefont {Su}, \citenamefont {Jung}, \citenamefont
  {Cacho}, \citenamefont {Wang}, \citenamefont {Li}, \citenamefont {Xu},
  \citenamefont {Guo}, \citenamefont {Yang}, \citenamefont {Liu}, \citenamefont
  {Chen},\ and\ \citenamefont {Jiang}}]{Xu2020}%
  \BibitemOpen
  \bibfield  {author} {\bibinfo {author} {\bibfnamefont {L.}~\bibnamefont
  {Xu}}, \bibinfo {author} {\bibfnamefont {Y.}~\bibnamefont {Mao}}, \bibinfo
  {author} {\bibfnamefont {H.}~\bibnamefont {Wang}}, \bibinfo {author}
  {\bibfnamefont {J.}~\bibnamefont {Li}}, \bibinfo {author} {\bibfnamefont
  {Y.}~\bibnamefont {Chen}}, \bibinfo {author} {\bibfnamefont {Y.}~\bibnamefont
  {Xia}}, \bibinfo {author} {\bibfnamefont {Y.}~\bibnamefont {Li}}, \bibinfo
  {author} {\bibfnamefont {D.}~\bibnamefont {Pei}}, \bibinfo {author}
  {\bibfnamefont {J.}~\bibnamefont {Zhang}}, \bibinfo {author} {\bibfnamefont
  {H.}~\bibnamefont {Zheng}}, \bibinfo {author} {\bibfnamefont
  {K.}~\bibnamefont {Huang}}, \bibinfo {author} {\bibfnamefont
  {C.}~\bibnamefont {Zhang}}, \bibinfo {author} {\bibfnamefont
  {S.}~\bibnamefont {Cui}}, \bibinfo {author} {\bibfnamefont {A.}~\bibnamefont
  {Liang}}, \bibinfo {author} {\bibfnamefont {W.}~\bibnamefont {Xia}}, \bibinfo
  {author} {\bibfnamefont {H.}~\bibnamefont {Su}}, \bibinfo {author}
  {\bibfnamefont {S.}~\bibnamefont {Jung}}, \bibinfo {author} {\bibfnamefont
  {C.}~\bibnamefont {Cacho}}, \bibinfo {author} {\bibfnamefont
  {M.}~\bibnamefont {Wang}}, \bibinfo {author} {\bibfnamefont {G.}~\bibnamefont
  {Li}}, \bibinfo {author} {\bibfnamefont {Y.}~\bibnamefont {Xu}}, \bibinfo
  {author} {\bibfnamefont {Y.}~\bibnamefont {Guo}}, \bibinfo {author}
  {\bibfnamefont {L.}~\bibnamefont {Yang}}, \bibinfo {author} {\bibfnamefont
  {Z.}~\bibnamefont {Liu}}, \bibinfo {author} {\bibfnamefont {Y.}~\bibnamefont
  {Chen}},\ and\ \bibinfo {author} {\bibfnamefont {M.}~\bibnamefont {Jiang}},\
  }\bibfield  {title} {\bibinfo {title} {{Persistent surface states with
  diminishing gap in
  ${\mathrm{MnBi}}_{2}{\mathrm{Te}}_{4}$/${\mathrm{Bi}}_{2}{\mathrm{Te}}_{3}$
  superlattice antiferromagnetic topological insulator}},\ }\href
  {https://doi.org/10.1016/j.scib.2020.07.032} {\bibfield  {journal} {\bibinfo
  {journal} {Sci. Bull.}\ }\textbf {\bibinfo {volume} {65}},\ \bibinfo {pages}
  {2086} (\bibinfo {year} {2020})}\BibitemShut {NoStop}%
\bibitem [{\citenamefont {Xie}\ \emph {et~al.}(2020)\citenamefont {Xie},
  \citenamefont {Wang}, \citenamefont {Cai}, \citenamefont {Chen},
  \citenamefont {Guo}, \citenamefont {Naveed}, \citenamefont {Zhang},
  \citenamefont {Zhang}, \citenamefont {Wang}, \citenamefont {Fei},
  \citenamefont {Zhang},\ and\ \citenamefont {Song}}]{Xie2020}%
  \BibitemOpen
  \bibfield  {author} {\bibinfo {author} {\bibfnamefont {H.}~\bibnamefont
  {Xie}}, \bibinfo {author} {\bibfnamefont {D.}~\bibnamefont {Wang}}, \bibinfo
  {author} {\bibfnamefont {Z.}~\bibnamefont {Cai}}, \bibinfo {author}
  {\bibfnamefont {B.}~\bibnamefont {Chen}}, \bibinfo {author} {\bibfnamefont
  {J.}~\bibnamefont {Guo}}, \bibinfo {author} {\bibfnamefont {M.}~\bibnamefont
  {Naveed}}, \bibinfo {author} {\bibfnamefont {S.}~\bibnamefont {Zhang}},
  \bibinfo {author} {\bibfnamefont {M.}~\bibnamefont {Zhang}}, \bibinfo
  {author} {\bibfnamefont {X.}~\bibnamefont {Wang}}, \bibinfo {author}
  {\bibfnamefont {F.}~\bibnamefont {Fei}}, \bibinfo {author} {\bibfnamefont
  {H.}~\bibnamefont {Zhang}},\ and\ \bibinfo {author} {\bibfnamefont
  {F.}~\bibnamefont {Song}},\ }\bibfield  {title} {\bibinfo {title} {{The
  mechanism exploration for zero-field ferromagnetism in intrinsic topological
  insulator ${\mathrm{MnBi}}_{2}{\mathrm{Te}}_{4}$ by
  ${\mathrm{Bi}}_{2}{\mathrm{Te}}_{3}$ intercalations}},\ }\href
  {https://doi.org/10.1063/5.0009085} {\bibfield  {journal} {\bibinfo
  {journal} {Appl. Phys. Lett.}\ }\textbf {\bibinfo {volume} {116}},\ \bibinfo
  {pages} {221902} (\bibinfo {year} {2020})}\BibitemShut {NoStop}%
\bibitem [{\citenamefont {Vidal}\ \emph {et~al.}(2021)\citenamefont {Vidal},
  \citenamefont {Bentmann}, \citenamefont {Facio}, \citenamefont {Heider},
  \citenamefont {Kagerer}, \citenamefont {Fornari}, \citenamefont {Peixoto},
  \citenamefont {Figgemeier}, \citenamefont {Jung}, \citenamefont {Cacho},
  \citenamefont {B{\"{u}}chner}, \citenamefont {{Van Den Brink}}, \citenamefont
  {Schneider}, \citenamefont {Plucinski}, \citenamefont {Schwier},
  \citenamefont {Shimada}, \citenamefont {Richter}, \citenamefont {Isaeva},\
  and\ \citenamefont {Reinert}}]{Vidal2021}%
  \BibitemOpen
  \bibfield  {author} {\bibinfo {author} {\bibfnamefont {R.~C.}\ \bibnamefont
  {Vidal}}, \bibinfo {author} {\bibfnamefont {H.}~\bibnamefont {Bentmann}},
  \bibinfo {author} {\bibfnamefont {J.~I.}\ \bibnamefont {Facio}}, \bibinfo
  {author} {\bibfnamefont {T.}~\bibnamefont {Heider}}, \bibinfo {author}
  {\bibfnamefont {P.}~\bibnamefont {Kagerer}}, \bibinfo {author} {\bibfnamefont
  {C.~I.}\ \bibnamefont {Fornari}}, \bibinfo {author} {\bibfnamefont {T.~R.}\
  \bibnamefont {Peixoto}}, \bibinfo {author} {\bibfnamefont {T.}~\bibnamefont
  {Figgemeier}}, \bibinfo {author} {\bibfnamefont {S.}~\bibnamefont {Jung}},
  \bibinfo {author} {\bibfnamefont {C.}~\bibnamefont {Cacho}}, \bibinfo
  {author} {\bibfnamefont {B.}~\bibnamefont {B{\"{u}}chner}}, \bibinfo {author}
  {\bibfnamefont {J.}~\bibnamefont {{Van Den Brink}}}, \bibinfo {author}
  {\bibfnamefont {C.~M.}\ \bibnamefont {Schneider}}, \bibinfo {author}
  {\bibfnamefont {L.}~\bibnamefont {Plucinski}}, \bibinfo {author}
  {\bibfnamefont {E.~F.}\ \bibnamefont {Schwier}}, \bibinfo {author}
  {\bibfnamefont {K.}~\bibnamefont {Shimada}}, \bibinfo {author} {\bibfnamefont
  {M.}~\bibnamefont {Richter}}, \bibinfo {author} {\bibfnamefont
  {A.}~\bibnamefont {Isaeva}},\ and\ \bibinfo {author} {\bibfnamefont
  {F.}~\bibnamefont {Reinert}},\ }\bibfield  {title} {\bibinfo {title}
  {{Orbital Complexity in Intrinsic Magnetic Topological Insulators
  ${\mathrm{MnBi}}_{4}{\mathrm{Te}}_{7}$ and
  ${\mathrm{MnBi}}_{6}{\mathrm{Te}}_{10}$}},\ }\href
  {https://doi.org/10.1103/PhysRevLett.126.176403} {\bibfield  {journal}
  {\bibinfo  {journal} {Phys. Rev. Lett.}\ }\textbf {\bibinfo {volume} {126}},\
  \bibinfo {pages} {176403} (\bibinfo {year} {2021})},\ \Eprint
  {https://arxiv.org/abs/2007.07637} {arXiv:2007.07637} \BibitemShut {NoStop}%
\bibitem [{\citenamefont {Jia}\ \emph {et~al.}(2021)\citenamefont {Jia},
  \citenamefont {Zhang}, \citenamefont {Ying}, \citenamefont {Xie},
  \citenamefont {Chen}, \citenamefont {Naveed}, \citenamefont {Fei},
  \citenamefont {Zhang}, \citenamefont {Pan},\ and\ \citenamefont
  {Song}}]{Jia2021}%
  \BibitemOpen
  \bibfield  {author} {\bibinfo {author} {\bibfnamefont {B.}~\bibnamefont
  {Jia}}, \bibinfo {author} {\bibfnamefont {S.}~\bibnamefont {Zhang}}, \bibinfo
  {author} {\bibfnamefont {Z.}~\bibnamefont {Ying}}, \bibinfo {author}
  {\bibfnamefont {H.}~\bibnamefont {Xie}}, \bibinfo {author} {\bibfnamefont
  {B.}~\bibnamefont {Chen}}, \bibinfo {author} {\bibfnamefont {M.}~\bibnamefont
  {Naveed}}, \bibinfo {author} {\bibfnamefont {F.}~\bibnamefont {Fei}},
  \bibinfo {author} {\bibfnamefont {M.}~\bibnamefont {Zhang}}, \bibinfo
  {author} {\bibfnamefont {D.}~\bibnamefont {Pan}},\ and\ \bibinfo {author}
  {\bibfnamefont {F.}~\bibnamefont {Song}},\ }\bibfield  {title} {\bibinfo
  {title} {{Unconventional anomalous Hall effect in magnetic topological
  insulator ${\mathrm{MnBi}}_{4}{\mathrm{Te}}_{7}$ device}},\ }\href
  {https://doi.org/10.1063/5.0041532} {\bibfield  {journal} {\bibinfo
  {journal} {Appl. Phys. Lett.}\ }\textbf {\bibinfo {volume} {118}},\ \bibinfo
  {pages} {083101} (\bibinfo {year} {2021})}\BibitemShut {NoStop}%
\bibitem [{\citenamefont {Chen}\ \emph {et~al.}(2021)\citenamefont {Chen},
  \citenamefont {Wang}, \citenamefont {Jiang}, \citenamefont {Zhang},
  \citenamefont {Cui}, \citenamefont {Guo}, \citenamefont {Xie}, \citenamefont
  {Zhang}, \citenamefont {Naveed}, \citenamefont {Du}, \citenamefont {Wang},
  \citenamefont {Zhang}, \citenamefont {Fei}, \citenamefont {Shen},
  \citenamefont {Sun},\ and\ \citenamefont {Song}}]{Chen2021}%
  \BibitemOpen
  \bibfield  {author} {\bibinfo {author} {\bibfnamefont {B.}~\bibnamefont
  {Chen}}, \bibinfo {author} {\bibfnamefont {D.}~\bibnamefont {Wang}}, \bibinfo
  {author} {\bibfnamefont {Z.}~\bibnamefont {Jiang}}, \bibinfo {author}
  {\bibfnamefont {B.}~\bibnamefont {Zhang}}, \bibinfo {author} {\bibfnamefont
  {S.}~\bibnamefont {Cui}}, \bibinfo {author} {\bibfnamefont {J.}~\bibnamefont
  {Guo}}, \bibinfo {author} {\bibfnamefont {H.}~\bibnamefont {Xie}}, \bibinfo
  {author} {\bibfnamefont {Y.}~\bibnamefont {Zhang}}, \bibinfo {author}
  {\bibfnamefont {M.}~\bibnamefont {Naveed}}, \bibinfo {author} {\bibfnamefont
  {Y.}~\bibnamefont {Du}}, \bibinfo {author} {\bibfnamefont {X.}~\bibnamefont
  {Wang}}, \bibinfo {author} {\bibfnamefont {H.}~\bibnamefont {Zhang}},
  \bibinfo {author} {\bibfnamefont {F.}~\bibnamefont {Fei}}, \bibinfo {author}
  {\bibfnamefont {D.}~\bibnamefont {Shen}}, \bibinfo {author} {\bibfnamefont
  {Z.}~\bibnamefont {Sun}},\ and\ \bibinfo {author} {\bibfnamefont
  {F.}~\bibnamefont {Song}},\ }\bibfield  {title} {\bibinfo {title}
  {{Coexistence of ferromagnetism and topology by charge carrier engineering in
  the intrinsic magnetic topological insulator
  ${\mathrm{MnBi}}_{4}{\mathrm{Te}}_{7}$}},\ }\href
  {https://doi.org/10.1103/PhysRevB.104.075134} {\bibfield  {journal} {\bibinfo
   {journal} {Phys. Rev. B}\ }\textbf {\bibinfo {volume} {104}},\ \bibinfo
  {pages} {075134} (\bibinfo {year} {2021})},\ \Eprint
  {https://arxiv.org/abs/2009.00039} {arXiv:2009.00039} \BibitemShut {NoStop}%
\bibitem [{\citenamefont {Lu}\ \emph {et~al.}(2021)\citenamefont {Lu},
  \citenamefont {Sun}, \citenamefont {Kumar}, \citenamefont {Wang},
  \citenamefont {Gu}, \citenamefont {Zeng}, \citenamefont {Hao}, \citenamefont
  {Li}, \citenamefont {Shao}, \citenamefont {Ma}, \citenamefont {Hao},
  \citenamefont {Zhang}, \citenamefont {Mansuer}, \citenamefont {Mei},
  \citenamefont {Zhao}, \citenamefont {Liu}, \citenamefont {Deng},
  \citenamefont {Huang}, \citenamefont {Shen}, \citenamefont {Shimada},
  \citenamefont {Schwier}, \citenamefont {Liu}, \citenamefont {Liu},\ and\
  \citenamefont {Chen}}]{Lu2021}%
  \BibitemOpen
  \bibfield  {author} {\bibinfo {author} {\bibfnamefont {R.}~\bibnamefont
  {Lu}}, \bibinfo {author} {\bibfnamefont {H.}~\bibnamefont {Sun}}, \bibinfo
  {author} {\bibfnamefont {S.}~\bibnamefont {Kumar}}, \bibinfo {author}
  {\bibfnamefont {Y.}~\bibnamefont {Wang}}, \bibinfo {author} {\bibfnamefont
  {M.}~\bibnamefont {Gu}}, \bibinfo {author} {\bibfnamefont {M.}~\bibnamefont
  {Zeng}}, \bibinfo {author} {\bibfnamefont {Y.~J.}\ \bibnamefont {Hao}},
  \bibinfo {author} {\bibfnamefont {J.}~\bibnamefont {Li}}, \bibinfo {author}
  {\bibfnamefont {J.}~\bibnamefont {Shao}}, \bibinfo {author} {\bibfnamefont
  {X.~M.}\ \bibnamefont {Ma}}, \bibinfo {author} {\bibfnamefont
  {Z.}~\bibnamefont {Hao}}, \bibinfo {author} {\bibfnamefont {K.}~\bibnamefont
  {Zhang}}, \bibinfo {author} {\bibfnamefont {W.}~\bibnamefont {Mansuer}},
  \bibinfo {author} {\bibfnamefont {J.}~\bibnamefont {Mei}}, \bibinfo {author}
  {\bibfnamefont {Y.}~\bibnamefont {Zhao}}, \bibinfo {author} {\bibfnamefont
  {C.}~\bibnamefont {Liu}}, \bibinfo {author} {\bibfnamefont {K.}~\bibnamefont
  {Deng}}, \bibinfo {author} {\bibfnamefont {W.}~\bibnamefont {Huang}},
  \bibinfo {author} {\bibfnamefont {B.}~\bibnamefont {Shen}}, \bibinfo {author}
  {\bibfnamefont {K.}~\bibnamefont {Shimada}}, \bibinfo {author} {\bibfnamefont
  {E.~F.}\ \bibnamefont {Schwier}}, \bibinfo {author} {\bibfnamefont
  {C.}~\bibnamefont {Liu}}, \bibinfo {author} {\bibfnamefont {Q.}~\bibnamefont
  {Liu}},\ and\ \bibinfo {author} {\bibfnamefont {C.}~\bibnamefont {Chen}},\
  }\bibfield  {title} {\bibinfo {title} {{Half-Magnetic Topological Insulator
  with Magnetization-Induced Dirac Gap at a Selected Surface}},\ }\href
  {https://doi.org/10.1103/PhysRevX.11.011039} {\bibfield  {journal} {\bibinfo
  {journal} {Phys. Rev. X}\ }\textbf {\bibinfo {volume} {11}},\ \bibinfo
  {pages} {011039} (\bibinfo {year} {2021})}\BibitemShut {NoStop}%
\bibitem [{\citenamefont {Shi}\ \emph {et~al.}(2020)\citenamefont {Shi},
  \citenamefont {Zhang}, \citenamefont {Yan}, \citenamefont {Feng},
  \citenamefont {Yang}, \citenamefont {Shi},\ and\ \citenamefont
  {Li}}]{Shi2020}%
  \BibitemOpen
  \bibfield  {author} {\bibinfo {author} {\bibfnamefont {G.}~\bibnamefont
  {Shi}}, \bibinfo {author} {\bibfnamefont {M.}~\bibnamefont {Zhang}}, \bibinfo
  {author} {\bibfnamefont {D.}~\bibnamefont {Yan}}, \bibinfo {author}
  {\bibfnamefont {H.}~\bibnamefont {Feng}}, \bibinfo {author} {\bibfnamefont
  {M.}~\bibnamefont {Yang}}, \bibinfo {author} {\bibfnamefont {Y.}~\bibnamefont
  {Shi}},\ and\ \bibinfo {author} {\bibfnamefont {Y.}~\bibnamefont {Li}},\
  }\bibfield  {title} {\bibinfo {title} {{Anomalous Hall Effect in Layered
  Ferrimagnet ${\mathrm{MnSb}}_{2}{\mathrm{Te}}_{4}$}},\ }\href
  {https://doi.org/10.1088/0256-307X/37/4/047301} {\bibfield  {journal}
  {\bibinfo  {journal} {Chinese Phys. Lett.}\ }\textbf {\bibinfo {volume}
  {37}},\ \bibinfo {pages} {047301} (\bibinfo {year} {2020})}\BibitemShut
  {NoStop}%
\bibitem [{\citenamefont {Rani}\ \emph {et~al.}(2019)\citenamefont {Rani},
  \citenamefont {Saxena}, \citenamefont {Sultana}, \citenamefont {Nagpal},
  \citenamefont {Islam}, \citenamefont {Patnaik},\ and\ \citenamefont
  {Awana}}]{Rani2019}%
  \BibitemOpen
  \bibfield  {author} {\bibinfo {author} {\bibfnamefont {P.}~\bibnamefont
  {Rani}}, \bibinfo {author} {\bibfnamefont {A.}~\bibnamefont {Saxena}},
  \bibinfo {author} {\bibfnamefont {R.}~\bibnamefont {Sultana}}, \bibinfo
  {author} {\bibfnamefont {V.}~\bibnamefont {Nagpal}}, \bibinfo {author}
  {\bibfnamefont {S.~S.}\ \bibnamefont {Islam}}, \bibinfo {author}
  {\bibfnamefont {S.}~\bibnamefont {Patnaik}},\ and\ \bibinfo {author}
  {\bibfnamefont {V.~P.}\ \bibnamefont {Awana}},\ }\bibfield  {title} {\bibinfo
  {title} {{Crystal Growth and Basic Transport and Magnetic Properties of
  ${\mathrm{MnBi}}_{2}{\mathrm{Te}}_{4}$}},\ }\href
  {https://doi.org/10.1007/s10948-019-05342-y} {\bibfield  {journal} {\bibinfo
  {journal} {J. Supercond. Novel Magn.}\ }\textbf {\bibinfo {volume} {32}},\
  \bibinfo {pages} {3705} (\bibinfo {year} {2019})},\ \Eprint
  {https://arxiv.org/abs/1906.09038} {arXiv:1906.09038} \BibitemShut {NoStop}%
\bibitem [{\citenamefont {Gong}\ \emph {et~al.}(2019)\citenamefont {Gong},
  \citenamefont {Guo}, \citenamefont {Li}, \citenamefont {Zhu}, \citenamefont
  {Liao}, \citenamefont {Liu}, \citenamefont {Zhang}, \citenamefont {Gu},
  \citenamefont {Tang}, \citenamefont {Feng}, \citenamefont {Zhang},
  \citenamefont {Li}, \citenamefont {Song}, \citenamefont {Wang}, \citenamefont
  {Yu}, \citenamefont {Chen}, \citenamefont {Wang}, \citenamefont {Yao},
  \citenamefont {Duan}, \citenamefont {Xu}, \citenamefont {Zhang},
  \citenamefont {Ma}, \citenamefont {Xue},\ and\ \citenamefont
  {He}}]{Gong2019}%
  \BibitemOpen
  \bibfield  {author} {\bibinfo {author} {\bibfnamefont {Y.}~\bibnamefont
  {Gong}}, \bibinfo {author} {\bibfnamefont {J.}~\bibnamefont {Guo}}, \bibinfo
  {author} {\bibfnamefont {J.}~\bibnamefont {Li}}, \bibinfo {author}
  {\bibfnamefont {K.}~\bibnamefont {Zhu}}, \bibinfo {author} {\bibfnamefont
  {M.}~\bibnamefont {Liao}}, \bibinfo {author} {\bibfnamefont {X.}~\bibnamefont
  {Liu}}, \bibinfo {author} {\bibfnamefont {Q.}~\bibnamefont {Zhang}}, \bibinfo
  {author} {\bibfnamefont {L.}~\bibnamefont {Gu}}, \bibinfo {author}
  {\bibfnamefont {L.}~\bibnamefont {Tang}}, \bibinfo {author} {\bibfnamefont
  {X.}~\bibnamefont {Feng}}, \bibinfo {author} {\bibfnamefont {D.}~\bibnamefont
  {Zhang}}, \bibinfo {author} {\bibfnamefont {W.}~\bibnamefont {Li}}, \bibinfo
  {author} {\bibfnamefont {C.}~\bibnamefont {Song}}, \bibinfo {author}
  {\bibfnamefont {L.}~\bibnamefont {Wang}}, \bibinfo {author} {\bibfnamefont
  {P.}~\bibnamefont {Yu}}, \bibinfo {author} {\bibfnamefont {X.}~\bibnamefont
  {Chen}}, \bibinfo {author} {\bibfnamefont {Y.}~\bibnamefont {Wang}}, \bibinfo
  {author} {\bibfnamefont {H.}~\bibnamefont {Yao}}, \bibinfo {author}
  {\bibfnamefont {W.}~\bibnamefont {Duan}}, \bibinfo {author} {\bibfnamefont
  {Y.}~\bibnamefont {Xu}}, \bibinfo {author} {\bibfnamefont {S.~C.}\
  \bibnamefont {Zhang}}, \bibinfo {author} {\bibfnamefont {X.}~\bibnamefont
  {Ma}}, \bibinfo {author} {\bibfnamefont {Q.~K.}\ \bibnamefont {Xue}},\ and\
  \bibinfo {author} {\bibfnamefont {K.}~\bibnamefont {He}},\ }\bibfield
  {title} {\bibinfo {title} {{Experimental Realization of an Intrinsic Magnetic
  Topological Insulator}},\ }\href
  {https://doi.org/10.1088/0256-307X/36/7/076801} {\bibfield  {journal}
  {\bibinfo  {journal} {Chinese Phys. Lett.}\ }\textbf {\bibinfo {volume}
  {36}},\ \bibinfo {pages} {076801} (\bibinfo {year} {2019})},\ \Eprint
  {https://arxiv.org/abs/1809.07926} {arXiv:1809.07926} \BibitemShut {NoStop}%
\bibitem [{\citenamefont {Qi}\ \emph {et~al.}(2020)\citenamefont {Qi},
  \citenamefont {Gao}, \citenamefont {Chang}, \citenamefont {Han},\ and\
  \citenamefont {Qiao}}]{Qi2020}%
  \BibitemOpen
  \bibfield  {author} {\bibinfo {author} {\bibfnamefont {S.}~\bibnamefont
  {Qi}}, \bibinfo {author} {\bibfnamefont {R.}~\bibnamefont {Gao}}, \bibinfo
  {author} {\bibfnamefont {M.}~\bibnamefont {Chang}}, \bibinfo {author}
  {\bibfnamefont {Y.}~\bibnamefont {Han}},\ and\ \bibinfo {author}
  {\bibfnamefont {Z.}~\bibnamefont {Qiao}},\ }\bibfield  {title} {\bibinfo
  {title} {{Pursuing the high-temperature quantum anomalous Hall effect in
  ${\mathrm{MnBi}}_{2}{\mathrm{Te}}_{4}$/${\mathrm{Sb}}_{2}{\mathrm{Te}}_{3}$
  heterostructures}},\ }\href {https://doi.org/10.1103/PhysRevB.101.014423}
  {\bibfield  {journal} {\bibinfo  {journal} {Phys. Rev. B}\ }\textbf {\bibinfo
  {volume} {101}},\ \bibinfo {pages} {014423} (\bibinfo {year} {2020})},\
  \Eprint {https://arxiv.org/abs/1908.00498} {arXiv:1908.00498} \BibitemShut
  {NoStop}%
\bibitem [{\citenamefont {He}(2020)}]{He2020}%
  \BibitemOpen
  \bibfield  {author} {\bibinfo {author} {\bibfnamefont {K.}~\bibnamefont
  {He}},\ }\bibfield  {title} {\bibinfo {title}
  {{${\mathrm{MnBi}}_{2}{\mathrm{Te}}_{4}$-family intrinsic magnetic
  topological materials}},\ }\href {https://doi.org/10.1038/s41535-020-00291-5}
  {\bibfield  {journal} {\bibinfo  {journal} {npj Quantum Mater.}\ }\textbf
  {\bibinfo {volume} {5}},\ \bibinfo {pages} {90} (\bibinfo {year}
  {2020})}\BibitemShut {NoStop}%
\bibitem [{\citenamefont {Wu}\ \emph {et~al.}(2019)\citenamefont {Wu},
  \citenamefont {Liu}, \citenamefont {Sasase}, \citenamefont {Ienaga},
  \citenamefont {Obata}, \citenamefont {Yukawa}, \citenamefont {Horiba},
  \citenamefont {Kumigashira}, \citenamefont {Okuma}, \citenamefont
  {Inoshita},\ and\ \citenamefont {Hosono}}]{Wu2019}%
  \BibitemOpen
  \bibfield  {author} {\bibinfo {author} {\bibfnamefont {J.}~\bibnamefont
  {Wu}}, \bibinfo {author} {\bibfnamefont {F.}~\bibnamefont {Liu}}, \bibinfo
  {author} {\bibfnamefont {M.}~\bibnamefont {Sasase}}, \bibinfo {author}
  {\bibfnamefont {K.}~\bibnamefont {Ienaga}}, \bibinfo {author} {\bibfnamefont
  {Y.}~\bibnamefont {Obata}}, \bibinfo {author} {\bibfnamefont
  {R.}~\bibnamefont {Yukawa}}, \bibinfo {author} {\bibfnamefont
  {K.}~\bibnamefont {Horiba}}, \bibinfo {author} {\bibfnamefont
  {H.}~\bibnamefont {Kumigashira}}, \bibinfo {author} {\bibfnamefont
  {S.}~\bibnamefont {Okuma}}, \bibinfo {author} {\bibfnamefont
  {T.}~\bibnamefont {Inoshita}},\ and\ \bibinfo {author} {\bibfnamefont
  {H.}~\bibnamefont {Hosono}},\ }\bibfield  {title} {\bibinfo {title} {{Natural
  van der Waals heterostructural single crystals with both magnetic and
  topological properties}},\ }\href {https://doi.org/10.1126/sciadv.aax9989}
  {\bibfield  {journal} {\bibinfo  {journal} {Sci. Adv.}\ }\textbf {\bibinfo
  {volume} {5}},\ \bibinfo {pages} {eaax9989} (\bibinfo {year}
  {2019})}\BibitemShut {NoStop}%
\bibitem [{\citenamefont {Vidal}\ \emph
  {et~al.}(2019{\natexlab{b}})\citenamefont {Vidal}, \citenamefont {Zeugner},
  \citenamefont {Facio}, \citenamefont {Ray}, \citenamefont {Haghighi},
  \citenamefont {Wolter}, \citenamefont {{Corredor Bohorquez}}, \citenamefont
  {Caglieris}, \citenamefont {Moser}, \citenamefont {Figgemeier}, \citenamefont
  {Peixoto}, \citenamefont {Vasili}, \citenamefont {Valvidares}, \citenamefont
  {Jung}, \citenamefont {Cacho}, \citenamefont {Alfonsov}, \citenamefont
  {Mehlawat}, \citenamefont {Kataev}, \citenamefont {Hess}, \citenamefont
  {Richter}, \citenamefont {B{\"{u}}chner}, \citenamefont {{Van Den Brink}},
  \citenamefont {Ruck}, \citenamefont {Reinert}, \citenamefont {Bentmann},\
  and\ \citenamefont {Isaeva}}]{Vidal2019}%
  \BibitemOpen
  \bibfield  {author} {\bibinfo {author} {\bibfnamefont {R.~C.}\ \bibnamefont
  {Vidal}}, \bibinfo {author} {\bibfnamefont {A.}~\bibnamefont {Zeugner}},
  \bibinfo {author} {\bibfnamefont {J.~I.}\ \bibnamefont {Facio}}, \bibinfo
  {author} {\bibfnamefont {R.}~\bibnamefont {Ray}}, \bibinfo {author}
  {\bibfnamefont {M.~H.}\ \bibnamefont {Haghighi}}, \bibinfo {author}
  {\bibfnamefont {A.~U.}\ \bibnamefont {Wolter}}, \bibinfo {author}
  {\bibfnamefont {L.~T.}\ \bibnamefont {{Corredor Bohorquez}}}, \bibinfo
  {author} {\bibfnamefont {F.}~\bibnamefont {Caglieris}}, \bibinfo {author}
  {\bibfnamefont {S.}~\bibnamefont {Moser}}, \bibinfo {author} {\bibfnamefont
  {T.}~\bibnamefont {Figgemeier}}, \bibinfo {author} {\bibfnamefont {T.~R.}\
  \bibnamefont {Peixoto}}, \bibinfo {author} {\bibfnamefont {H.~B.}\
  \bibnamefont {Vasili}}, \bibinfo {author} {\bibfnamefont {M.}~\bibnamefont
  {Valvidares}}, \bibinfo {author} {\bibfnamefont {S.}~\bibnamefont {Jung}},
  \bibinfo {author} {\bibfnamefont {C.}~\bibnamefont {Cacho}}, \bibinfo
  {author} {\bibfnamefont {A.}~\bibnamefont {Alfonsov}}, \bibinfo {author}
  {\bibfnamefont {K.}~\bibnamefont {Mehlawat}}, \bibinfo {author}
  {\bibfnamefont {V.}~\bibnamefont {Kataev}}, \bibinfo {author} {\bibfnamefont
  {C.}~\bibnamefont {Hess}}, \bibinfo {author} {\bibfnamefont {M.}~\bibnamefont
  {Richter}}, \bibinfo {author} {\bibfnamefont {B.}~\bibnamefont
  {B{\"{u}}chner}}, \bibinfo {author} {\bibfnamefont {J.}~\bibnamefont {{Van
  Den Brink}}}, \bibinfo {author} {\bibfnamefont {M.}~\bibnamefont {Ruck}},
  \bibinfo {author} {\bibfnamefont {F.}~\bibnamefont {Reinert}}, \bibinfo
  {author} {\bibfnamefont {H.}~\bibnamefont {Bentmann}},\ and\ \bibinfo
  {author} {\bibfnamefont {A.}~\bibnamefont {Isaeva}},\ }\bibfield  {title}
  {\bibinfo {title} {{Topological Electronic Structure and Intrinsic
  Magnetization in ${\mathrm{MnBi}}_{4}{\mathrm{Te}}_{7}$: A
  ${\mathrm{Bi}}_{2}{\mathrm{Te}}_{3}$ Derivative with a Periodic Mn
  Sublattice}},\ }\href {https://doi.org/10.1103/PhysRevX.9.041065} {\bibfield
  {journal} {\bibinfo  {journal} {Phys. Rev. X}\ }\textbf {\bibinfo {volume}
  {91}},\ \bibinfo {pages} {041065} (\bibinfo {year} {2019}{\natexlab{b}})},\
  \Eprint {https://arxiv.org/abs/1906.08394} {arXiv:1906.08394} \BibitemShut
  {NoStop}%
\bibitem [{\citenamefont {Sun}\ and\ \citenamefont {He}(2021)}]{Sun2021}%
  \BibitemOpen
  \bibfield  {author} {\bibinfo {author} {\bibfnamefont {H.-M.}\ \bibnamefont
  {Sun}}\ and\ \bibinfo {author} {\bibfnamefont {Q.-L.}\ \bibnamefont {He}},\
  }\bibfield  {title} {\bibinfo {title} {{Physical problems and experimental
  progress in layered magnetic topological materials}},\ }\href
  {https://doi.org/10.7498/aps.70.20210133} {\bibfield  {journal} {\bibinfo
  {journal} {ACTA PHYS. SIN-CH ED}\ }\textbf {\bibinfo {volume} {70}},\
  \bibinfo {pages} {127302} (\bibinfo {year} {2021})}\BibitemShut {NoStop}%
\bibitem [{\citenamefont {Yan}\ \emph {et~al.}(2019)\citenamefont {Yan},
  \citenamefont {Okamoto}, \citenamefont {McGuire}, \citenamefont {May},
  \citenamefont {McQueeney},\ and\ \citenamefont
  {Sales}}]{PhysRevB.100.104409}%
  \BibitemOpen
  \bibfield  {author} {\bibinfo {author} {\bibfnamefont {J.-Q.}\ \bibnamefont
  {Yan}}, \bibinfo {author} {\bibfnamefont {S.}~\bibnamefont {Okamoto}},
  \bibinfo {author} {\bibfnamefont {M.~A.}\ \bibnamefont {McGuire}}, \bibinfo
  {author} {\bibfnamefont {A.~F.}\ \bibnamefont {May}}, \bibinfo {author}
  {\bibfnamefont {R.~J.}\ \bibnamefont {McQueeney}},\ and\ \bibinfo {author}
  {\bibfnamefont {B.~C.}\ \bibnamefont {Sales}},\ }\bibfield  {title} {\bibinfo
  {title} {{Evolution of structural, magnetic, and transport properties in
  ${\mathrm{MnBi}}_{2\ensuremath{-}x}{\mathrm{Sb}}_{x}{\mathrm{Te}}_{4}$}},\
  }\href {https://doi.org/10.1103/PhysRevB.100.104409} {\bibfield  {journal}
  {\bibinfo  {journal} {Phys. Rev. B}\ }\textbf {\bibinfo {volume} {100}},\
  \bibinfo {pages} {104409} (\bibinfo {year} {2019})}\BibitemShut {NoStop}%
\bibitem [{\citenamefont {Chen}\ \emph {et~al.}(2020)\citenamefont {Chen},
  \citenamefont {Chuang}, \citenamefont {Lee}, \citenamefont {Zhu},
  \citenamefont {Honz}, \citenamefont {Guan}, \citenamefont {Wang},
  \citenamefont {Wang}, \citenamefont {Mao}, \citenamefont {Zhu}, \citenamefont
  {Heikes}, \citenamefont {Quarterman}, \citenamefont {Zajdel}, \citenamefont
  {Borchers},\ and\ \citenamefont {Ratcliff}}]{PhysRevMaterials.4.064411}%
  \BibitemOpen
  \bibfield  {author} {\bibinfo {author} {\bibfnamefont {Y.}~\bibnamefont
  {Chen}}, \bibinfo {author} {\bibfnamefont {Y.-W.}\ \bibnamefont {Chuang}},
  \bibinfo {author} {\bibfnamefont {S.~H.}\ \bibnamefont {Lee}}, \bibinfo
  {author} {\bibfnamefont {Y.}~\bibnamefont {Zhu}}, \bibinfo {author}
  {\bibfnamefont {K.}~\bibnamefont {Honz}}, \bibinfo {author} {\bibfnamefont
  {Y.}~\bibnamefont {Guan}}, \bibinfo {author} {\bibfnamefont {Y.}~\bibnamefont
  {Wang}}, \bibinfo {author} {\bibfnamefont {K.}~\bibnamefont {Wang}}, \bibinfo
  {author} {\bibfnamefont {Z.}~\bibnamefont {Mao}}, \bibinfo {author}
  {\bibfnamefont {J.}~\bibnamefont {Zhu}}, \bibinfo {author} {\bibfnamefont
  {C.}~\bibnamefont {Heikes}}, \bibinfo {author} {\bibfnamefont
  {P.}~\bibnamefont {Quarterman}}, \bibinfo {author} {\bibfnamefont
  {P.}~\bibnamefont {Zajdel}}, \bibinfo {author} {\bibfnamefont {J.~A.}\
  \bibnamefont {Borchers}},\ and\ \bibinfo {author} {\bibfnamefont
  {W.}~\bibnamefont {Ratcliff}},\ }\bibfield  {title} {\bibinfo {title}
  {{Ferromagnetism in van der Waals compound
  $\mathrm{MnS}{\mathrm{b}}_{1.8}\mathrm{B}{\mathrm{i}}_{0.2}\mathrm{T}{\mathrm{e}}_{4}$}},\
  }\href {https://doi.org/10.1103/PhysRevMaterials.4.064411} {\bibfield
  {journal} {\bibinfo  {journal} {Phys. Rev. Mater.}\ }\textbf {\bibinfo
  {volume} {4}},\ \bibinfo {pages} {064411} (\bibinfo {year}
  {2020})}\BibitemShut {NoStop}%
\bibitem [{\citenamefont {Hou}\ and\ \citenamefont {Wu}(2021)}]{Hou2021}%
  \BibitemOpen
  \bibfield  {author} {\bibinfo {author} {\bibfnamefont {Y.~S.}\ \bibnamefont
  {Hou}}\ and\ \bibinfo {author} {\bibfnamefont {R.~Q.}\ \bibnamefont {Wu}},\
  }\bibfield  {title} {\bibinfo {title} {{Alloying vanadium in
  ${\mathrm{MnBi}}_{2}{\mathrm{Te}}_{4}$ for robust ferromagnetic coupling and
  quantum anomalous Hall effect}},\ }\href
  {https://doi.org/10.1103/PhysRevB.103.064412} {\bibfield  {journal} {\bibinfo
   {journal} {Phys. Rev. B}\ }\textbf {\bibinfo {volume} {103}},\ \bibinfo
  {pages} {064412} (\bibinfo {year} {2021})},\ \Eprint
  {https://arxiv.org/abs/2004.04862} {arXiv:2004.04862} \BibitemShut {NoStop}%
\bibitem [{\citenamefont {Zhang}\ \emph {et~al.}(2021)\citenamefont {Zhang},
  \citenamefont {Yang}, \citenamefont {Wang},\ and\ \citenamefont
  {Xu}}]{Zhang2021}%
  \BibitemOpen
  \bibfield  {author} {\bibinfo {author} {\bibfnamefont {H.}~\bibnamefont
  {Zhang}}, \bibinfo {author} {\bibfnamefont {W.}~\bibnamefont {Yang}},
  \bibinfo {author} {\bibfnamefont {Y.}~\bibnamefont {Wang}},\ and\ \bibinfo
  {author} {\bibfnamefont {X.}~\bibnamefont {Xu}},\ }\bibfield  {title}
  {\bibinfo {title} {{Tunable topological states in layered magnetic materials
  of ${\mathrm{MnSb}}_{2}{\mathrm{Te}}_{4}$,
  ${\mathrm{MnBi}}_{2}{\mathrm{Se}}_{4}$, and
  ${\mathrm{MnSb}}_{2}{\mathrm{Se}}_{4}$}},\ }\href
  {https://doi.org/10.1103/PhysRevB.103.094433} {\bibfield  {journal} {\bibinfo
   {journal} {Phys. Rev. B}\ }\textbf {\bibinfo {volume} {103}},\ \bibinfo
  {pages} {094433} (\bibinfo {year} {2021})}\BibitemShut {NoStop}%
\bibitem [{\citenamefont {Golias}\ \emph {et~al.}(2021)\citenamefont {Golias},
  \citenamefont {Weschke}, \citenamefont {Flanagan}, \citenamefont {Schierle},
  \citenamefont {Richardella}, \citenamefont {Rienks}, \citenamefont {Mandal},
  \citenamefont {Varykhalov}, \citenamefont {S{\'{a}}nchez-Barriga},
  \citenamefont {Radu}, \citenamefont {Samarth},\ and\ \citenamefont
  {Rader}}]{Golias2021}%
  \BibitemOpen
  \bibfield  {author} {\bibinfo {author} {\bibfnamefont {E.}~\bibnamefont
  {Golias}}, \bibinfo {author} {\bibfnamefont {E.}~\bibnamefont {Weschke}},
  \bibinfo {author} {\bibfnamefont {T.}~\bibnamefont {Flanagan}}, \bibinfo
  {author} {\bibfnamefont {E.}~\bibnamefont {Schierle}}, \bibinfo {author}
  {\bibfnamefont {A.}~\bibnamefont {Richardella}}, \bibinfo {author}
  {\bibfnamefont {E.~D.}\ \bibnamefont {Rienks}}, \bibinfo {author}
  {\bibfnamefont {P.~S.}\ \bibnamefont {Mandal}}, \bibinfo {author}
  {\bibfnamefont {A.}~\bibnamefont {Varykhalov}}, \bibinfo {author}
  {\bibfnamefont {J.}~\bibnamefont {S{\'{a}}nchez-Barriga}}, \bibinfo {author}
  {\bibfnamefont {F.}~\bibnamefont {Radu}}, \bibinfo {author} {\bibfnamefont
  {N.}~\bibnamefont {Samarth}},\ and\ \bibinfo {author} {\bibfnamefont
  {O.}~\bibnamefont {Rader}},\ }\bibfield  {title} {\bibinfo {title}
  {{Magnetization relaxation and search for the magnetic gap in bulk-insulating
  V-doped ${\mathrm{(Bi, Sb)}}_{2}{\mathrm{Te}}_{3}$}},\ }\href
  {https://doi.org/10.1063/5.0070557} {\bibfield  {journal} {\bibinfo
  {journal} {Appl. Phys. Lett.}\ }\textbf {\bibinfo {volume} {119}},\ \bibinfo
  {pages} {132404} (\bibinfo {year} {2021})},\ \Eprint
  {https://arxiv.org/abs/2010.07083} {arXiv:2010.07083} \BibitemShut {NoStop}%
\bibitem [{\citenamefont {Wimmer}\ \emph {et~al.}(2021)\citenamefont {Wimmer},
  \citenamefont {S{\'{a}}nchez-Barriga}, \citenamefont {K{\"{u}}ppers},
  \citenamefont {Ney}, \citenamefont {Schierle}, \citenamefont {Freyse},
  \citenamefont {Caha}, \citenamefont {Michali{\v{c}}ka}, \citenamefont
  {Liebmann}, \citenamefont {Primetzhofer}, \citenamefont {Hoffman},
  \citenamefont {Ernst}, \citenamefont {Otrokov}, \citenamefont {Bihlmayer},
  \citenamefont {Weschke}, \citenamefont {Lake}, \citenamefont {Chulkov},
  \citenamefont {Morgenstern}, \citenamefont {Bauer}, \citenamefont
  {Springholz},\ and\ \citenamefont {Rader}}]{Wimmer2021}%
  \BibitemOpen
  \bibfield  {author} {\bibinfo {author} {\bibfnamefont {S.}~\bibnamefont
  {Wimmer}}, \bibinfo {author} {\bibfnamefont {J.}~\bibnamefont
  {S{\'{a}}nchez-Barriga}}, \bibinfo {author} {\bibfnamefont {P.}~\bibnamefont
  {K{\"{u}}ppers}}, \bibinfo {author} {\bibfnamefont {A.}~\bibnamefont {Ney}},
  \bibinfo {author} {\bibfnamefont {E.}~\bibnamefont {Schierle}}, \bibinfo
  {author} {\bibfnamefont {F.}~\bibnamefont {Freyse}}, \bibinfo {author}
  {\bibfnamefont {O.}~\bibnamefont {Caha}}, \bibinfo {author} {\bibfnamefont
  {J.}~\bibnamefont {Michali{\v{c}}ka}}, \bibinfo {author} {\bibfnamefont
  {M.}~\bibnamefont {Liebmann}}, \bibinfo {author} {\bibfnamefont
  {D.}~\bibnamefont {Primetzhofer}}, \bibinfo {author} {\bibfnamefont
  {M.}~\bibnamefont {Hoffman}}, \bibinfo {author} {\bibfnamefont
  {A.}~\bibnamefont {Ernst}}, \bibinfo {author} {\bibfnamefont {M.~M.}\
  \bibnamefont {Otrokov}}, \bibinfo {author} {\bibfnamefont {G.}~\bibnamefont
  {Bihlmayer}}, \bibinfo {author} {\bibfnamefont {E.}~\bibnamefont {Weschke}},
  \bibinfo {author} {\bibfnamefont {B.}~\bibnamefont {Lake}}, \bibinfo {author}
  {\bibfnamefont {E.~V.}\ \bibnamefont {Chulkov}}, \bibinfo {author}
  {\bibfnamefont {M.}~\bibnamefont {Morgenstern}}, \bibinfo {author}
  {\bibfnamefont {G.}~\bibnamefont {Bauer}}, \bibinfo {author} {\bibfnamefont
  {G.}~\bibnamefont {Springholz}},\ and\ \bibinfo {author} {\bibfnamefont
  {O.}~\bibnamefont {Rader}},\ }\bibfield  {title} {\bibinfo {title} {{Mn-Rich
  ${\mathrm{MnSb}}_{2}{\mathrm{Te}}_{4}$: A Topological Insulator with Magnetic
  Gap Closing at High Curie Temperatures of 45-50 K}},\ }\href
  {https://doi.org/10.1002/adma.202102935} {\bibfield  {journal} {\bibinfo
  {journal} {Adv. Mater.}\ }\textbf {\bibinfo {volume} {33}},\ \bibinfo {pages}
  {2102935} (\bibinfo {year} {2021})},\ \Eprint
  {https://arxiv.org/abs/2011.07052} {arXiv:2011.07052} \BibitemShut {NoStop}%
\bibitem [{\citenamefont {Ovchinnikov}\ \emph {et~al.}(2021)\citenamefont
  {Ovchinnikov}, \citenamefont {Huang}, \citenamefont {Lin}, \citenamefont
  {Fei}, \citenamefont {Cai}, \citenamefont {Song}, \citenamefont {He},
  \citenamefont {Jiang}, \citenamefont {Wang}, \citenamefont {Li},
  \citenamefont {Wang}, \citenamefont {Wu}, \citenamefont {Xiao}, \citenamefont
  {Chu}, \citenamefont {Yan}, \citenamefont {Chang}, \citenamefont {Cui},\ and\
  \citenamefont {Xu}}]{doi:10.1021/acs.nanolett.0c05117}%
  \BibitemOpen
  \bibfield  {author} {\bibinfo {author} {\bibfnamefont {D.}~\bibnamefont
  {Ovchinnikov}}, \bibinfo {author} {\bibfnamefont {X.}~\bibnamefont {Huang}},
  \bibinfo {author} {\bibfnamefont {Z.}~\bibnamefont {Lin}}, \bibinfo {author}
  {\bibfnamefont {Z.}~\bibnamefont {Fei}}, \bibinfo {author} {\bibfnamefont
  {J.}~\bibnamefont {Cai}}, \bibinfo {author} {\bibfnamefont {T.}~\bibnamefont
  {Song}}, \bibinfo {author} {\bibfnamefont {M.}~\bibnamefont {He}}, \bibinfo
  {author} {\bibfnamefont {Q.}~\bibnamefont {Jiang}}, \bibinfo {author}
  {\bibfnamefont {C.}~\bibnamefont {Wang}}, \bibinfo {author} {\bibfnamefont
  {H.}~\bibnamefont {Li}}, \bibinfo {author} {\bibfnamefont {Y.}~\bibnamefont
  {Wang}}, \bibinfo {author} {\bibfnamefont {Y.}~\bibnamefont {Wu}}, \bibinfo
  {author} {\bibfnamefont {D.}~\bibnamefont {Xiao}}, \bibinfo {author}
  {\bibfnamefont {J.-H.}\ \bibnamefont {Chu}}, \bibinfo {author} {\bibfnamefont
  {J.}~\bibnamefont {Yan}}, \bibinfo {author} {\bibfnamefont {C.-Z.}\
  \bibnamefont {Chang}}, \bibinfo {author} {\bibfnamefont {Y.-T.}\ \bibnamefont
  {Cui}},\ and\ \bibinfo {author} {\bibfnamefont {X.}~\bibnamefont {Xu}},\
  }\bibfield  {title} {\bibinfo {title} {{Intertwined Topological and Magnetic
  Orders in Atomically Thin Chern Insulator
  ${\mathrm{MnBi}}_{2}{\mathrm{Te}}_{4}$}},\ }\href
  {https://doi.org/10.1021/acs.nanolett.0c05117} {\bibfield  {journal}
  {\bibinfo  {journal} {Nano Lett.}\ }\textbf {\bibinfo {volume} {21}},\
  \bibinfo {pages} {2544} (\bibinfo {year} {2021})},\ \bibinfo {note} {pMID:
  33710884}\BibitemShut {NoStop}%
\bibitem [{\citenamefont {Ge}\ \emph {et~al.}(2020)\citenamefont {Ge},
  \citenamefont {Liu}, \citenamefont {Li}, \citenamefont {Li}, \citenamefont
  {Luo}, \citenamefont {Wu}, \citenamefont {Xu},\ and\ \citenamefont
  {Wang}}]{10.1093/nsr/nwaa089}%
  \BibitemOpen
  \bibfield  {author} {\bibinfo {author} {\bibfnamefont {J.}~\bibnamefont
  {Ge}}, \bibinfo {author} {\bibfnamefont {Y.}~\bibnamefont {Liu}}, \bibinfo
  {author} {\bibfnamefont {J.}~\bibnamefont {Li}}, \bibinfo {author}
  {\bibfnamefont {H.}~\bibnamefont {Li}}, \bibinfo {author} {\bibfnamefont
  {T.}~\bibnamefont {Luo}}, \bibinfo {author} {\bibfnamefont {Y.}~\bibnamefont
  {Wu}}, \bibinfo {author} {\bibfnamefont {Y.}~\bibnamefont {Xu}},\ and\
  \bibinfo {author} {\bibfnamefont {J.}~\bibnamefont {Wang}},\ }\bibfield
  {title} {\bibinfo {title} {{High-Chern-number and high-temperature quantum
  Hall effect without Landau levels}},\ }\href
  {https://doi.org/10.1093/nsr/nwaa089} {\bibfield  {journal} {\bibinfo
  {journal} {National Science Review}\ }\textbf {\bibinfo {volume} {7}},\
  \bibinfo {pages} {1280} (\bibinfo {year} {2020})},\ \Eprint
  {https://arxiv.org/abs/https://academic.oup.com/nsr/article-pdf/7/8/1280/38881797/nwaa089.pdf}
  {https://academic.oup.com/nsr/article-pdf/7/8/1280/38881797/nwaa089.pdf}
  \BibitemShut {NoStop}%
\bibitem [{\citenamefont {Singh}\ \emph {et~al.}(2021)\citenamefont {Singh},
  \citenamefont {Kumar}, \citenamefont {Alam}, \citenamefont {Gangwar},
  \citenamefont {Ghosh}, \citenamefont {Pal}, \citenamefont {Singh},
  \citenamefont {Shahi}, \citenamefont {Chaudhary}, \citenamefont {Shimada},\
  and\ \citenamefont {Chatterjee}}]{Singh2021}%
  \BibitemOpen
  \bibfield  {author} {\bibinfo {author} {\bibfnamefont {M.}~\bibnamefont
  {Singh}}, \bibinfo {author} {\bibfnamefont {S.}~\bibnamefont {Kumar}},
  \bibinfo {author} {\bibfnamefont {M.}~\bibnamefont {Alam}}, \bibinfo {author}
  {\bibfnamefont {V.~K.}\ \bibnamefont {Gangwar}}, \bibinfo {author}
  {\bibfnamefont {L.}~\bibnamefont {Ghosh}}, \bibinfo {author} {\bibfnamefont
  {D.}~\bibnamefont {Pal}}, \bibinfo {author} {\bibfnamefont {R.}~\bibnamefont
  {Singh}}, \bibinfo {author} {\bibfnamefont {P.}~\bibnamefont {Shahi}},
  \bibinfo {author} {\bibfnamefont {P.}~\bibnamefont {Chaudhary}}, \bibinfo
  {author} {\bibfnamefont {K.}~\bibnamefont {Shimada}},\ and\ \bibinfo {author}
  {\bibfnamefont {S.}~\bibnamefont {Chatterjee}},\ }\bibfield  {title}
  {\bibinfo {title} {{Evidence of surface and bulk magnetic ordering in Fe and
  Mn doped ${\mathrm{Bi}}_{2}{\mathrm{(SeS)}}_{3}$topological insulator}},\
  }\href {https://doi.org/10.1063/5.0035433} {\bibfield  {journal} {\bibinfo
  {journal} {Appl. Phys. Lett.}\ }\textbf {\bibinfo {volume} {118}},\ \bibinfo
  {pages} {132409} (\bibinfo {year} {2021})}\BibitemShut {NoStop}%
\bibitem [{\citenamefont {Jiang}\ \emph {et~al.}(2021)\citenamefont {Jiang},
  \citenamefont {Liu}, \citenamefont {Li},\ and\ \citenamefont
  {Mi}}]{Jiang2021}%
  \BibitemOpen
  \bibfield  {author} {\bibinfo {author} {\bibfnamefont {J.}~\bibnamefont
  {Jiang}}, \bibinfo {author} {\bibfnamefont {X.}~\bibnamefont {Liu}}, \bibinfo
  {author} {\bibfnamefont {R.}~\bibnamefont {Li}},\ and\ \bibinfo {author}
  {\bibfnamefont {W.}~\bibnamefont {Mi}},\ }\bibfield  {title} {\bibinfo
  {title} {{Topological spin textures in a two-dimensional
  ${\mathrm{MnBi}}_{2}{\mathrm{(Se,Te)}}_{4}$Janus material}},\ }\href
  {https://doi.org/10.1063/5.0057794} {\bibfield  {journal} {\bibinfo
  {journal} {Appl. Phys. Lett.}\ }\textbf {\bibinfo {volume} {119}},\ \bibinfo
  {pages} {072401} (\bibinfo {year} {2021})}\BibitemShut {NoStop}%
\bibitem [{\citenamefont {Watanabe}\ \emph {et~al.}(2022)\citenamefont
  {Watanabe}, \citenamefont {Yoshimi}, \citenamefont {Kawamura}, \citenamefont
  {Kaneko}, \citenamefont {Takahashi}, \citenamefont {Tsukazaki}, \citenamefont
  {Kawasaki},\ and\ \citenamefont {Tokura}}]{Watanabe2022}%
  \BibitemOpen
  \bibfield  {author} {\bibinfo {author} {\bibfnamefont {R.}~\bibnamefont
  {Watanabe}}, \bibinfo {author} {\bibfnamefont {R.}~\bibnamefont {Yoshimi}},
  \bibinfo {author} {\bibfnamefont {M.}~\bibnamefont {Kawamura}}, \bibinfo
  {author} {\bibfnamefont {Y.}~\bibnamefont {Kaneko}}, \bibinfo {author}
  {\bibfnamefont {K.~S.}\ \bibnamefont {Takahashi}}, \bibinfo {author}
  {\bibfnamefont {A.}~\bibnamefont {Tsukazaki}}, \bibinfo {author}
  {\bibfnamefont {M.}~\bibnamefont {Kawasaki}},\ and\ \bibinfo {author}
  {\bibfnamefont {Y.}~\bibnamefont {Tokura}},\ }\bibfield  {title} {\bibinfo
  {title} {{Enhancement of anomalous Hall effect in epitaxial thin films of
  intrinsic magnetic topological insulator
  ${\mathrm{MnBi}}_{2}{\mathrm{Te}}_{4}$ with Fermi-level tuning}},\ }\href
  {https://doi.org/10.1063/5.0067893} {\bibfield  {journal} {\bibinfo
  {journal} {Appl. Phys. Lett.}\ }\textbf {\bibinfo {volume} {120}},\ \bibinfo
  {pages} {031901} (\bibinfo {year} {2022})}\BibitemShut {NoStop}%
\bibitem [{\citenamefont {Zhu}\ \emph {et~al.}(2020)\citenamefont {Zhu},
  \citenamefont {Song}, \citenamefont {Liao}, \citenamefont {Zhou},
  \citenamefont {Bai}, \citenamefont {Zhou},\ and\ \citenamefont
  {Pan}}]{Zhu2020}%
  \BibitemOpen
  \bibfield  {author} {\bibinfo {author} {\bibfnamefont {W.}~\bibnamefont
  {Zhu}}, \bibinfo {author} {\bibfnamefont {C.}~\bibnamefont {Song}}, \bibinfo
  {author} {\bibfnamefont {L.}~\bibnamefont {Liao}}, \bibinfo {author}
  {\bibfnamefont {Z.}~\bibnamefont {Zhou}}, \bibinfo {author} {\bibfnamefont
  {H.}~\bibnamefont {Bai}}, \bibinfo {author} {\bibfnamefont {Y.}~\bibnamefont
  {Zhou}},\ and\ \bibinfo {author} {\bibfnamefont {F.}~\bibnamefont {Pan}},\
  }\bibfield  {title} {\bibinfo {title} {{Quantum anomalous Hall insulator
  state in ferromagnetically ordered
  ${\mathrm{MnBi}}_{2}{\mathrm{Te}}_{4}$/${\mathrm{VBi}}_{2}{\mathrm{Te}}_{4}$
  heterostructures}},\ }\href {https://doi.org/10.1103/PhysRevB.102.085111}
  {\bibfield  {journal} {\bibinfo  {journal} {Phys. Rev. B}\ }\textbf {\bibinfo
  {volume} {102}},\ \bibinfo {pages} {085111} (\bibinfo {year}
  {2020})}\BibitemShut {NoStop}%
\bibitem [{\citenamefont {Fu}\ \emph {et~al.}(2020)\citenamefont {Fu},
  \citenamefont {Liu},\ and\ \citenamefont {Yan}}]{Fu2020}%
  \BibitemOpen
  \bibfield  {author} {\bibinfo {author} {\bibfnamefont {H.}~\bibnamefont
  {Fu}}, \bibinfo {author} {\bibfnamefont {C.~X.}\ \bibnamefont {Liu}},\ and\
  \bibinfo {author} {\bibfnamefont {B.}~\bibnamefont {Yan}},\ }\bibfield
  {title} {\bibinfo {title} {{Exchange bias and quantum anomalous nomalous Hall
  effect in the
  ${\mathrm{MnBi}}_{2}{\mathrm{Te}}_{4}$/${\mathrm{Cr}}{\mathrm{I}}_{3}$
  heterostructure}},\ }\href {https://doi.org/10.1126/sciadv.aaz0948}
  {\bibfield  {journal} {\bibinfo  {journal} {Sci. Adv.}\ }\textbf {\bibinfo
  {volume} {6}},\ \bibinfo {pages} {eaaz0948} (\bibinfo {year}
  {2020})}\BibitemShut {NoStop}%
\bibitem [{\citenamefont {Gao}\ \emph {et~al.}(2021{\natexlab{a}})\citenamefont
  {Gao}, \citenamefont {Qin}, \citenamefont {Qi}, \citenamefont {Qiao},\ and\
  \citenamefont {Ren}}]{Gao2021}%
  \BibitemOpen
  \bibfield  {author} {\bibinfo {author} {\bibfnamefont {R.}~\bibnamefont
  {Gao}}, \bibinfo {author} {\bibfnamefont {G.}~\bibnamefont {Qin}}, \bibinfo
  {author} {\bibfnamefont {S.}~\bibnamefont {Qi}}, \bibinfo {author}
  {\bibfnamefont {Z.}~\bibnamefont {Qiao}},\ and\ \bibinfo {author}
  {\bibfnamefont {W.}~\bibnamefont {Ren}},\ }\bibfield  {title} {\bibinfo
  {title} {{Quantum anomalous Hall effect in
  ${\mathrm{MnBi}}_{2}{\mathrm{Te}}_{4}$ van der Waals heterostructures}},\
  }\href {https://doi.org/10.1103/PhysRevMaterials.5.114201} {\bibfield
  {journal} {\bibinfo  {journal} {Phys. Rev. Mater.}\ }\textbf {\bibinfo
  {volume} {5}},\ \bibinfo {pages} {114201} (\bibinfo {year}
  {2021}{\natexlab{a}})}\BibitemShut {NoStop}%
\bibitem [{\citenamefont {Li}\ \emph {et~al.}(2020)\citenamefont {Li},
  \citenamefont {Li}, \citenamefont {He}, \citenamefont {Wan}, \citenamefont
  {Duan},\ and\ \citenamefont {Xu}}]{Li2020}%
  \BibitemOpen
  \bibfield  {author} {\bibinfo {author} {\bibfnamefont {Z.}~\bibnamefont
  {Li}}, \bibinfo {author} {\bibfnamefont {J.}~\bibnamefont {Li}}, \bibinfo
  {author} {\bibfnamefont {K.}~\bibnamefont {He}}, \bibinfo {author}
  {\bibfnamefont {X.}~\bibnamefont {Wan}}, \bibinfo {author} {\bibfnamefont
  {W.}~\bibnamefont {Duan}},\ and\ \bibinfo {author} {\bibfnamefont
  {Y.}~\bibnamefont {Xu}},\ }\bibfield  {title} {\bibinfo {title} {{Tunable
  interlayer magnetism and band topology in van der Waals heterostructures of
  ${\mathrm{MnBi}}_{2}{\mathrm{Te}}_{4}$-family materials}},\ }\href
  {https://doi.org/10.1103/PhysRevB.102.081107} {\bibfield  {journal} {\bibinfo
   {journal} {Phys. Rev. B}\ }\textbf {\bibinfo {volume} {102}},\ \bibinfo
  {pages} {081107(R)} (\bibinfo {year} {2020})},\ \Eprint
  {https://arxiv.org/abs/2003.13485} {arXiv:2003.13485} \BibitemShut {NoStop}%
\bibitem [{\citenamefont {Yan}\ \emph {et~al.}(2021)\citenamefont {Yan},
  \citenamefont {Jia}, \citenamefont {Shi}, \citenamefont {Dong},\ and\
  \citenamefont {Xu}}]{Yan2021}%
  \BibitemOpen
  \bibfield  {author} {\bibinfo {author} {\bibfnamefont {Z.}~\bibnamefont
  {Yan}}, \bibinfo {author} {\bibfnamefont {X.}~\bibnamefont {Jia}}, \bibinfo
  {author} {\bibfnamefont {X.}~\bibnamefont {Shi}}, \bibinfo {author}
  {\bibfnamefont {X.}~\bibnamefont {Dong}},\ and\ \bibinfo {author}
  {\bibfnamefont {X.}~\bibnamefont {Xu}},\ }\bibfield  {title} {\bibinfo
  {title} {{Barrier-dependent electronic transport propertieBarrier-dependent
  electronic transport propertiess in two-dimensional
  ${\mathrm{MnBi}}_{2}{\mathrm{Te}}_{4}$-based van der Waals magnetic tunnel
  junctions}},\ }\href {https://doi.org/10.1063/5.0052720} {\bibfield
  {journal} {\bibinfo  {journal} {Appl.Phys. Lett.}\ }\textbf {\bibinfo
  {volume} {118}},\ \bibinfo {pages} {223503} (\bibinfo {year}
  {2021})}\BibitemShut {NoStop}%
\bibitem [{\citenamefont {Hao}\ \emph {et~al.}(2019)\citenamefont {Hao},
  \citenamefont {Liu}, \citenamefont {Feng}, \citenamefont {Ma}, \citenamefont
  {Schwier}, \citenamefont {Arita}, \citenamefont {Kumar}, \citenamefont {Hu},
  \citenamefont {Lu}, \citenamefont {Zeng}, \citenamefont {Wang}, \citenamefont
  {Hao}, \citenamefont {Sun}, \citenamefont {Zhang}, \citenamefont {Mei},
  \citenamefont {Ni}, \citenamefont {Wu}, \citenamefont {Shimada},
  \citenamefont {Chen}, \citenamefont {Liu},\ and\ \citenamefont
  {Liu}}]{PhysRevX.9.041038}%
  \BibitemOpen
  \bibfield  {author} {\bibinfo {author} {\bibfnamefont {Y.-J.}\ \bibnamefont
  {Hao}}, \bibinfo {author} {\bibfnamefont {P.}~\bibnamefont {Liu}}, \bibinfo
  {author} {\bibfnamefont {Y.}~\bibnamefont {Feng}}, \bibinfo {author}
  {\bibfnamefont {X.-M.}\ \bibnamefont {Ma}}, \bibinfo {author} {\bibfnamefont
  {E.~F.}\ \bibnamefont {Schwier}}, \bibinfo {author} {\bibfnamefont
  {M.}~\bibnamefont {Arita}}, \bibinfo {author} {\bibfnamefont
  {S.}~\bibnamefont {Kumar}}, \bibinfo {author} {\bibfnamefont
  {C.}~\bibnamefont {Hu}}, \bibinfo {author} {\bibfnamefont {R.}~\bibnamefont
  {Lu}}, \bibinfo {author} {\bibfnamefont {M.}~\bibnamefont {Zeng}}, \bibinfo
  {author} {\bibfnamefont {Y.}~\bibnamefont {Wang}}, \bibinfo {author}
  {\bibfnamefont {Z.}~\bibnamefont {Hao}}, \bibinfo {author} {\bibfnamefont
  {H.-Y.}\ \bibnamefont {Sun}}, \bibinfo {author} {\bibfnamefont
  {K.}~\bibnamefont {Zhang}}, \bibinfo {author} {\bibfnamefont
  {J.}~\bibnamefont {Mei}}, \bibinfo {author} {\bibfnamefont {N.}~\bibnamefont
  {Ni}}, \bibinfo {author} {\bibfnamefont {L.}~\bibnamefont {Wu}}, \bibinfo
  {author} {\bibfnamefont {K.}~\bibnamefont {Shimada}}, \bibinfo {author}
  {\bibfnamefont {C.}~\bibnamefont {Chen}}, \bibinfo {author} {\bibfnamefont
  {Q.}~\bibnamefont {Liu}},\ and\ \bibinfo {author} {\bibfnamefont
  {C.}~\bibnamefont {Liu}},\ }\bibfield  {title} {\bibinfo {title} {{Gapless
  Surface Dirac Cone in Antiferromagnetic Topological Insulator
  ${\mathrm{MnBi}}_{2}{\mathrm{Te}}_{4}$}},\ }\href
  {https://doi.org/10.1103/PhysRevX.9.041038} {\bibfield  {journal} {\bibinfo
  {journal} {Phys. Rev. X}\ }\textbf {\bibinfo {volume} {9}},\ \bibinfo {pages}
  {041038} (\bibinfo {year} {2019})}\BibitemShut {NoStop}%
\bibitem [{\citenamefont {Chen}\ \emph {et~al.}(2019)\citenamefont {Chen},
  \citenamefont {Xu}, \citenamefont {Li}, \citenamefont {Li}, \citenamefont
  {Wang}, \citenamefont {Zhang}, \citenamefont {Li}, \citenamefont {Wu},
  \citenamefont {Liang}, \citenamefont {Chen}, \citenamefont {Jung},
  \citenamefont {Cacho}, \citenamefont {Mao}, \citenamefont {Liu},
  \citenamefont {Wang}, \citenamefont {Guo}, \citenamefont {Xu}, \citenamefont
  {Liu}, \citenamefont {Yang},\ and\ \citenamefont {Chen}}]{PhysRevX.9.041040}%
  \BibitemOpen
  \bibfield  {author} {\bibinfo {author} {\bibfnamefont {Y.~J.}\ \bibnamefont
  {Chen}}, \bibinfo {author} {\bibfnamefont {L.~X.}\ \bibnamefont {Xu}},
  \bibinfo {author} {\bibfnamefont {J.~H.}\ \bibnamefont {Li}}, \bibinfo
  {author} {\bibfnamefont {Y.~W.}\ \bibnamefont {Li}}, \bibinfo {author}
  {\bibfnamefont {H.~Y.}\ \bibnamefont {Wang}}, \bibinfo {author}
  {\bibfnamefont {C.~F.}\ \bibnamefont {Zhang}}, \bibinfo {author}
  {\bibfnamefont {H.}~\bibnamefont {Li}}, \bibinfo {author} {\bibfnamefont
  {Y.}~\bibnamefont {Wu}}, \bibinfo {author} {\bibfnamefont {A.~J.}\
  \bibnamefont {Liang}}, \bibinfo {author} {\bibfnamefont {C.}~\bibnamefont
  {Chen}}, \bibinfo {author} {\bibfnamefont {S.~W.}\ \bibnamefont {Jung}},
  \bibinfo {author} {\bibfnamefont {C.}~\bibnamefont {Cacho}}, \bibinfo
  {author} {\bibfnamefont {Y.~H.}\ \bibnamefont {Mao}}, \bibinfo {author}
  {\bibfnamefont {S.}~\bibnamefont {Liu}}, \bibinfo {author} {\bibfnamefont
  {M.~X.}\ \bibnamefont {Wang}}, \bibinfo {author} {\bibfnamefont {Y.~F.}\
  \bibnamefont {Guo}}, \bibinfo {author} {\bibfnamefont {Y.}~\bibnamefont
  {Xu}}, \bibinfo {author} {\bibfnamefont {Z.~K.}\ \bibnamefont {Liu}},
  \bibinfo {author} {\bibfnamefont {L.~X.}\ \bibnamefont {Yang}},\ and\
  \bibinfo {author} {\bibfnamefont {Y.~L.}\ \bibnamefont {Chen}},\ }\bibfield
  {title} {\bibinfo {title} {{Topological Electronic Structure and Its
  Temperature Evolution in Antiferromagnetic Topological Insulator
  ${\mathrm{MnBi}}_{2}{\mathrm{Te}}_{4}$}},\ }\href
  {https://doi.org/10.1103/PhysRevX.9.041040} {\bibfield  {journal} {\bibinfo
  {journal} {Phys. Rev. X}\ }\textbf {\bibinfo {volume} {9}},\ \bibinfo {pages}
  {041040} (\bibinfo {year} {2019})}\BibitemShut {NoStop}%
\bibitem [{\citenamefont {Swatek}\ \emph {et~al.}(2020)\citenamefont {Swatek},
  \citenamefont {Wu}, \citenamefont {Wang}, \citenamefont {Lee}, \citenamefont
  {Schrunk}, \citenamefont {Yan},\ and\ \citenamefont
  {Kaminski}}]{PhysRevB.101.161109}%
  \BibitemOpen
  \bibfield  {author} {\bibinfo {author} {\bibfnamefont {P.}~\bibnamefont
  {Swatek}}, \bibinfo {author} {\bibfnamefont {Y.}~\bibnamefont {Wu}}, \bibinfo
  {author} {\bibfnamefont {L.-L.}\ \bibnamefont {Wang}}, \bibinfo {author}
  {\bibfnamefont {K.}~\bibnamefont {Lee}}, \bibinfo {author} {\bibfnamefont
  {B.}~\bibnamefont {Schrunk}}, \bibinfo {author} {\bibfnamefont
  {J.}~\bibnamefont {Yan}},\ and\ \bibinfo {author} {\bibfnamefont
  {A.}~\bibnamefont {Kaminski}},\ }\bibfield  {title} {\bibinfo {title}
  {{Gapless Dirac surface states in the antiferromagnetic topological insulator
  ${\mathrm{MnBi}}_{2}{\mathrm{Te}}_{4}$}},\ }\href
  {https://doi.org/10.1103/PhysRevB.101.161109} {\bibfield  {journal} {\bibinfo
   {journal} {Phys. Rev. B}\ }\textbf {\bibinfo {volume} {101}},\ \bibinfo
  {pages} {161109} (\bibinfo {year} {2020})}\BibitemShut {NoStop}%
\bibitem [{\citenamefont {Yuan}\ \emph {et~al.}(2020)\citenamefont {Yuan},
  \citenamefont {Wang}, \citenamefont {Li}, \citenamefont {Li}, \citenamefont
  {Ji}, \citenamefont {Hao}, \citenamefont {Wu}, \citenamefont {He},
  \citenamefont {Wang}, \citenamefont {Xu}, \citenamefont {Duan}, \citenamefont
  {Li},\ and\ \citenamefont {Xue}}]{doi:10.1021/acs.nanolett.0c00031}%
  \BibitemOpen
  \bibfield  {author} {\bibinfo {author} {\bibfnamefont {Y.}~\bibnamefont
  {Yuan}}, \bibinfo {author} {\bibfnamefont {X.}~\bibnamefont {Wang}}, \bibinfo
  {author} {\bibfnamefont {H.}~\bibnamefont {Li}}, \bibinfo {author}
  {\bibfnamefont {J.}~\bibnamefont {Li}}, \bibinfo {author} {\bibfnamefont
  {Y.}~\bibnamefont {Ji}}, \bibinfo {author} {\bibfnamefont {Z.}~\bibnamefont
  {Hao}}, \bibinfo {author} {\bibfnamefont {Y.}~\bibnamefont {Wu}}, \bibinfo
  {author} {\bibfnamefont {K.}~\bibnamefont {He}}, \bibinfo {author}
  {\bibfnamefont {Y.}~\bibnamefont {Wang}}, \bibinfo {author} {\bibfnamefont
  {Y.}~\bibnamefont {Xu}}, \bibinfo {author} {\bibfnamefont {W.}~\bibnamefont
  {Duan}}, \bibinfo {author} {\bibfnamefont {W.}~\bibnamefont {Li}},\ and\
  \bibinfo {author} {\bibfnamefont {Q.-K.}\ \bibnamefont {Xue}},\ }\bibfield
  {title} {\bibinfo {title} {Electronic states and magnetic response of
  ${\mathrm{mnbi}}_{2}{\mathrm{te}}_{4}$ by scanning tunneling microscopy and
  spectroscopy},\ }\href {https://doi.org/10.1021/acs.nanolett.0c00031}
  {\bibfield  {journal} {\bibinfo  {journal} {Nano Lett.}\ }\textbf {\bibinfo
  {volume} {20}},\ \bibinfo {pages} {3271} (\bibinfo {year} {2020})},\ \bibinfo
  {note} {pMID: 32298117}\BibitemShut {NoStop}%
\bibitem [{\citenamefont {Huan}\ \emph {et~al.}(2021)\citenamefont {Huan},
  \citenamefont {Wang}, \citenamefont {Su}, \citenamefont {Wang}, \citenamefont
  {Wang}, \citenamefont {Yu}, \citenamefont {Zou}, \citenamefont {Zhang},\ and\
  \citenamefont {Guo}}]{Huan2021}%
  \BibitemOpen
  \bibfield  {author} {\bibinfo {author} {\bibfnamefont {S.}~\bibnamefont
  {Huan}}, \bibinfo {author} {\bibfnamefont {D.}~\bibnamefont {Wang}}, \bibinfo
  {author} {\bibfnamefont {H.}~\bibnamefont {Su}}, \bibinfo {author}
  {\bibfnamefont {H.}~\bibnamefont {Wang}}, \bibinfo {author} {\bibfnamefont
  {X.}~\bibnamefont {Wang}}, \bibinfo {author} {\bibfnamefont {N.}~\bibnamefont
  {Yu}}, \bibinfo {author} {\bibfnamefont {Z.}~\bibnamefont {Zou}}, \bibinfo
  {author} {\bibfnamefont {H.}~\bibnamefont {Zhang}},\ and\ \bibinfo {author}
  {\bibfnamefont {Y.}~\bibnamefont {Guo}},\ }\bibfield  {title} {\bibinfo
  {title} {{Magnetism-induced ideal Weyl state in bulk van der Waals crystal
  ${\mathrm{MnSb}}_{2}{\mathrm{Te}}_{4}$}},\ }\href
  {https://doi.org/10.1063/5.0047438} {\bibfield  {journal} {\bibinfo
  {journal} {Appl. Phys. Lett.}\ }\textbf {\bibinfo {volume} {118}},\ \bibinfo
  {pages} {192105} (\bibinfo {year} {2021})}\BibitemShut {NoStop}%
\bibitem [{\citenamefont {Eremeev}\ \emph {et~al.}(2017)\citenamefont
  {Eremeev}, \citenamefont {Otrokov},\ and\ \citenamefont
  {Chulkov}}]{EREMEEV2017172}%
  \BibitemOpen
  \bibfield  {author} {\bibinfo {author} {\bibfnamefont {S.}~\bibnamefont
  {Eremeev}}, \bibinfo {author} {\bibfnamefont {M.}~\bibnamefont {Otrokov}},\
  and\ \bibinfo {author} {\bibfnamefont {E.}~\bibnamefont {Chulkov}},\
  }\bibfield  {title} {\bibinfo {title} {Competing rhombohedral and monoclinic
  crystal structures in ${\mathrm{mnpn}}_{2}{\mathrm{ch}}_{4}$ compounds: An
  ab-initio study},\ }\href
  {https://doi.org/https://doi.org/10.1016/j.jallcom.2017.03.121} {\bibfield
  {journal} {\bibinfo  {journal} {J. Alloy. Compd.}\ }\textbf {\bibinfo
  {volume} {709}},\ \bibinfo {pages} {172} (\bibinfo {year}
  {2017})}\BibitemShut {NoStop}%
\bibitem [{\citenamefont {Murakami}\ \emph {et~al.}(2019)\citenamefont
  {Murakami}, \citenamefont {Nambu}, \citenamefont {Koretsune}, \citenamefont
  {Xiangyu}, \citenamefont {Yamamoto}, \citenamefont {Brown},\ and\
  \citenamefont {Kageyama}}]{PhysRevB.100.195103}%
  \BibitemOpen
  \bibfield  {author} {\bibinfo {author} {\bibfnamefont {T.}~\bibnamefont
  {Murakami}}, \bibinfo {author} {\bibfnamefont {Y.}~\bibnamefont {Nambu}},
  \bibinfo {author} {\bibfnamefont {T.}~\bibnamefont {Koretsune}}, \bibinfo
  {author} {\bibfnamefont {G.}~\bibnamefont {Xiangyu}}, \bibinfo {author}
  {\bibfnamefont {T.}~\bibnamefont {Yamamoto}}, \bibinfo {author}
  {\bibfnamefont {C.~M.}\ \bibnamefont {Brown}},\ and\ \bibinfo {author}
  {\bibfnamefont {H.}~\bibnamefont {Kageyama}},\ }\bibfield  {title} {\bibinfo
  {title} {{Realization of interlayer ferromagnetic interaction in
  $\mathrm{MnS}{\mathrm{b}}_{2}\mathrm{T}{\mathrm{e}}_{4}$ toward the magnetic
  Weyl semimetal state}},\ }\href {https://doi.org/10.1103/PhysRevB.100.195103}
  {\bibfield  {journal} {\bibinfo  {journal} {Phys. Rev. B}\ }\textbf {\bibinfo
  {volume} {100}},\ \bibinfo {pages} {195103} (\bibinfo {year}
  {2019})}\BibitemShut {NoStop}%
\bibitem [{\citenamefont {Saxena}\ \emph {et~al.}(2020)\citenamefont {Saxena},
  \citenamefont {Rani}, \citenamefont {Nagpal}, \citenamefont {Patnaik},
  \citenamefont {Felner},\ and\ \citenamefont {Awana}}]{Saxena2020}%
  \BibitemOpen
  \bibfield  {author} {\bibinfo {author} {\bibfnamefont {A.}~\bibnamefont
  {Saxena}}, \bibinfo {author} {\bibfnamefont {P.}~\bibnamefont {Rani}},
  \bibinfo {author} {\bibfnamefont {V.}~\bibnamefont {Nagpal}}, \bibinfo
  {author} {\bibfnamefont {S.}~\bibnamefont {Patnaik}}, \bibinfo {author}
  {\bibfnamefont {I.}~\bibnamefont {Felner}},\ and\ \bibinfo {author}
  {\bibfnamefont {V.~P.}\ \bibnamefont {Awana}},\ }\bibfield  {title} {\bibinfo
  {title} {{Crystal Growth and Characterization of Possible New Magnetic
  Topological Insulators ${\mathrm{FeBi}}_{2}{\mathrm{Te}}_{4}$}},\ }\href
  {https://doi.org/10.1007/s10948-020-05531-0} {\bibfield  {journal} {\bibinfo
  {journal} {J. Supercond. Nov. Magn.}\ }\textbf {\bibinfo {volume} {33}},\
  \bibinfo {pages} {2251} (\bibinfo {year} {2020})},\ \Eprint
  {https://arxiv.org/abs/2004.13584} {arXiv:2004.13584} \BibitemShut {NoStop}%
\bibitem [{\citenamefont {Petrov}\ \emph {et~al.}(2021)\citenamefont {Petrov},
  \citenamefont {Ernst}, \citenamefont {Menshchikova},\ and\ \citenamefont
  {Chulkov}}]{doi:10.1021/acs.jpclett.1c02396}%
  \BibitemOpen
  \bibfield  {author} {\bibinfo {author} {\bibfnamefont {E.~K.}\ \bibnamefont
  {Petrov}}, \bibinfo {author} {\bibfnamefont {A.}~\bibnamefont {Ernst}},
  \bibinfo {author} {\bibfnamefont {T.~V.}\ \bibnamefont {Menshchikova}},\ and\
  \bibinfo {author} {\bibfnamefont {E.~V.}\ \bibnamefont {Chulkov}},\
  }\bibfield  {title} {\bibinfo {title} {Intrinsic magnetic topological
  insulator state induced by the jahn-teller effect},\ }\href
  {https://doi.org/10.1021/acs.jpclett.1c02396} {\bibfield  {journal} {\bibinfo
   {journal} {J. Phys. Chem. Lett.}\ }\textbf {\bibinfo {volume} {12}},\
  \bibinfo {pages} {9076} (\bibinfo {year} {2021})},\ \bibinfo {note} {pMID:
  34516740}\BibitemShut {NoStop}%
\bibitem [{\citenamefont {Kobia\l{}ka}\ \emph {et~al.}(2022)\citenamefont
  {Kobia\l{}ka}, \citenamefont {Sternik},\ and\ \citenamefont
  {Ptok}}]{PhysRevB.105.214304}%
  \BibitemOpen
  \bibfield  {author} {\bibinfo {author} {\bibfnamefont {A.}~\bibnamefont
  {Kobia\l{}ka}}, \bibinfo {author} {\bibfnamefont {M.}~\bibnamefont
  {Sternik}},\ and\ \bibinfo {author} {\bibfnamefont {A.}~\bibnamefont
  {Ptok}},\ }\bibfield  {title} {\bibinfo {title} {{Dynamical properties of the
  magnetic topological insulator $T{\mathrm{Bi}}_{2}{\mathrm{Te}}_{4}
  (T=\mathrm{Mn},\mathrm{Fe})$: Phonons dispersion, Raman active modes, and
  chiral phonons study}},\ }\href {https://doi.org/10.1103/PhysRevB.105.214304}
  {\bibfield  {journal} {\bibinfo  {journal} {Phys. Rev. B}\ }\textbf {\bibinfo
  {volume} {105}},\ \bibinfo {pages} {214304} (\bibinfo {year}
  {2022})}\BibitemShut {NoStop}%
\bibitem [{\citenamefont {Wang}\ \emph {et~al.}(2023)\citenamefont {Wang},
  \citenamefont {Zhao}, \citenamefont {Wu}, \citenamefont {Zhou}, \citenamefont
  {Li}, \citenamefont {Edmonds},\ and\ \citenamefont
  {Yang}}]{10.1088/1674-1056/acd522}%
  \BibitemOpen
  \bibfield  {author} {\bibinfo {author} {\bibfnamefont {Q.}~\bibnamefont
  {Wang}}, \bibinfo {author} {\bibfnamefont {J.}~\bibnamefont {Zhao}}, \bibinfo
  {author} {\bibfnamefont {W.}~\bibnamefont {Wu}}, \bibinfo {author}
  {\bibfnamefont {Y.}~\bibnamefont {Zhou}}, \bibinfo {author} {\bibfnamefont
  {Q.}~\bibnamefont {Li}}, \bibinfo {author} {\bibfnamefont {M.~T.}\
  \bibnamefont {Edmonds}},\ and\ \bibinfo {author} {\bibfnamefont {S.~A.}\
  \bibnamefont {Yang}},\ }\bibfield  {title} {\bibinfo {title} {Magnetic and
  electronic properties of bulk and two-dimensional febi2te4: A
  first-principles study},\ }\href
  {http://iopscience.iop.org/article/10.1088/1674-1056/acd522} {\bibfield
  {journal} {\bibinfo  {journal} {Chinese Physics B}\ } (\bibinfo {year}
  {2023})}\BibitemShut {NoStop}%
\bibitem [{\citenamefont {Kresse}\ and\ \citenamefont
  {Hafner}(1993)}]{PhysRevB.48.13115}%
  \BibitemOpen
  \bibfield  {author} {\bibinfo {author} {\bibfnamefont {G.}~\bibnamefont
  {Kresse}}\ and\ \bibinfo {author} {\bibfnamefont {J.}~\bibnamefont
  {Hafner}},\ }\bibfield  {title} {\bibinfo {title} {Ab initio molecular
  dynamics for open-shell transition metals},\ }\href
  {https://doi.org/10.1103/PhysRevB.48.13115} {\bibfield  {journal} {\bibinfo
  {journal} {Phys. Rev. B}\ }\textbf {\bibinfo {volume} {48}},\ \bibinfo
  {pages} {13115} (\bibinfo {year} {1993})}\BibitemShut {NoStop}%
\bibitem [{\citenamefont {Kresse}\ and\ \citenamefont
  {Furthm\"uller}(1996)}]{PhysRevB.54.11169}%
  \BibitemOpen
  \bibfield  {author} {\bibinfo {author} {\bibfnamefont {G.}~\bibnamefont
  {Kresse}}\ and\ \bibinfo {author} {\bibfnamefont {J.}~\bibnamefont
  {Furthm\"uller}},\ }\bibfield  {title} {\bibinfo {title} {Efficient iterative
  schemes for ab initio total-energy calculations using a plane-wave basis
  set},\ }\href {https://doi.org/10.1103/PhysRevB.54.11169} {\bibfield
  {journal} {\bibinfo  {journal} {Phys. Rev. B}\ }\textbf {\bibinfo {volume}
  {54}},\ \bibinfo {pages} {11169} (\bibinfo {year} {1996})}\BibitemShut
  {NoStop}%
\bibitem [{\citenamefont {Bl\"ochl}(1994)}]{PhysRevB.50.17953}%
  \BibitemOpen
  \bibfield  {author} {\bibinfo {author} {\bibfnamefont {P.~E.}\ \bibnamefont
  {Bl\"ochl}},\ }\bibfield  {title} {\bibinfo {title} {Projector augmented-wave
  method},\ }\href {https://doi.org/10.1103/PhysRevB.50.17953} {\bibfield
  {journal} {\bibinfo  {journal} {Phys. Rev. B}\ }\textbf {\bibinfo {volume}
  {50}},\ \bibinfo {pages} {17953} (\bibinfo {year} {1994})}\BibitemShut
  {NoStop}%
\bibitem [{\citenamefont {Perdew}\ \emph {et~al.}(1996)\citenamefont {Perdew},
  \citenamefont {Burke},\ and\ \citenamefont
  {Ernzerhof}}]{PhysRevLett.77.3865}%
  \BibitemOpen
  \bibfield  {author} {\bibinfo {author} {\bibfnamefont {J.~P.}\ \bibnamefont
  {Perdew}}, \bibinfo {author} {\bibfnamefont {K.}~\bibnamefont {Burke}},\ and\
  \bibinfo {author} {\bibfnamefont {M.}~\bibnamefont {Ernzerhof}},\ }\bibfield
  {title} {\bibinfo {title} {Generalized gradient approximation made simple},\
  }\href {https://doi.org/10.1103/PhysRevLett.77.3865} {\bibfield  {journal}
  {\bibinfo  {journal} {Phys. Rev. Lett.}\ }\textbf {\bibinfo {volume} {77}},\
  \bibinfo {pages} {3865} (\bibinfo {year} {1996})}\BibitemShut {NoStop}%
\bibitem [{\citenamefont {Grimme}\ \emph {et~al.}(2010)\citenamefont {Grimme},
  \citenamefont {Antony}, \citenamefont {Ehrlich},\ and\ \citenamefont
  {Krieg}}]{10.1063/1.3382344}%
  \BibitemOpen
  \bibfield  {author} {\bibinfo {author} {\bibfnamefont {S.}~\bibnamefont
  {Grimme}}, \bibinfo {author} {\bibfnamefont {J.}~\bibnamefont {Antony}},
  \bibinfo {author} {\bibfnamefont {S.}~\bibnamefont {Ehrlich}},\ and\ \bibinfo
  {author} {\bibfnamefont {H.}~\bibnamefont {Krieg}},\ }\bibfield  {title}
  {\bibinfo {title} {{A consistent and accurate ab initio parametrization of
  density functional dispersion correction (DFT-D) for the 94 elements H-Pu}},\
  }\href {https://doi.org/10.1063/1.3382344} {\bibfield  {journal} {\bibinfo
  {journal} {The Journal of Chemical Physics}\ }\textbf {\bibinfo {volume}
  {132}},\ \bibinfo {pages} {154104} (\bibinfo {year} {2010})}\BibitemShut
  {NoStop}%
\bibitem [{\citenamefont {Marzari}\ and\ \citenamefont
  {Vanderbilt}(1997)}]{PhysRevB.56.12847}%
  \BibitemOpen
  \bibfield  {author} {\bibinfo {author} {\bibfnamefont {N.}~\bibnamefont
  {Marzari}}\ and\ \bibinfo {author} {\bibfnamefont {D.}~\bibnamefont
  {Vanderbilt}},\ }\bibfield  {title} {\bibinfo {title} {Maximally localized
  generalized wannier functions for composite energy bands},\ }\href
  {https://doi.org/10.1103/PhysRevB.56.12847} {\bibfield  {journal} {\bibinfo
  {journal} {Phys. Rev. B}\ }\textbf {\bibinfo {volume} {56}},\ \bibinfo
  {pages} {12847} (\bibinfo {year} {1997})}\BibitemShut {NoStop}%
\bibitem [{\citenamefont {Souza}\ \emph {et~al.}(2001)\citenamefont {Souza},
  \citenamefont {Marzari},\ and\ \citenamefont
  {Vanderbilt}}]{souza2001maximally}%
  \BibitemOpen
  \bibfield  {author} {\bibinfo {author} {\bibfnamefont {I.}~\bibnamefont
  {Souza}}, \bibinfo {author} {\bibfnamefont {N.}~\bibnamefont {Marzari}},\
  and\ \bibinfo {author} {\bibfnamefont {D.}~\bibnamefont {Vanderbilt}},\
  }\bibfield  {title} {\bibinfo {title} {Maximally localized wannier functions
  for entangled energy bands},\ }\href
  {https://doi.org/10.1103/PhysRevB.65.035109} {\bibfield  {journal} {\bibinfo
  {journal} {Phys. Rev. B}\ }\textbf {\bibinfo {volume} {65}},\ \bibinfo
  {pages} {035109} (\bibinfo {year} {2001})}\BibitemShut {NoStop}%
\bibitem [{\citenamefont {Lee}\ and\ \citenamefont
  {Joannopoulos}(1981{\natexlab{a}})}]{lee1981simple1}%
  \BibitemOpen
  \bibfield  {author} {\bibinfo {author} {\bibfnamefont {D.}~\bibnamefont
  {Lee}}\ and\ \bibinfo {author} {\bibfnamefont {J.}~\bibnamefont
  {Joannopoulos}},\ }\bibfield  {title} {\bibinfo {title} {Simple scheme for
  surface-band calculations. i},\ }\href
  {https://doi.org/10.1103/PhysRevB.23.4988} {\bibfield  {journal} {\bibinfo
  {journal} {Phys. Rev. B}\ }\textbf {\bibinfo {volume} {23}},\ \bibinfo
  {pages} {4988} (\bibinfo {year} {1981}{\natexlab{a}})}\BibitemShut {NoStop}%
\bibitem [{\citenamefont {Lee}\ and\ \citenamefont
  {Joannopoulos}(1981{\natexlab{b}})}]{lee1981simple2}%
  \BibitemOpen
  \bibfield  {author} {\bibinfo {author} {\bibfnamefont {D.}~\bibnamefont
  {Lee}}\ and\ \bibinfo {author} {\bibfnamefont {J.}~\bibnamefont
  {Joannopoulos}},\ }\bibfield  {title} {\bibinfo {title} {Simple scheme for
  surface-band calculations. ii. the green's function},\ }\href
  {https://doi.org/10.1103/PhysRevB.23.4997} {\bibfield  {journal} {\bibinfo
  {journal} {Phys. Rev. B}\ }\textbf {\bibinfo {volume} {23}},\ \bibinfo
  {pages} {4997} (\bibinfo {year} {1981}{\natexlab{b}})}\BibitemShut {NoStop}%
\bibitem [{\citenamefont {Sancho}\ \emph {et~al.}(1984)\citenamefont {Sancho},
  \citenamefont {Sancho},\ and\ \citenamefont {Rubio}}]{sancho1984quick}%
  \BibitemOpen
  \bibfield  {author} {\bibinfo {author} {\bibfnamefont {M.~L.}\ \bibnamefont
  {Sancho}}, \bibinfo {author} {\bibfnamefont {J.~L.}\ \bibnamefont {Sancho}},\
  and\ \bibinfo {author} {\bibfnamefont {J.}~\bibnamefont {Rubio}},\ }\bibfield
   {title} {\bibinfo {title} {Quick iterative scheme for the calculation of
  transfer matrices: application to mo (100)},\ }\href
  {https://doi.org/10.1088/0305-4608/14/5/016} {\bibfield  {journal} {\bibinfo
  {journal} {J. Phys. F: Met. Phys.}\ }\textbf {\bibinfo {volume} {14}},\
  \bibinfo {pages} {1205} (\bibinfo {year} {1984})}\BibitemShut {NoStop}%
\bibitem [{\citenamefont {Sancho}\ \emph {et~al.}(1985)\citenamefont {Sancho},
  \citenamefont {Sancho}, \citenamefont {Sancho},\ and\ \citenamefont
  {Rubio}}]{sancho1985highly}%
  \BibitemOpen
  \bibfield  {author} {\bibinfo {author} {\bibfnamefont {M.~L.}\ \bibnamefont
  {Sancho}}, \bibinfo {author} {\bibfnamefont {J.~L.}\ \bibnamefont {Sancho}},
  \bibinfo {author} {\bibfnamefont {J.~L.}\ \bibnamefont {Sancho}},\ and\
  \bibinfo {author} {\bibfnamefont {J.}~\bibnamefont {Rubio}},\ }\bibfield
  {title} {\bibinfo {title} {Highly convergent schemes for the calculation of
  bulk and surface green functions},\ }\href
  {https://doi.org/10.1088/0305-4608/15/4/009} {\bibfield  {journal} {\bibinfo
  {journal} {J. Phys. F: Met. Phys.}\ }\textbf {\bibinfo {volume} {15}},\
  \bibinfo {pages} {851} (\bibinfo {year} {1985})}\BibitemShut {NoStop}%
\bibitem [{\citenamefont {Wu}\ \emph {et~al.}(2018)\citenamefont {Wu},
  \citenamefont {Zhang}, \citenamefont {Song}, \citenamefont {Troyer},\ and\
  \citenamefont {Soluyanov}}]{wu2018wanniertools}%
  \BibitemOpen
  \bibfield  {author} {\bibinfo {author} {\bibfnamefont {Q.}~\bibnamefont
  {Wu}}, \bibinfo {author} {\bibfnamefont {S.}~\bibnamefont {Zhang}}, \bibinfo
  {author} {\bibfnamefont {H.-F.}\ \bibnamefont {Song}}, \bibinfo {author}
  {\bibfnamefont {M.}~\bibnamefont {Troyer}},\ and\ \bibinfo {author}
  {\bibfnamefont {A.~A.}\ \bibnamefont {Soluyanov}},\ }\bibfield  {title}
  {\bibinfo {title} {Wanniertools: An open-source software package for novel
  topological materials},\ }\href {https://doi.org/10.1016/j.cpc.2017.09.033}
  {\bibfield  {journal} {\bibinfo  {journal} {Comput. Phys. Commun.}\ }\textbf
  {\bibinfo {volume} {224}},\ \bibinfo {pages} {405} (\bibinfo {year}
  {2018})}\BibitemShut {NoStop}%
\bibitem [{\citenamefont {Gao}\ \emph {et~al.}(2021{\natexlab{b}})\citenamefont
  {Gao}, \citenamefont {Wu}, \citenamefont {Persson},\ and\ \citenamefont
  {Wang}}]{GAO2021107760}%
  \BibitemOpen
  \bibfield  {author} {\bibinfo {author} {\bibfnamefont {J.}~\bibnamefont
  {Gao}}, \bibinfo {author} {\bibfnamefont {Q.}~\bibnamefont {Wu}}, \bibinfo
  {author} {\bibfnamefont {C.}~\bibnamefont {Persson}},\ and\ \bibinfo {author}
  {\bibfnamefont {Z.}~\bibnamefont {Wang}},\ }\bibfield  {title} {\bibinfo
  {title} {Irvsp: To obtain irreducible representations of electronic states in
  the vasp},\ }\href
  {https://doi.org/https://doi.org/10.1016/j.cpc.2020.107760} {\bibfield
  {journal} {\bibinfo  {journal} {Comput. Phys. Commun.}\ }\textbf {\bibinfo
  {volume} {261}},\ \bibinfo {pages} {107760} (\bibinfo {year}
  {2021}{\natexlab{b}})}\BibitemShut {NoStop}%
\bibitem [{\citenamefont {Gao}\ \emph {et~al.}(2022)\citenamefont {Gao},
  \citenamefont {Guo}, \citenamefont {Weng},\ and\ \citenamefont
  {Wang}}]{PhysRevB.106.035150}%
  \BibitemOpen
  \bibfield  {author} {\bibinfo {author} {\bibfnamefont {J.}~\bibnamefont
  {Gao}}, \bibinfo {author} {\bibfnamefont {Z.}~\bibnamefont {Guo}}, \bibinfo
  {author} {\bibfnamefont {H.}~\bibnamefont {Weng}},\ and\ \bibinfo {author}
  {\bibfnamefont {Z.}~\bibnamefont {Wang}},\ }\bibfield  {title} {\bibinfo
  {title} {Magnetic band representations, fu-kane-like symmetry indicators, and
  magnetic topological materials},\ }\href
  {https://doi.org/10.1103/PhysRevB.106.035150} {\bibfield  {journal} {\bibinfo
   {journal} {Phys. Rev. B}\ }\textbf {\bibinfo {volume} {106}},\ \bibinfo
  {pages} {035150} (\bibinfo {year} {2022})}\BibitemShut {NoStop}%
\bibitem [{\citenamefont {Soluyanov}\ \emph {et~al.}(2015)\citenamefont
  {Soluyanov}, \citenamefont {Gresch}, \citenamefont {Wang}, \citenamefont
  {Wu}, \citenamefont {Troyer}, \citenamefont {Dai},\ and\ \citenamefont
  {Bernevig}}]{Soluyanov2015}%
  \BibitemOpen
  \bibfield  {author} {\bibinfo {author} {\bibfnamefont {A.~A.}\ \bibnamefont
  {Soluyanov}}, \bibinfo {author} {\bibfnamefont {D.}~\bibnamefont {Gresch}},
  \bibinfo {author} {\bibfnamefont {Z.}~\bibnamefont {Wang}}, \bibinfo {author}
  {\bibfnamefont {Q.}~\bibnamefont {Wu}}, \bibinfo {author} {\bibfnamefont
  {M.}~\bibnamefont {Troyer}}, \bibinfo {author} {\bibfnamefont
  {X.}~\bibnamefont {Dai}},\ and\ \bibinfo {author} {\bibfnamefont {B.~A.}\
  \bibnamefont {Bernevig}},\ }\bibfield  {title} {\bibinfo {title} {{Type-II
  Weyl semimetals}},\ }\href {https://doi.org/10.1038/nature15768} {\bibfield
  {journal} {\bibinfo  {journal} {Nature}\ }\textbf {\bibinfo {volume} {527}},\
  \bibinfo {pages} {495} (\bibinfo {year} {2015})},\ \Eprint
  {https://arxiv.org/abs/1507.01603} {arXiv:1507.01603} \BibitemShut {NoStop}%
\bibitem [{\citenamefont {Zhang}\ \emph {et~al.}(2009)\citenamefont {Zhang},
  \citenamefont {Liu}, \citenamefont {Qi}, \citenamefont {Dai}, \citenamefont
  {Fang},\ and\ \citenamefont {Zhang}}]{TB1}%
  \BibitemOpen
  \bibfield  {author} {\bibinfo {author} {\bibfnamefont {H.}~\bibnamefont
  {Zhang}}, \bibinfo {author} {\bibfnamefont {C.-X.}\ \bibnamefont {Liu}},
  \bibinfo {author} {\bibfnamefont {X.-L.}\ \bibnamefont {Qi}}, \bibinfo
  {author} {\bibfnamefont {X.}~\bibnamefont {Dai}}, \bibinfo {author}
  {\bibfnamefont {Z.}~\bibnamefont {Fang}},\ and\ \bibinfo {author}
  {\bibfnamefont {S.-C.}\ \bibnamefont {Zhang}},\ }\bibfield  {title} {\bibinfo
  {title} {{Topological insulators in $Bi_{2}Se_{3}$, $Bi_{2}Te_{3}$ and
  $Sb_{2}Te_{3}$ with a single Dirac cone on the surface}},\ }\href
  {https://doi.org/10.1038/nphys1270} {\bibfield  {journal} {\bibinfo
  {journal} {Nat. Phys.}\ }\textbf {\bibinfo {volume} {5}},\ \bibinfo {pages}
  {438} (\bibinfo {year} {2009})}\BibitemShut {NoStop}%
\end{thebibliography}%


\end{document}